\documentclass[letterpaper,twocolumn,10pt]{article}
\usepackage{usenix2019_v3.1}

\usepackage[english]{babel}
\usepackage{blindtext}

\usepackage{hyperref}

\usepackage{color}
\usepackage{tikz}
\usepackage[tight]{subfigure}
\usepackage{comment}
\usepackage{amsmath}
\usepackage{enumerate}
\usepackage{multirow}
\usepackage{hhline}

\usepackage{enumerate}
\usepackage{enumitem}

\usepackage{slashbox}

\usepackage{amsmath}
\usepackage{xspace}
\usepackage{algorithm}
\usepackage[font=small]{caption}
\usepackage[noend]{algpseudocode}
\usepackage[capitalize]{cleveref}
\makeatletter
\def\BState{\State\hskip-\ALG@thistlm}
\makeatother

\algdef{SE}[SUBALG]{Indent}{EndIndent}{}{\algorithmicend\ }%
\algtext*{Indent}
\algtext*{EndIndent}
\algrenewcommand\algorithmicindent{1.0em}%

\newif\ifcomm
\ifcomm
\else
\commfalse
\fi
\ifcomm
	\newcommand{\mycomm}[3]{{\footnotesize{{\color{#2} \textbf{[#1: #3]}}}}}
	\newcommand{\CRdel}[1]{\textcolor{red}{\sout{#1}}}
        
        \newcommand{\revise}[1]{{\color{blue}{#1}}}
\else
    \newcommand{\mycomm}[3]{}
    \newcommand{\CRdel}[1]{}
    
    \newcommand{\revise}[1]{\color{black}{#1}}
\fi

\newcommand{\MM}[1]{\mycomm{MM}{red}{#1}} 
 
\newcommand{\RBB}[1]{\mycomm{Ran}{green}{#1}} 
\newcommand{\ran}[1]{\RBB{#1}}
 
\newcommand{\minlan}[1]{\mycomm{Minlan}{green}{#1}}
\newcommand{\Gcomment}[1]{\mycomm{Gregory}{red}{#1}}

\newcommand{\Wenchen}[1]{\mycomm{Wenchen}{Wenchen}{#1}}

\definecolor{Wenchen}{RGB}{200,0,200}

\hypersetup{pdfstartview=FitH,pdfpagelayout=SinglePage}

\newcommand{\subp}{\noindent\textbf}

\newcommand{\ie}{\textit{i.e.}}
\newcommand{\eg}{\textit{e.g.}}

\newcommand{\sysname}{FRANCIS\xspace}

\newcommand{\langname}{$\mu$-FRANCIS\xspace}
\renewcommand{\langname}{\sysname}

\renewcommand{\eg}{e.g.}

\begin{document}



\title{FRANCIS:  
Fast Reaction Algorithms for Network Coordination In Switches}

\newif\ifauthor
\authortrue
\ifauthor
\else
\authorfalse
\fi

\ifauthor
\author{
{\rm Wenchen Han}\\
Peking University
\and
{\rm Vic Feng}\\
Harvard University
 \and
 {\rm Gregory Schwartzman}\\
JAIST
 \and
 {\rm Yuliang Li} \\
 Google
 \and
 {\rm Michael Mitzenmacher}\\
Harvard University
 \and
 {\rm Minlan Yu}\\
Harvard University
 \and
 {\rm Ran Ben-Basat}\\
University College London
} 
\else
\author{\color{red}{Paper under review at NSDI '25. Do not distribute.}}
\fi










\maketitle

\begin{abstract}

Optimizing the reaction to network events, which is critical in tasks such as clock synchronization, multicast, and routing, becomes increasingly challenging as networks grow larger.
To improve the reaction time compared to centralized solutions, the theory community has made significant progress in the design of \textit{message-passing} algorithms that leverage all nodes for distributed computation, and the advent of programmable switches makes it now possible to materialize them.

We propose FRANCIS, a framework and associated libraries for running message-passing algorithms on programmable switches.
It features primitives that allow easy integration of such algorithms for quickly reacting to network events while optimizing resource consumption.
We use FRANCIS to implement event reaction solutions that improve clock synchronization, source-routed multicast, and routing and demonstrate up to $18\times$ reduction in reaction time.
\end{abstract}


\vspace{-0.1cm}
\section{Introduction}
\vspace{-0.1cm}

Quick reaction to network events is critical to ensure the performance of many applications such as distributed transactional databases~\cite{corbett2013spanner, shamis2019fast} and multicast-based machine learning~\cite{li2023gleam}. 
For such reactions, these applications commonly rely on lower-level distributed networking protocols such as spanning tree~\cite{rfc4318}, shortest path routing~\cite{10.17487/RFC1058,rfc2328}, clock synchronization~\cite{rfc5905}, and networked consensus protocols~\cite{rfc7787}. 
Such protocols must be optimized as networks grow larger (e.g., connecting 100K servers) and faster (e.g., with a few microseconds RTT~\cite{lowRTT}) and as applications require ultra {low latency (e.g., distributed transactional database~\cite{corbett2013spanner, shamis2019fast}, memcached~\cite{fitzpatrick2004distributed}, or
RAMCloud~\cite{ousterhout2010case,ousterhout2015ramcloud}).
Compared to centralized SDN approaches~\cite{hong2018b4, jain2013b4}, distributed algorithms inherently adapt more quickly to transient network state changes such as link failures and short traffic bursts.


To leverage the distributed nature of networks, the theoretical computer science community has extensively studied the design of distributed message-passing algorithms for a variety of fundamental problems, including those mentioned above. Here, by message-passing algorithms, we mean 
a computational approach where nodes exchange short messages with their neighbors to coordinate and solve problems in a decentralized manner. This model, in which the goal is to minimize the number of messaging rounds, is called the CONGEST model.
Many core issues to the networking community have been explored within this framework, such as spanning tree protocols~\cite{10.1145/3380546,pandurangan2019time}, shortest paths computation~\cite{agarwal2020faster,chechik2020single}, consensus~\cite{pease1980reaching}, and graph queries~\cite{fischer2018possibilities,korhonen2018deterministic}.
However, the intricate nature of these solutions, combined with the constant emergence of new advancements, has made them challenging to implement on traditional hardware switches.

The recent emergence of programmable switches~\cite{Tofino, broadcom} provides a natural platform for deploying message-passing algorithms for low-latency distributed systems. Today, the data plane of programmable switches can provide nanosecond-level packet processing time and terabit-level bandwidth.
Moreover, switches are first to observe network events 
such as link failure or load imbalance and are thus best placed to provide fast reaction for \mbox{network protocols~\cite{DBLP:journals/ton/ChiesaSABKNS21,DBLP:conf/nsdi/HolterbachMADVV19}.}

In this paper, we present a general framework for distributed reaction to network events, which we dub FRANCIS (Fast Reaction Algorithms for Network
Coordination In Switches). \sysname enables easier conversion of current and future theoretical results in message-passing algorithms into viable real-world distributed implementations on networked switches. 
It introduces the following three-phase paradigm of distributed reaction. 
First, the switches detect a network event in the data plane, initiating the reaction protocol. 
Next, a fast recovery phase is triggered, in which we run distributed message-passing algorithms on the switches to restore connectivity and functionality.
Finally, we distributedly optimize the solution to bring our system back to a stable state that provides better efficiency. 

\sysname makes the following contributions towards switch implementations of message-passing algorithms. First, we establish a framework (Section~\ref{sec:distributed-framework}) with algorithmic abstractions and a library of primitives to allow easier specification of reaction algorithms. Second, we provide an execution engine (Section~\ref{sec:execution-engine}) that facilitates implementation on switches by overcoming hardware limitations, handling network unreliability, managing resource constraints, and introducing optimizations to speed up reactions.

%
%


%
We demonstrate the benefit of \sysname in the context of clock synchronization, source-routed multicast, and routing, by implementing Tofino~\cite{Tofino} and simulation prototypes for each use case respectively.
For clock synchronization, we compare with Sundial~\cite{Sundial} and PTP~\cite{PTP}, and demonstrate how \sysname prototype quickly recovers from multiple failures, improves the time uncertainty bound by up to $9.3\times$, and lowers the clock drift. For multicast, compared to Elmo~\cite{Elmo} and Orca~\cite{orca}, our prototype uses packet headers that are smaller by up to $52.0\%$ and achieves up to $18\times$ faster reaction to network events. For routing, we adapt Contra using \sysname, achieving approximately $10\times$ \mbox{faster reaction speed than the original Contra protocol.}

The rest of the paper is organized as follows. Section~\ref{sec:background} describes \sysname's objectives for network event reaction, surveys the most relevant works, and justifies our choice to focus on synchronous algorithms. Section~\ref{sec:distributed-framework} explains our three-phase reaction framework and how \sysname's 
primitives allow modular and easy algorithmic specification of a reaction scheme. Section~\ref{sec:execution-engine} introduces \sysname's message-passing execution engine that facilitates implementation on programmable switches. Section~\ref{sec:use_cases} showcases how \sysname is deployed in real-world examples. Sections~\ref{sec:implementation} and \ref{sec:eval} cover implementation details and evaluation of \sysname. We then provide an extended description of related work in Section~\ref{sec:related} and discuss future work in Section~\ref{sec:discussion}.
Finally, many more details and discussions appear in the appendices.

\vspace{-1mm}
\section{Background and motivation}
\vspace{-1mm}
\label{sec:background}
In modern data centers, with faster and larger networks~\cite{Jupiter} and greater demand for low latency for distributed user applications, it becomes increasingly crucial to design a system that reacts efficiently to network events such as (possibly multiple concurrent) link and switch failures, packet loss, sudden increases in link utilization, connections to new switches, etc.
%


\vspace{-1mm}
\subsection{\mbox{Objectives for reacting to network events}}
\label{subsec:bg-objective}

We describe \sysname's general objectives for reacting to network events: fast reaction, low resource overhead, and resiliency to concurrent events. 

\vspace{0.05in}
\subp{Fast reaction.} 
%
Because many applications have little tolerance for even a short period of disturbance or disconnection caused by a network event, fast reaction and recovery are critical.
Events such as link failures can happen as frequently as once every few minutes in a large-scale data center network~\cite{gill2011understanding, net-fail-study}. These and other more transient events, such as traffic bursts, are becoming more frequent as networks grow larger~\cite{Jupiter}, increasing the need for fast reactions. We justify the importance of fast reaction with the following examples (which are expanded in Section~\ref{sec:use_cases}).

\smallskip
\noindent\textit{Clock synchronization}. Clock synchronization~\cite{PTP, Sundial,geng2018exploiting, dptp, yu2022orbweaver} is a key component for applications such as distributed transactional databases~\cite{corbett2013spanner, shamis2019fast}. These applications require the synchronization system to provide accurate timestamps to minimize their transaction delays. 
Failures which is recovered slowly cause severe drifts $\Delta t$ between clocks, and accordingly, as transactions must wait for $\Delta t$ to guarantee consistency~\cite{corbett2013spanner}, such drifts lead to degraded 
transaction delays.

\smallskip
\noindent\textit{Multicast}. Multicast~\cite{scaling-ip-multicast, ip-multicast-rfc, Elmo, orca} is a common one-to-many transmission pattern that is widely deployed.
As the number of multicast groups (rules) can be orders of magnitude larger than the number of links~\cite{Elmo}, a switch or link failure can affect the connectivity of many groups. Applications like distributed machine learning systems~\cite{li2023gleam} running on top of these failed groups can suffer from degraded performance during the reaction time, which in turn leads to increased training costs and longer training time.

\smallskip
\subp{Low resource overhead.} When reacting to network events, whether operating in the data plane or in the control plane, the message overhead generated by the distributed protocols should be minimized to reduce congestion. Further, as switches need to support other operations, it is also essential to optimize the amount of other resources such as the memory and compute operations.

\smallskip
\subp{Resiliency to concurrent events.} Network events are sometimes coupled with each other, in which case a single incident takes down multiple devices or links~\cite{gill2011understanding}. This poses the challenge of being resilient to handling multiple concurrent events, ideally without adding assumptions to the pattern of such concurrent events and being applicable to any topology~\cite{fat-tree, Jupiter, Jellyfish}. We note that concurrent events of an incident can happen asynchronously, \ie, new events happen while \sysname reacts to earlier events of the same incident.

\vspace*{-1mm}
\subsection{Existing network event reaction systems}
\vspace*{-1mm}
\label{subsec:prior-solutions}
We now summarize previous works on reacting to network events and point out their limitations on why they fail to fulfill our objectives in Section~\ref{subsec:bg-objective}.

\vspace{0.05in}
\subp{SDN-based methods.}  A common practice is to compute reaction configurations to network events in the centralized~\cite{google-control} SDN control plane. While centralized methods bring benefits such as separated management and optimized configuration solutions, the \textit{reaction time} is orders of magnitude slower than distributed methods~\cite{dSDN, Sundial, hsu2020contra}. The bottlenecks as identified in~\cite{dSDN} include time to gather and transmit relevant information for all switches through out-of-band networks, and to install new rules at the switch level.
For example, for controller-based network event reaction solutions such as Sundial~\cite{Sundial} in clock synchronization and Elmo~\cite{Elmo} in source-routed multicast, it is reported to require 10s to 100s of milliseconds in data center networks, unacceptable for many applications.

Decentralized SDN-based solution~\cite{dSDN} seems promising in reducing the reaction time in WAN. While \sysname's framework could also be extended to running message-passing in switches' control-plane (CP), we argue that data-plane (DP) message-passing achieves much lower latency. The high latency for distributed CP~\cite{zeng2022tiara} mainly comes from low processing speed of switches' CPU and extra delay in delivering packets from DP to CP ($100K$ops and $\sim 10\mu$s for Tofino~\cite{Tofino}).

\vspace{0.05in}
\subp{Backup plan based solutions.} Other works propose to handle network events via a backup (failover) plan computed by an SDN controller and installed on the switches. When an event happens, each impacted switch executes the backup plan directly at the data plane, thus optimizing the reaction time. However, such solutions are typically not resilient to multiple concurrent events, as the large number of event combinations requires an excessive amount of time to compute and excessive memory at the switches to encode each plan.
For example, Sundial~\cite{Sundial} for clock synchronization proposes to react to a \textit{single} failure by using the backup plan to form a new synchronization tree that replaces the affected one. 
However, \textit{when a second failure happens}, this may require falling back to a significantly slower controller-based solution to compute a new backup plan. Other solutions using backup plans, such as F10's rerouting protocol~\cite{liu2013f10}, also require additional assumptions, such as having a Clos topology and falling back to the controller if there is \textit{more than one failure}.

\vspace{0.05in}
\subp{Periodically run solutions.} Another line of work includes solutions that execute periodically, regardless of whether there is a network event or not.  Essentially, these solutions are event-agnostic and thus commonly have \textit{higher message overheads}.
Contra~\cite{hsu2020contra} (which generalizes a distance-vector routing protocol for unicast to support custom performance metrics) and PTP~\cite{PTP} (which runs the BMCA algorithm in distributed control plane for clock synchronization tree construction) are two examples. Such protocols are generally inefficient as running the algorithm periodically wastes bandwidth if there is no network event.

\vspace{-6mm}
\subsection{Synchronous versus asynchronous}\label{subsec:why-synchronous-algorithms}
\vspace{-1mm}



Message-passing algorithms can be either synchronous or asynchronous.  \sysname is mainly designed and optimized for synchronous algorithms; these are generally easier to design and optimize, as we next explain.


For synchronous message-passing algorithms, communication takes place over rounds. At the beginning of each round, each node receives all messages sent by its neighbors in the previous round. Next, each node can perform some computation and update its local state. At the end of the round, each node can send messages to its neighbors (potentially a different message for each neighbor), which are delivered at the next round. This framework has been formalized theoretically in, e.g., the CONGEST model~\cite{Linial92, peleg2000distributed}, where the packet size is restricted to be $O(\log n)$ bits where $n$ is the number of network devices (e.g., switches). 
Synchrony can be achieved by adding synchronizers~\cite{Awerbuch85} to an \mbox{asynchronous network, as we elaborate on in Section \ref{subsec:backend-synchronization}.}

In contrast, asynchronous algorithms act on a per-message basis: incoming messages trigger updates in the switch, which performs some computation and may send new messages to one or more neighbors. As mentioned, Contra~\cite{hsu2020contra} is essentially an asynchronous protocol where a switch's shortest-path entries are updated upon receiving probe messages, and the resulting messages may {be sent to neighbors for \mbox{further updates.}}

We motivate \sysname's focus on supporting synchronous protocols by comparing them in terms of convergence time, bandwidth usage, and other properties.
A potential benefit of asynchronous algorithms is that they can send messages with updated information immediately after a message arrives, without waiting until the subsequent communication round. 
Synchronous protocols have to wait for all messages in a round to arrive, potentially leading to straggler issues that may increase convergence time e.g., due to packet loss or local computations. We expect this to be less of an issue in practice, as \sysname is meant for a data center setting where loss is minimal and uses lightweight local computations.  
On the other hand, asynchronous protocols typically use additional bandwidth by sending intermediate values that may not have been necessary to send because of later received information.  With restrictions on the bandwidth usage, this effect can lead to longer convergence time. (Of course, one can reduce wasted bandwidth by introducing waiting to make an asynchronous algorithm more like a synchronous one.)
For example, we adapt Contra's asynchronous routing algorithm into a synchronous version, and our results show that our basic synchronous version converges $1.87\times$ faster (Table~\ref{tab:routing-performance}).

We additionally argue that synchrony allows for easier protocol design and optimization. 
Specifically, the round-based granularity of synchronous distributed algorithms allows us to easily build high-level algorithmic abstractions and primitives that simplify users' design of their network event reaction algorithms (see Section~\ref{sec:distributed-framework}). Similarly, round-based granularity appears more amenable to certain optimizations in our framework.
One example is message packing, which takes advantage of the fact that we may have multiple algorithm instances running the same algorithm in parallel, so that instances can be synchronized together to pack many messages into a single packet per port, as \mbox{we describe in detail in Section~\ref{subsubsec:efficient-communication}. }


\begin{table*}
\centering
\resizebox{1.00099\linewidth}{!}{
\begin{tabular}{|l|ll|l|}
\hline
\textbf{Phase}
& \multicolumn{1}{l|}{\textbf{Clock synchronization}}                                                                                                                                                        & \textbf{Source-routed multicast} &  \textbf{Routing}                                                                                                                                                                                         \\ \hline
\textbf{Detection}                                                                 & \multicolumn{1}{c|}{\begin{tabular}[c]{@{}c@{}} Passive, based on the clock synchronization messages \end{tabular}} & \multicolumn{1}{l|}{\begin{tabular}[l]{@{}l@{}} Active \end{tabular}} & Active\\ \hline
\multirow{2}{*}{\textbf{\begin{tabular}[c]{@{}l@{}}Fast \\ recovery\end{tabular}}} &
\multirow{2}{*}{\begin{tabular}[c]{@{}l@{}}Construct a flooding spanning tree \\ and initializing the synchronizer.\end{tabular}} &  \multicolumn{1}{|l|}{Flooding spanning tree construction and initialization. Then}  &\multirow{2}{*}{\begin{tabular}[c]{@{}l@{}}Run F10~\cite{liu2013f10}'s Pushback Flow Redirection algorithm \\ to find new routes for affected flows efficiently.\end{tabular}}\\ 
&
\multicolumn{1}{l|}{}    & 
Synchronously build a multicast subtree for each group. &  \\ \hline
\textbf{\begin{tabular}[c]{@{}l@{}}Distributed\\ optimization\end{tabular}}        & \multicolumn{1}{l|}{\begin{tabular}[c]{@{}l@{}}Run BFS from multiple roots. For each tree,\\ calculate its center and the depth if we select\\ it as the root. Elect the min-depth tree.\end{tabular}} & \begin{tabular}[c]{@{}l@{}}Construct a new graph spanner, and build  a shortest-path\\ tree on the spanner for each group. Use greedy set-cover\\ to minimize the number of edges in each tree. (Note that\\ this is possible as the tree does not span the whole graph.)\end{tabular} & \multicolumn{1}{l|}{\begin{tabular}[c]{@{}l@{}} Run a synchronous Contra version. In each \\ round, wait for the neighbors in the product \\ graph to send updates. Forward the updated \\entries to neighbors at the end of the round.\end{tabular}}  \\ \hline
\end{tabular}
}
\vspace{-0.105in}
\caption{The three reaction phases of our three use cases.}\label{tbl:frameworkPhases}
\vspace{-0.18in}
\end{table*}

\vspace{-0.07in}
\section{The \sysname Framework} \label{sec:distributed-framework}
\vspace{-0.11cm}
\revise{This section presents the conceptual and algorithmic contributions of \sysname towards our proposed objectives in Section~\ref{subsec:bg-objective}. Section~\ref{subsec: overview} introduces the three-phase distributed reaction framework. Specifically, \sysname's detector and its mechanism of reinvoking reaction address handling \textit{concurrent events} with any temporal and topological patterns; the fast recovery phase prioritizes \textit{fast reaction} using more efficient message-passing algorithms. Section~\ref{subsec:designing-reactions} elaborates the algorithmic abstractions and primitives to facilitate specifying message-passing reactions. It also enables explicit specification of resource constraints to achieve low resource overhead.
}

\begin{figure}
	\centering
	\vspace*{-0.mm}
	\includegraphics[width=\linewidth]{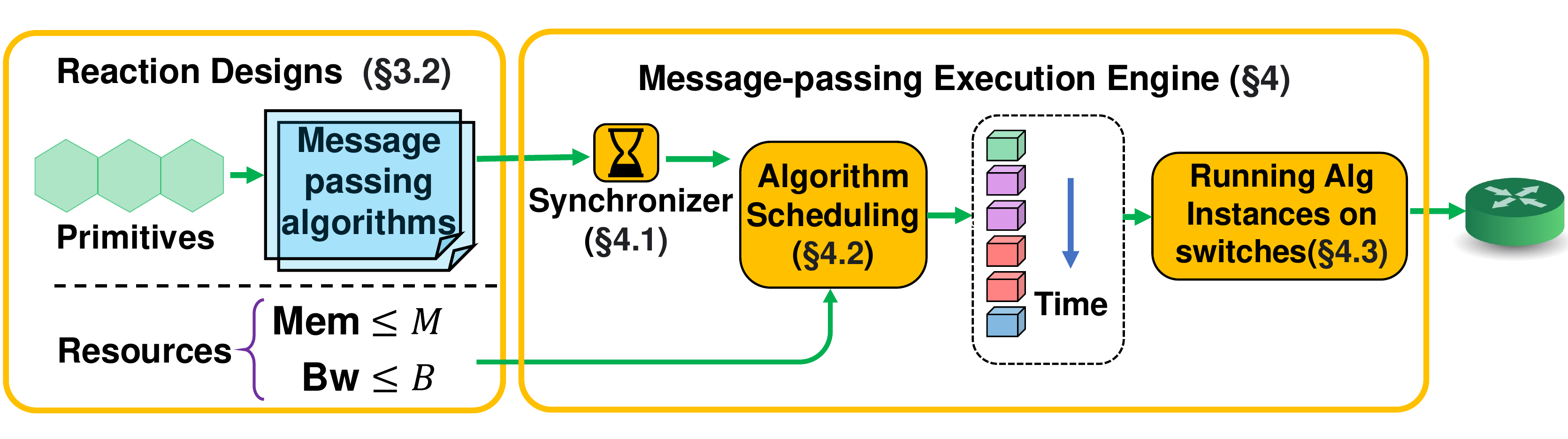}
	\vspace{-8mm}
	\caption{An overview of \sysname. Algorithm scheduling is a module that generates \textit{at runtime} the schedule to perform the next computation or communication actions (shown as cubes).} 
	\vspace{-0.2mm}
	\label{fig:overview-architecture}
\end{figure}


\vspace{-0.1in}
\subsection{Three phase distributed reaction}\label{subsec: overview}
\vspace{-0.0in}

\revise{
We now describe our detection phase and justify why we need both fast recovery and distributed optimization for reaction. The high-level description of our reaction phases for our use cases is summarized in Table~\ref{tbl:frameworkPhases}.
}

\smallskip
\subp{Detection.} Our detector listens to events such as failures and congested links through timer-based approaches, which periodically exchange messages between neighboring switches and report an event upon a timeout. \sysname works with different protocols~\cite{802.1ag-2007, BFD, DP83867IR, Sundial}, including passive detection (which listens to some existing periodic traffic, as in~\cite{Sundial}) and active detection (which generates probe packets, as in ~\cite{BFD}). 

\smallskip
\noindent\textit{Handling multiple concurrent events.} 
One assumption that simplifies the design of distributed reaction algorithms is to ensure a static network, \ie, no events or failures occur during the execution of the reaction algorithms. This does not hold in a real network, and we thus propose to \textit{reinvoke} the reaction routine from scratch upon new events and invalidate previous runs. This way, the single validated run of reactions always operates on the static network state after the latest network event, where the assumption still holds. \sysname's detector achieves this by listening to any new events and reinvoking a new run of reactions upon a new event.


\smallskip
\subp{Fast recovery.} 
To achieve fast reaction, we prioritize restoring basic functionality or connectivity quickly, without necessarily finding optimized solutions or network configurations. The reaction algorithm could be an approximation or heuristic algorithm that converges quickly. For example, if a link that is a part of a clock synchronization tree goes down, we compute a new spanning tree for synchronization quickly through flooding, even though the flooding tree is usually suboptimal and can lead to a high time uncertainty bound. 

\smallskip
\subp{Distributed optimization.} 
After restoring functionality, \sysname aims to further improve the quality of the solution for better long-term performance of applications. 
For example, 
as the clock drift is proportional to the synchronization tree depth,
we run the algorithm listed in Table~\ref{tbl:frameworkPhases} to improve it. The distributed optimization needs not to yield an optimal solution; if desired, a later step may apply \mbox{further centralized optimization of the solution.}

\vspace{-0.1in}
\subsection{Designing \sysname reactions}
\vspace{-0.1cm}
\label{subsec:designing-reactions}

A synchronous message-passing reaction scheme includes two basic components: the specification of message-passing algorithms for reaction and their resource constraints. Here, we assume that synchrony is guaranteed by our synchronizer module (as detailed in Section~\ref{subsec:backend-synchronization}).

The specification of a message-passing algorithm includes communications with neighbors and switch-local computations. For communication, the switch sends up to one message per neighbor for each round; for computation, the switch processes incoming messages either in a streaming or batched fashion. 
With \textit{streaming}, the switch updates its state upon the arrival of each message. However, some algorithms are not amenable to streaming computation. In that case, we can use \textit{batching}, where we store messages in memory and perform computations upon receiving all messages from neighbors. The downside of batching is the need for recirculation to process messages from all neighbors, which increases the total reaction time. {We elaborate on the two options in Appendix~\ref{sec:pseudocodes}.}

We note that a single message-passing algorithm may correspond to the simultaneous execution of multiple algorithm instances. For example, in multicast, we run the same (but many instances of) multicast tree recovery algorithm for each affected multicast group. As discussed in the following section, \sysname makes parallel execution of multiple instances feasible, which not only reduces the reaction time but also allows further optimizations such as message packing. 




To allow simpler development of distributed reaction algorithms, we provide a library of some commonly used message-passing algorithm primitives, as summarized in~\cref{tbl:primitive-calls}. Each of them is useful for one or more \sysname's use cases in Section~\ref{sec:use_cases}. For example, \texttt{shortest path tree} serves as a single-source-multi-destination communication structure that minimizes the latency to every destination, which is useful for tasks such as clock synchronization and multicast. 
\texttt{Asynchronous global flooding} runs the communication of flooding to broadcast messages and can optionally bind with functions on top such as running the synchronizer's initialization across the networks and constructing a (flooding) tree with minimum time; it is useful for all three use cases.

\begin{table}
\centering
\resizebox{1.\linewidth}{!}{
\begin{tabular}{|l|l|}
\hline

\textbf{Algorithmic primitives} & \textbf{Description}\\ \hline

\begin{tabular}[]{@{}l@{}} \textbf{\texttt{Asynchronous global flooding}} \\ (Asynchronous extension)\end{tabular} & \begin{tabular}[c]{@{}l@{}} Use flooding to 1) broadcast a short message to all switches \\ 2) construct a flooding spanning tree, \\ 3) initializing the synchronizer throughout the network.
\end{tabular} \\ \hline

\begin{tabular}[]{@{}l@{}} \textbf{\texttt{Bottom-up tree aggregation}} \end{tabular}  & \begin{tabular}[c]{@{}l@{}} Aggregate values from the leaves of a spanning tree \\ to the  root. \end{tabular}  \\ \hline

\begin{tabular}[]{@{}l@{}} \textbf{\texttt{Conditional broadcast}} \\ \textbf{\texttt{and aggregation}} (as \\ detailed in Algorithm~\ref{alg:cond-broadcast-agg}) \end{tabular} & \begin{tabular}[c]{@{}l@{}} Maintain a single value $v$ for each switch. Aggregate $v$ \\ with those received from neighbors for each round. Then \\ broadcast the updated value next round provided that \\ the condition function is met. \end{tabular}  \\ \hline

\begin{tabular}[]{@{}l@{}} \textbf{\texttt{Shortest path tree}}\end{tabular} & \begin{tabular}[c]{@{}l@{}} Construct a spanning tree layer-by-layer in a BFS fashion. \\ Each round the current leaf switches look for unvisited \\ switches to be added as a new layer of leaf switches. \end{tabular} \\ \hline

\begin{tabular}[]{@{}l@{}} \textbf{\texttt{Set cover on a shortest}} \\ \textbf{\texttt{path graph}} \\ (as detailed in Appendix~\ref{app:set_cover})\end{tabular} & \begin{tabular}[c]{@{}l@{}} In the shortest path graph, greedily select a minimal number 
\\ of switches in layer $i$ to cover layer $i + 1$, \ie, every switch \\ in layer $i + 1$ has a selected neighboring switch in layer $i$. \end{tabular} \\ \hline

\begin{tabular}[]{@{}l@{}} \textbf{\texttt{Synchronous shortest path}} \end{tabular}  & \begin{tabular}[c]{@{}l@{}} Synchronously compute shortest path routes w.r.t. a \\ routing policy.\end{tabular} \\  \hline


\end{tabular}
}
\vspace{-0.2cm}
\caption{Descriptions of some \sysname's algorithmic primitives. The full version of the table is shown in Table~\ref{tbl:primitive-calls-full}.}\label{tbl:primitive-calls}

\end{table}

Switch memory and bandwidth are two scarce resources, and \sysname as a control program should not consume too much of them. Therefore, we allow users to define the resources dedicated to \sysname for network event reaction. \sysname's execution engine then achieves algorithm scheduling (\eg, rate-limiting, batch  execution) at runtime to meet the resource constraints (Section~\ref{subsubsec:algorithm-scheduling}).

\vspace{-0.3cm}
\section{Message-passing execution engine}
\vspace{-0.1cm}
\label{sec:execution-engine}

\revise{\sysname proposes to leverage hardware advances of programmable switches~\cite{Tofino, broadcom} and Ethernet~\cite{8207825} and execute message-passing in the switch's data plane to achieve towards fast reaction.} Despite data plane limitations, we enable the efficient execution of many message-passing algorithms on switches, subject to resource constraints and network unreliability.
Here we identify the following general challenges, provide platform-agnostic primitives to address them, and implement a proof of concept system on Tofino~\cite{Tofino}.

First, many message-passing algorithms do not consider network unreliability, such as heterogeneity in packet delay, packet order uncertainty, and transient packet losses.~\footnote{Persistent losses are treated as a new failure and handled in Section~\ref{sec:distributed-framework}.}
Accordingly, our solutions use synchronization protocols (Section~\ref{subsec:backend-synchronization}) and include a mechanism for handling packet loss (Section~\ref{subsubsec:efficient-communication}). Another challenge is scheduling the execution of multiple algorithm instances under the constraints on memory and bandwidth, minimizing the reaction time. 
We address this problem in Section~\ref{subsubsec:algorithm-scheduling}. 
A further challenge is to introduce optimizations for improved performance on computationally restrictive programmable switches. We elaborate on the two techniques, namely dynamic multicasting and message packing for efficient communication~\footnote{Message-packing improves the bandwidth utilization rate and is a key to our objective of low bandwidth overhead.}, but also integrate arithmetic approximation and for-loop elimination in \sysname for achieving switch compatibility (Section~\ref{subsubsec:p4-execution}).

\vspace{-0.4cm}
\subsection{Algorithm synchronization}
\vspace{-0.05in}
\label{subsec:backend-synchronization}
\sysname's synchronizer module is a critical component to allow message-passing algorithms to execute synchronously (allowing all switches to know when a round begins) in the presence of network unreliability (e.g., packet loss, uneven queues, etc.).
Different synchronizers present different tradeoffs between message overheads, convergence, and synchrony properties. Here we explain the $\alpha$ synchronizer we implemented, and defer discussion on other choices to Appendix~\ref{app:sync}.

Before a switch starts its $r$th round, it has to confirm that all neighboring switches have finished their $(r - 1)$th round. Hence each switch $u$ broadcasts a short synchronization message to all their neighbors, telling them that it has completed its $(r-1)$st round. The neighboring switches perform similarly, and $u$ listens to all such synchronization messages until it receives messages from all neighbors. This indicates the completion of round $r - 1$ of all its neighbors, and switch $u$ can then proceed to round $r$. Finally, at the beginning of the (synchronous) algorithm execution, we invoke the \texttt{Asynchronous global flooding} primitive (Table~\ref{tbl:primitive-calls}) to flood messages to all switches and bootstrap the synchronizer on each switch, which initializes its local round number to $1$ and begins the message-passing execution.

\subsection{Algorithm scheduling}\label{subsubsec:algorithm-scheduling}

As mentioned in Section~\ref{subsec:designing-reactions}, \sysname aims to limit both bandwidth and switch memory consumption. It also aims to support running multiple algorithm instances in parallel, which means that we have to split these resources between the instances. This is achieved by our scheduling module.

\vspace{0.05in}
\subp{Memory constraints.} Ideally, full parallelization of algorithm instances achieves the best reaction time, since different instances are bottlenecked by the bandwidth of different subsets of links. However, the allocated memory for \sysname can be insufficient to support this.
To that end, at runtime, \sysname picks the maximal number of algorithm instances that have not yet been completed and can fit within the switch's memory constraint. Such a set of instances is called a batch. When a batch terminates, our scheduler picks the next batch of instances to run. We call this batch execution. 
We note that most message-passing algorithms, including those in our use cases, have a fixed memory requirement known at compile time, and we assume this is the case for all instances here. As a result, there will never be a need to preempt an algorithm instance due to insufficient memory.

\vspace{0.05in}
\subp{Bandwidth constraints.} Next, we need to rate-limit the scheduled instances at runtime to ensure we adhere to the bandwidth constraint.  
This is done by storing messages in the switches' memory whose transmission is postponed to avoid exceeding the link's bandwidth usage constraint. 
We then use the packet generator to generate packets that retrieve stored messages at a frequency that adheres to this constraint.
%
Here we do not need dynamic memory allocation since each synchronous message-passing algorithm (and the \texttt{Asynchronous global flooding} algorithm primitive) instance only needs to cache up to a fixed number of messages (the number of ports of the switch).
To {{share bandwidth}} among different algorithm instances, we choose which instances {next} transmit on each link in a fair-sharing way. 
We note that this is \mbox{a design choice, and one can program other policies.}


\vspace{-0.2cm}
\subsection{Switch compatibility and optimizations}
\label{subsubsec:p4-execution}
\vspace{-0.1cm}

We now discuss challenges in running \sysname algorithms in programmable switches.
Figure~\ref{fig:p4-layer} depicts the key designs, in which we highlight our main contributions.

\smallskip
\noindent\textbf{Computation inside switches}
\label{sec: p4 computation}
While not all message-passing algorithms are implementable in the data plane of programmable switches, the following primitives ease adapting such algorithms.

\noindent\textit{1. Approximating complex arithmetic operations.}
\underline{Challenge}. Many distributed algorithms involve operations like multiplication and floating-point arithmetic, which may not be natively supported by the switch's architecture. 

\noindent\underline{Solution}.
We enable complex arithmetic via approximate arithmetic operations, following previous works~\cite{ben2020pint, 10.1145/3318464.3389698}.

%
%

\noindent\textit{2. Eliminating for-loops.} 
\underline{Challenge}. 
Message-passing's computation is specified as a subroutine that executes after all messages from neighbors arrive, but switches are generally not capable of looping \mbox{over all of them in a single pipeline pass.}


\begin{figure}
	\centering
	\vspace*{-1mm}
	\includegraphics[width=0.465\textwidth]{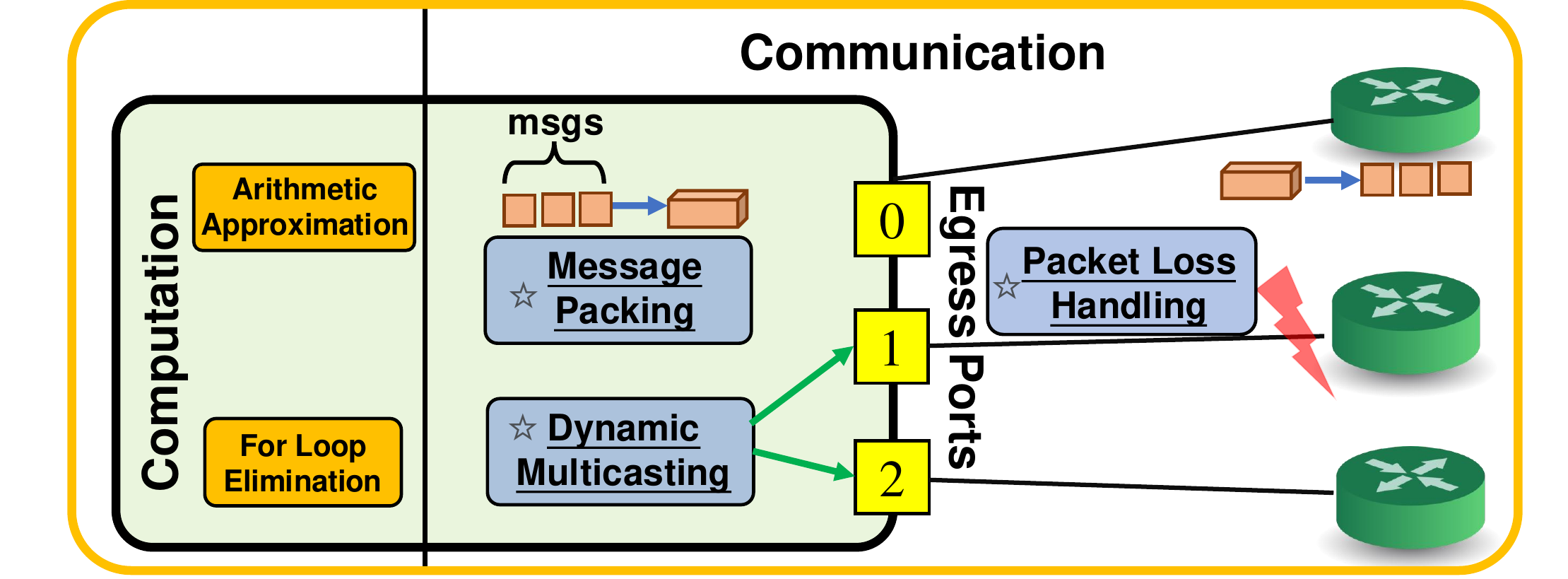}
	\vspace{-3mm}
	\caption{Running algorithm instances on programmable switches.} 
	\label{fig:p4-layer}
\end{figure}

\noindent\underline{Solution}. 
The computation of many of the algorithms of interest can be done in a streaming way, one packet at a time. While other methods can be implemented using recirculation, in all of our use cases, we did not require it. 


\textbf{Efficient communication across switches}
\label{subsubsec:efficient-communication}

\noindent\textit{1. Enabling dynamic multicast.} 
\noindent\underline{Challenge}.
The theoretical models allow a switch to send different messages to different subsets of neighbors. 
To achieve this, we establish dynamic multicast rules \textit{on the fly}, i.e., at runtime, we may send different packets to an arbitrary set of destination ports as determined by the algorithm.
However, in 
programmable devices like Tofino~\cite{Tofino}, 
the switches’ CPU can only configure a limited number of \textit{static} multicast rules ($2^{16}$ for Tofino) before running the distributed algorithm.
In an $\ell > 16$ port switch, there are $2^\ell > 2^{16}$ potential port subsets, which means that we cannot pre-configure all groups ahead of time. As using the CPU to modify the groups at runtime is costly (for Tofino, about 10$\mu$s to reach the CPU~\cite{PCIe} and another \mbox{10$\mu$s to update~\cite{zeng2022tiara}), we require a different approach.} 



\noindent\underline{Solution}.
To support general (any neighbor subset) and scalable (suitable for switches with many ports) messaging, we introduce \emph{dynamic} multicasting. 
In Tofino, we store Boolean filter vectors in egress SRAM registers, and modify them at runtime according to the algorithm. 
To multicast packets to neighbors, we \emph{broadcast} it to all egress ports, and then drop those that are filtered.
This approach inevitably generates wasted packets from the traffic manager to the egress pipeline.
However, since our multicast is only used on a small portion of the traffic used by the distributed algorithms, this appears quite manageable. Thus, we implement \mbox{dynamic multicasting for scalability and generality}.


\smallskip
\noindent\textit{2. Handling packet loss.}\label{subsec:handle-packet-loss} \noindent\underline{Challenge}.
Most message-passing algorithms assume reliable networks but, in practice, we have packet loss.  We cannot fully implement reliable transmission protocols, such as TCP, on switches' data plane, as switches are not capable of supporting dynamic memory management required for software TCP, and they may not intrinsically support timeouts and complex retransmission primitives.

\noindent\underline{Solution}. We propose a simple retransmission protocol that works on programmable switches.
Following our approach for handling memory and bandwidth constraints (Section~\ref{subsubsec:algorithm-scheduling}), the switch keeps a copy of packets (the total number is bounded thanks to our algorithm scheduling) in memory until receiving an ACK message, at which point it can free the memory.
In Tofino, we leverage the packet generator to trigger probes to the egress pipeline that actively check (in a round-robin fashion) if a packet timed out and performs retransmissions.

\smallskip
\noindent\textit{3. Message packing.} \underline{Challenge}. Generally, messages in CONGEST algorithms are small in size, ranging from a single bit to several bytes. However, Ethernet frames carrying the messages have a minimum size of $64$ bytes.  Dedicating an Ethernet frame to a single message can therefore yield a low bandwidth utilization rate, with only a small portion of the frame being the actual message and a large portion being paddings and headers. In our use cases of multicast and Contra routing~\cite{hsu2020contra}, we launch thousands of algorithm instances to react to a network event. In such cases, a low bandwidth \mbox{utilization would result in a longer reaction time.}

\noindent\underline{Solution.} Our message packing technique works by concatenating multiple messages from different algorithm instances destined for the same neighbor into a single packet, so that the receiver can process multiple messages at a time.
%
Further, we would ideally like to perform as few arithmetic operations as possible, which minimizes the number of packets that traverse the switch's pipeline, given that the switches' arithmetic and pipeline stages are limited.
We therefore aim to pack values in such a way that we can perform operations on multiple values with a single arithmetic operation.  For example, we could pack two $16$-bit values $a_1$ and $a_2$ into a single $32$-bit word, so that incrementing both $a_1$ and $a_2$ by $1$ can be achieved by \mbox{adding $2^{16}+1$ to $a_1*2^{16}+a_2$ (assuming no overflows).}


Additionally, the receiver needs to know which algorithm instances of the messages are packed in the current packet. The sender therefore adds values $\ell,r$ and a bitmap to each packet indicating that it carries messages from instances IDs $\subseteq \{\ell,\ell+1,\ldots,r\}$.
Again, message packing is designed specifically for synchronous algorithms, as it would be harder and take more bandwidth to apply it to asynchronous algorithms that act on each message arrival, whose patterns are irregular.

\vspace{-0.2cm}
\section{Use cases}\label{sec:use_cases}
\vspace{-0.2cm}

We next show how \sysname can be used to easily implement distributed algorithms using three real-world use cases of our framework: clock synchronization, source-routed multicast, and routing. For the first two use cases, we propose our own data-plane reaction protocols to handle network events. For the routing case, we adapt and optimize Contra~\cite{hsu2020contra}, a state-of-the-art data-plane protocol, to use \sysname.  For these applications, we focus on link failures and switch failures as \mbox{the important relevant network events.}





\vspace{-0.2cm}
\subsection{Clock synchronization}\label{subsec:clock-sync}

{\bf Background}. 
Clock synchronization is achieved by periodically exchanging timestamp information between adjacent devices. 
In state-of-the-art designs (e.g.,~\cite{Sundial, PTP}), the information is carried by \emph{synchronization packets} through the \emph{synchronization spanning tree}. The
tree's root acts as the `grandmaster clock' that serves as the time reference to other clocks. 
The precision of the clock of device $k$ can be measured by its \textit{time uncertainty} $\epsilon^{(k)}(t)$~\cite{Sundial} at time $t$, calculated as
\setlength{\abovedisplayskip}{2pt}
\setlength{\belowdisplayskip}{1pt}
\setlength{\abovedisplayshortskip}{1pt}
\setlength{\belowdisplayshortskip}{1pt}
\setlength{\jot}{1pt}
\begin{align}\label{eq:epsilon}
    \epsilon^{(k)}(t) = \epsilon_0 \cdot \text{depth}^{(k)} + (\mathit{t} - \tau_{\mathit{sync}}^{(k)}) \cdot \texttt{max\_drift\_rate} \ .
\end{align}


Here, $\tau_{\mathit{sync}}^{(k)}$  is the \textit{last} time device $k$ received a synchronization message, $\texttt{max\_drift\_rate}$ is the maximum drift rate of a clock, $\epsilon_0$ is the noise inherent in hardware, and depth$^{(k)}$ is the number of hops between device $k$ and $\mathit{root}$. 
%
Finally, the overall \textit{time uncertainty bound} is defined as $\epsilon(t)=\max_{k} \epsilon^{(k)}(t)$, which measures the maximal discrepancy between any pair of clocks. 
During normal operation,~\footnote{We use $\epsilon$ for $\epsilon(t)$ when convenient where $t$ is implied.} $\epsilon$ is primarily determined by the \textit{tree depth}; when failures break connectivity of the synchronization tree, $\epsilon$ grows proportionally to the failure time $\mathit{t} - \tau_{\mathit{sync}}^{(k)}$ until a new tree is reconstructed.

\smallskip
\textbf{Main objectives}. As suggested by Sundial~\cite{Sundial}, to achieve fault-tolerant clock synchronization, we aim to minimize the peak $\epsilon$ in case of failures, and keep $\epsilon$ consistently small during normal times (without failures).  We therefore think of $\epsilon$ in two phases: the peak $\epsilon$ obtained during event reaction, and $\epsilon$ after the distributed optimization. During fast recovery, we aim to construct a clock synchronization tree as quickly as possible to prevent large peak $\epsilon$. We then further reduce $\epsilon$ by optimizing the depth of the synchronization tree in the distributed optimization phase.


%

%
  



\begin{table}
\vspace{-0.0in}
    \centering
    \resizebox{.9\linewidth}{!}{
    \begin{tabular}{|c|c|c|c|} \hline
        Topology &  FatTree & Jupiter & Jellyfish\\ \hline
        Number of Affected Groups & 1344 & 2147 & 1175 \\ \hline
    \end{tabular}
    }
    \vspace{-0.05in}
    \caption{The maximal number of multicast groups affected by a link failure over different datasets.}
    
    \label{tab:mcast_affected_group}
\end{table}

\smallskip
\textit{Algorithm design.}
We now describe our proposed algorithms, following the key ideas summarized in Table~\ref{tbl:frameworkPhases}.

\smallskip
\noindent{\bf (1) The detection phase}. We adopt Sundial's~\cite{Sundial} passive detection protocol where we listen to synchronization messages and detect a failure on timeout. The original protocol, however, does not find which specific devices have failed, but only concludes that some upstream links or switches have failed. To determine the specific failed devices, we propose a simple ping-on-timeout technique, as discussed in Appendix~\ref{appendix:impl-details}.

\smallskip
\noindent{\bf (2) The fast recovery phase}. Upon detecting a failure, 
we restore connectivity by constructing
a fast recovery spanning tree using asynchronous flooding. This spanning tree can then deliver synchronization messages until we complete the distributed optimization phase. The algorithm can be easily implemented leveraging \langname's \texttt{Asynchronous global flooding} algorithmic primitive (Table~\ref{tbl:primitive-calls}). This primitive also initializes the synchronizer for running synchronous algorithms in the next phase.
Concurrent failures are handled according to the approach specified in Section~\ref{subsec: overview}.

\smallskip
\noindent{\bf (3) The distributed optimization phase}. 
 In the fast recovery phase, we also randomly sample several roots and broadcast them via \texttt{Asynchronous global flooding} to every switch. Then we run the BFS-based \texttt{shortest path tree} algorithmic primitive (Table~\ref{tbl:primitive-calls}) to construct the shortest path spanning tree for each root. 
We then run the \texttt{Bottom-up tree aggregation} primitive to calculate the depth for each subtree. 
Following this, we move the root to the tree's center. 
Finally, we pick the tree with minimal tree depth via leader election as our final synchronization tree, leveraging the \texttt{Conditional broadcast and aggregation} primitive. 
We give the pseudo code as Algorithm~\ref{alg:pseudocode-clock-sync} in Appendix~\ref{app: pseudocode}.

The shortest path tree algorithm only returns an optimal-depth spanning tree for a given root. In practice, one may choose which and how many roots to use. We observe from our experiments (Appendix~\ref{appendix:flex-reliable-exps}) that sampling a few candidate roots often suffices for finding an optimal-depth tree with $>99\%$ probability in data center topologies.
We note that running shortest path tree on all nodes creates excessive network overhead and takes significant time.
Optionally, after the distributed optimization, the controller could compute an optimal depth synchronization tree and inform the switches.






\vspace*{-1mm}
\subsection{Source-routed multicast}\label{subsec:multicast}
\vspace*{-1mm}


\subp{Background}. 
Given a multicast group $\text{group}_i = (src_i, dsts_i)$, where $src_i$ is the single sender host and $dsts_i$ is the set of receiver hosts, the goal is to deliver the packet to $dsts_i$ through a \emph{multicast tree}, $T_i$, duplicating the packet at tree branches as needed. In this section, we base our design on \emph{source-routed} multicast. In this setting, source hosts \emph{encode} the multicast tree as rules that are added to the headers of multicast packets. The switches then forward packets according to the rules in the packets. A network failure can affect thousands of multicast groups by breaking the connectivity of their corresponding multicast trees (Table~\ref{tab:mcast_affected_group}), requiring reconstructing a multicast tree and rules for each affected group.

\smallskip
\subp{Main objectives}. We classify the goals for a fault-tolerant source-routed multicast system as follows.  In the fast recovery phase, the system should ideally restore functionality for thousands of affected groups in $100$s of microseconds. In the distributed optimization phase, we wish to minimize packet delay.
The delay is determined mainly by the depth of the multicast tree, so we prefer shallow trees.
Finally, we aim to minimize traffic overhead that is determined by both the size of the \mbox{multicast header and the size of the multicast tree.}


\smallskip
\textit{Algorithm design}.
We now describe our approach, following the key ideas summarized in Table~\ref{tbl:frameworkPhases}.

\smallskip
\noindent{\bf (1) The detection phase}. We use active detection of failures.

\smallskip
\noindent{\bf (2) The fast recovery phase}. 
With limited bandwidth, fixing thousands of broken multicast trees in parallel incurs a large message overhead and requires significant time. 
Instead, we build a single fast recovery spanning tree $T_0$ and reconstruct multicast trees on top of it. This is achieved via flooding (see Table~\ref{tbl:primitive-calls}). 
At the same time, switches adjacent to the failure(s) retrieve the affected groups stored in their memory and inform the Top-of-Rack switches of the affected groups. 

Once $T_0$ is constructed, connectivity is guaranteed. 
The remaining task is to compute multicast trees that are subtrees of $T_0$, such that the subtree connects all destinations for each group. We propose the following synchronous algorithm that works in a bottom-up manner on $T_0$. Starting from the destination switches, we determine whether switch $i$ is on the multicast tree of group $j$ (it is if one of $i$'s children on $T_0$ belongs to group $j$'s tree) and record switch $i$'s children for group $j$. However, to limit the bandwidth, we do not compute each such subtree individually.
To that end, we instead propagate participation bitmaps for all groups simultaneously, where each switch $i$ marks whether it belongs to the multicast tree of group $j$. For a switch that receives participation bitmaps from its children in $T_0$, it updates its own bitmap accordingly using bitwise operations. To further reduce overhead, we leverage the message packing technique to place multiple such bitmaps into each packet. We defer further details to Appendix~\ref{appendix:construct-fast-recovery-multicast-tree}.

%
%


\smallskip
\noindent{\bf (3) The distributed optimization phase}.
In state-of-the-art source-routed multicast solutions like Elmo~\cite{Elmo}, senders encode propagation rules on each multicast packet, thus requiring a header overhead driven by topology size and distribution of destinations. 
To reduce these overheads, \sysname leverages \emph{spanners} to encode rules over a small sub-topology.
A spanner is a sub-topology $G' \subset G$ of the original topology $G$ such that the $\texttt{stretch} = \frac{dist_{G'}(u, v)}{dist_G(u, v)}$ is bounded. 

We use the Baswana-Sen algorithm~\cite{spanner}, that given a user-defined $\texttt{stretch}$ parameter, computes a spanner with this stretch in $O((\texttt{stretch})^2)$ rounds. The benefit of higher $\texttt{stretch}$ values is that they allow more pruning of the topology, thereby decreasing the needed multicast packet header size, the overall latency of multicast traffic, and the total consumed network bandwidth.


Accordingly, after computing the spanner, we reconstruct a shortest path tree on top of it for each affected group.
To that end, we use the algorithmic primitive \texttt{Shortest path tree}. Using this tree, we compute the distance of each destination from the source of the group.
Ideally, the multicast tree for the group should then be the one with the minimal number of nodes such that each destination is connected to the source via a shortest path.
However, finding such a tree is NP-hard, as we show in Appendix~\ref{app:SPT_hardness}. 
Instead, we leverage the primitive \texttt{Set cover on a shortest path tree}, which is an approximation algorithm to construct a small multicast tree for each group. 
%
%
%
%
%
Optionally, the controller can then spend more time optimizing the spanner and set covers.


\vspace{-0.1cm}
\subsection{Contra for Routing}\label{subsec:contra}
\vspace{-0.05cm}

We also adapt Contra's routing protocol~\cite{hsu2020contra} using \sysname,
which we later show can reduce the reaction time to network events by up to $90\%$ (Section ~\ref{subsubsec:exp-contra}). Motivated by the limitations of Contra mentioned in Section~\ref{subsec:why-synchronous-algorithms}, we design a wait-and-merge approach to convert Contra's asynchronous protocol to an equivalent synchronous message-passing protocol that can be implemented in \sysname.  This conversion also allows us to apply the message packing technique.  Both of these approaches save bandwidth, and under bandwidth constraints \mbox{can lead to significantly faster reaction}.

\textbf{Contra's adaptation to \sysname}. We have designed a synchronous version of Contra's shortest path algorithm, as detailed in Algorithm~\ref{alg:contra} in Appendix~\ref{app: pseudocode}. This implementation is part of 
\sysname's library, and we provide an algorithmic primitive for it. Our modified Contra algorithm \textit{waits-and-merges} information from different neighbors that correspond to the same routing entry during a round. Intuitively, the modified protocol prevents the intermediate update results from being sent early, saving bandwidth.
We empirically verify this in Section~\ref{subsubsec:exp-contra}.
Using message packing on top of the synchronous message-passing algorithm offers further optimization over the original asynchronous Contra implementation. We regard the algorithm that reconstructs the routes of a given destination $dst$ as one algorithm instance. \sysname packs $8$ messages with consecutive instance IDs and the same transmitted state into a packet, and \sysname \mbox{processes such a packet in a single pass}. 

\vspace{-0.25cm}
\section{Prototype implementation}\label{sec:implementation}
\vspace{-0.2cm}

We have implemented \sysname's prototypes both in Tofino P4~\cite{Tofino, p416} and in simulation on top of our own discrete event simulator. The Tofino implementation validates \sysname's design in small-scale network topologies and better indicates the real-world performance of \sysname, while simulation is ideal for efficiently executing large-scale experiments. 

\smallskip
\subp{Tofino emulation.} Clock synchronization for Tofino using \sysname is deployed onto a testbed with a Wedge 100BF-32X Tofino switch. Inspired by prior works~\cite{dptp, ben2020pint}, our implementation is able to virtualize multiple logical switches on the single physical switch that form a 3-4-ary FatTree topology (as specified in Table~\ref{tab:dc-topologies}). The virtualization is achieved directly in the data plane and thus incurs no additional overhead. Specifically, each logical switch is assigned dedicated ports for each corresponding logical link, and sending packets to a neighboring switch is achieved via looping back traffic to its neighbor's port. 

\smallskip
\subp{Resource consumption.} We evaluate the resource consumption of \sysname's Tofino prototypes, as shown in Table~\ref{tab:p4-resource-usage}. We note that the SRAM usage is not explicitly restricted and the algorithm scheduling chooses full parallelization accordingly. We fit our clock synchronization prototype into the Tofino architecture and our multicast and Contra prototypes into Tofino 2~\footnote{The physical switch of our testbed is in Tofino; multicast and Contra that cannot fit in Tofino are examined with Intel-modeled Tofino 2 architecture.}. We see that all use cases incur minimal consumption of $\le 3.07\%$ of Tofino's SRAM budget even without memory restrictions, and $\le 2.43\%$ of the TCAM budget.

\vspace{-1mm}
\section{Evaluation}
\vspace{-1mm}
\label{sec:eval}

We evaluate \sysname on the three use cases detailed in Section~\ref{sec:use_cases}. Additionally, we implement and evaluate F10 routing~\cite{liu2013f10} using \sysname and discuss results in Appendix~\ref{appendix:f10}. We first elaborate on the testbed evaluation (Section~\ref{subsec:testbed}) comparing \sysname Tofino against state-of-the-art systems. We then proceed to large-scale simulation experiments in Section~\ref{subsec:eval-w-state-of-the-art} to study \sysname's performance at scale. In Section~\ref{subsec:exp-combined}, we evaluate running the three applications concurrently in \sysname.  
Finally, we show in \cref{subsec:param-settings} how different parameter configurations affect  performance.
}


\vspace{-1mm}
\subsection{Testbed Evaluation}\label{subsec:testbed}

\begin{table}[]
\vspace{-0.03in}
\centering
\resizebox{.95\linewidth}{!}{
    \begin{tabular}{|c|c|c|c|c|}
        \hline
        \textbf{Application} & \textbf{LoC in P4} & \textbf{Stages} & \textbf{SRAM usage}   & \textbf{TCAM} \\ \hline
        Clock Synchronization & 1731 & 12 & $3.07\%$  & $2.43\%$ \\ \hline
        Multicast & 1616 & 19 & $1.48\%$  & $1.88\%$ \\ \hline
        Contra & 1150& 18 & $1.78\%$ & $1.30\%$  \\ \hline
    \end{tabular}
    }
    \vspace{-0.1cm}
    \caption{Lines of code of \sysname's three use cases and their resource usage on Tofino switches for a 64-ary FatTree topology. Note that results for clock synchronization are for Tofino architecture while results for other applications are for Tofino 2.}
    \label{tab:p4-resource-usage}
\end{table}

We first evaluate \sysname's performance in clock synchronization over the Tofino testbed. We compare against PTP's~\cite{PTP} best master clock algorithm (BMCA), a state-of-the-art approach that automatically rebuilds a clock synchronization topology upon a network event. We choose 3-4-ary FatTree (\cref{tab:dc-topologies}, explained in Appendix~\ref{appendix:exp-setups}) as the logical topology that we could fit into our Tofino switch.

\begin{table}[]
\centering
\resizebox{.89\linewidth}{!}{
    \begin{tabular}{|c|c|c|c|c|}
        \hline
        & 3-4-ary FatTree & 64-ary FatTree  & Jupiter & Jellyfish \\ \hline
        \begin{tabular}[c]{@{}l@{}}
            \#Switches ($n$) 
        \end{tabular} & 
        15 &
        5120 & 
        22528 & 
        4000 \\ \hline
        \begin{tabular}[c]{@{}l@{}}
        \#Edges ($m$)
        \end{tabular} & 
        21 &
        131072 & 
        163840 & 
        126950 \\ \hline
        Diameter ($D$) & 4 & 4 & 8 & 3  \\ \hline
    \end{tabular}
}
    \vspace{-0.2cm}
    \caption{Characteristics of the evaluated topologies.}
    
    \label{tab:dc-topologies}
\end{table}

\begin{figure}
 
         \begin{minipage}[t]{0.57\linewidth}{
		\vspace{-0.00in}
		\begin{center}
		\includegraphics[width=\textwidth, ]{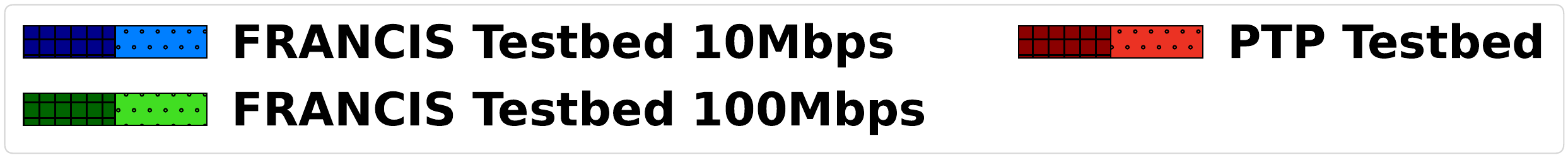}
		\end{center}
		}
		\end{minipage}

        \hspace{-0.2cm}
	\subfigure[Reaction time]{
		\begin{minipage}[t]{0.3\linewidth}{
		\vspace{-0.08in}
		\begin{center}
		\includegraphics[width=\textwidth, ]{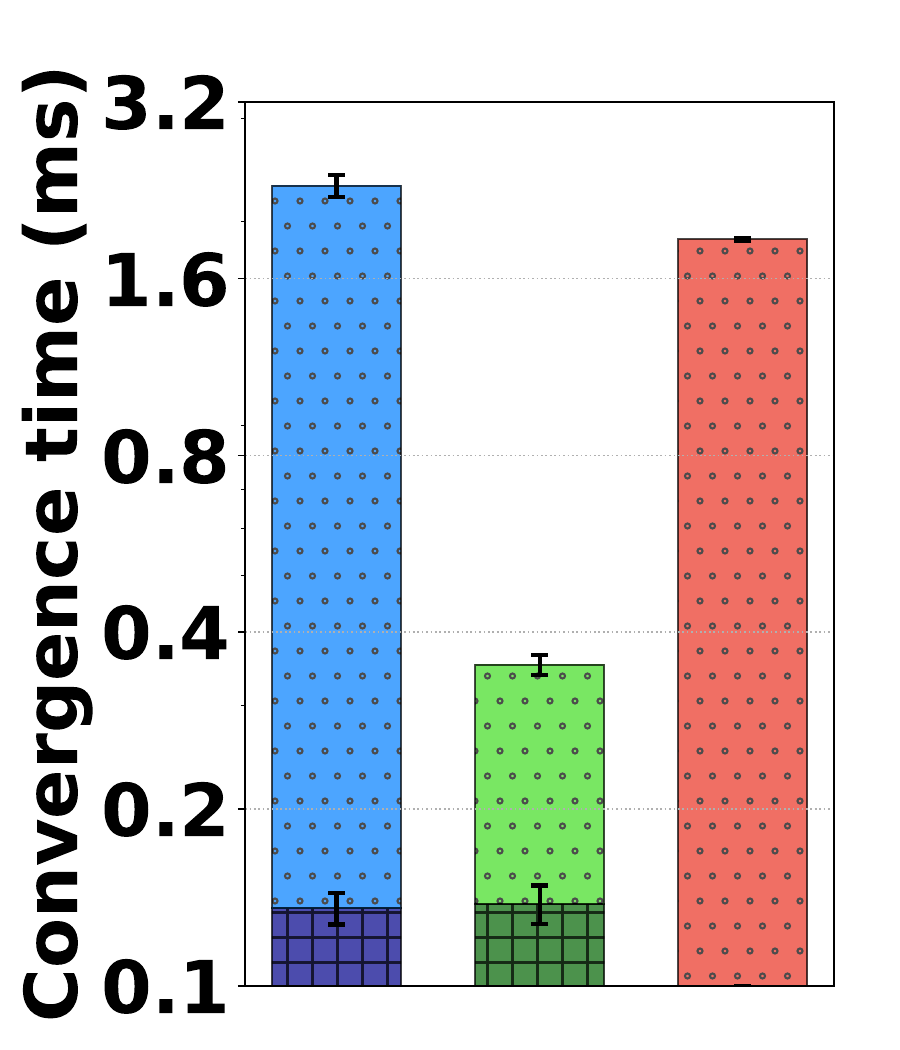}
		\end{center}
            \vspace{-0.10165in}
		}
		\label{subfig:testbed_conv_time2}
		\end{minipage}
	}
	\hspace{-0.5cm}
	\subfigure[Peak $\epsilon$]{
		\begin{minipage}[t]{0.3\linewidth}{
		\vspace{-0.08in}
		\begin{center}
		\includegraphics[width=\textwidth, ]{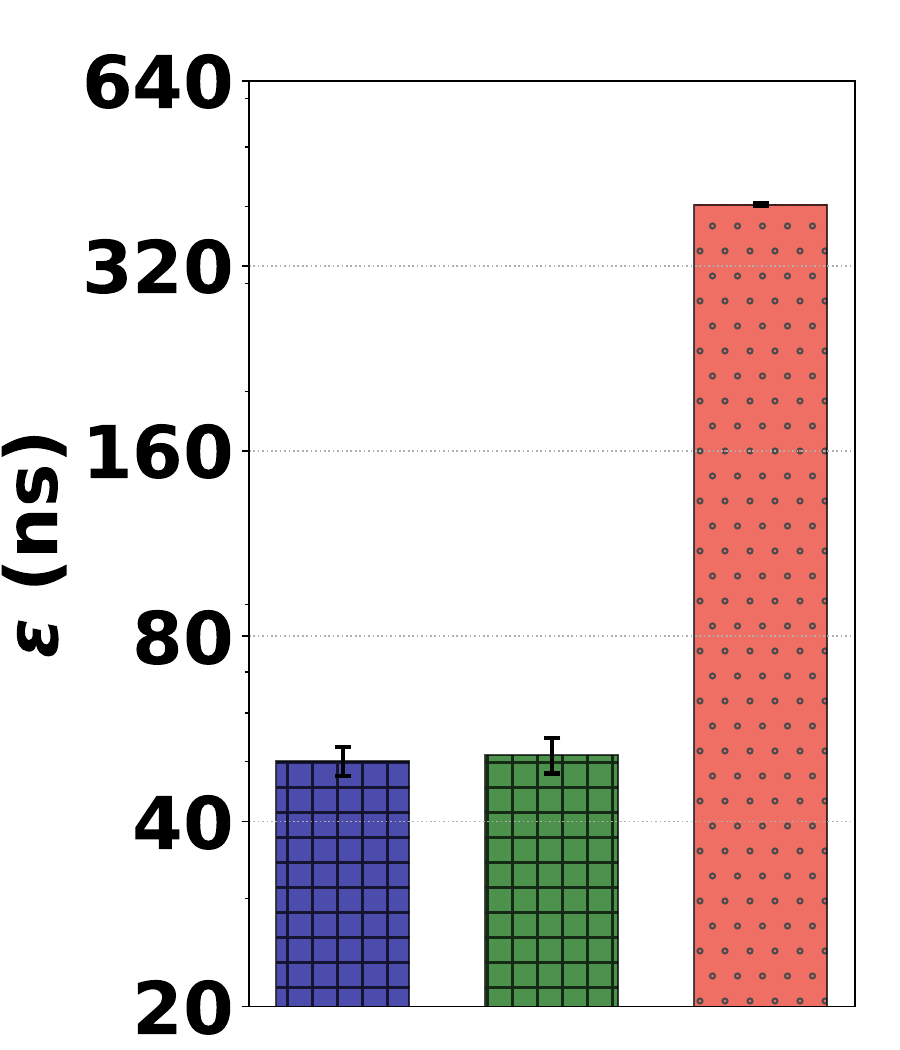}
		\end{center}
            \vspace{-0.10165in}
		}
		\label{subfig:testbed_peak_epsilon2}
		\end{minipage}
	}
        \hspace{-0.4cm}
        \subfigure[Time to $\epsilon$]{
		\begin{minipage}[t]{0.36\linewidth}{
		\vspace{-0.9cm}
		\begin{center}
		\includegraphics[width=\textwidth, ]{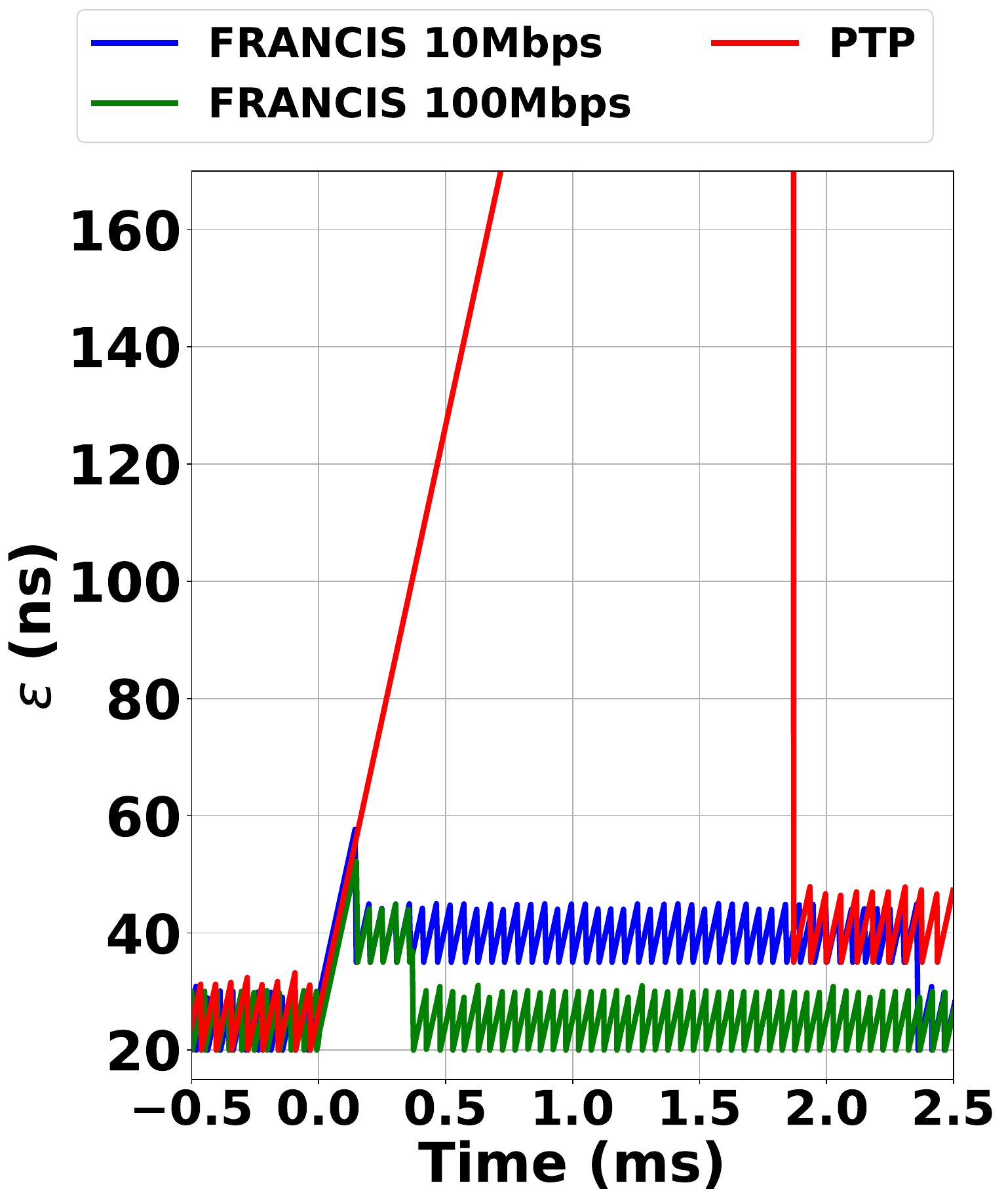}
		\end{center}
            \vspace{-0.10165in}
		}
		\label{subfig:testbed_tte2}
		\end{minipage}
	}
        \vspace{-0.3cm}
	\caption{Performance of \sysname's Tofino prototype in failure recovery for clock synchronization, compared with PTP~\cite{PTP}'s best master clock algorithm to handle failures. The dark, cross-patched bars represent fast recovery and the light, dotted bars correspond to distributed optimization. A single link failure happens at $0$ms.}
	\label{fig: testbed2}
\end{figure}

\MM{Repeats the 3-4 virtualized FatTree already described in previous paragraph; remove.}\Wenchen{Trying to emphasize "logical/virtualized topology" on switches and virtualization on hosts are achieved differently.}

\smallskip
\subp{Testbed deployment of PTP.} State-of-the-art PTP systems generally run BMCA in software, and we follow the setup to run LinuxPTP~\cite{linuxptp} with the same topology being virtualized on our server. The typical announce interval for BMCA is in order of seconds, but for testing purposes, we push it to the limit of $122\mu$s, the minimum we could achieve. Setting a shorter interval reduces the reaction time and thus peak $\epsilon$ (according to Eq.~\ref{eq:epsilon}), but incurs more computation and communication costs - an $122\mu$s interval corresponds to a total of $\sim 100$K packets processed per second over all neighbors.

\smallskip
\subp{Other setups.} 
We generate a single link failure at time $0$. For \sysname, we limit via algorithm scheduling (Section~\ref{subsubsec:algorithm-scheduling}) per-link bandwidth usage to be $100$Mbps and $10$Mbps respectively. Other settings are discussed further in Appendix~\ref{appendix:exp-setups}. 

\smallskip
\subp{Comparisons with PTP.} As illustrated in Figure~\ref{subfig:testbed_conv_time2}, \sysname achieves prompt reaction to the link failure and successfully restores connectivity of the clock synchronization tree within $170\mu$s in the fast recovery phase. The results do not depend on the bandwidth limitation, as the \texttt{Asynchronous global flooding} for fast recovery generates only one prioritized packet per link, for which no extra delay is incurred to meet the bandwidth constraint (Section~\ref{subsubsec:algorithm-scheduling}). The subsequent distributed optimization phase further optimizes the longer-term time uncertainty bound $\epsilon$ by reducing the tree depth (as shown in Figure~\ref{subfig:testbed_tte2}). The overall reaction time as shown is $352\mu$s for $100$Mbps which is $
>5\times$ faster than PTP with the aforementioned, already optimized setups. 

As suggested by Eq.~\ref{eq:epsilon}, the peak $\epsilon$ is proportional to the time when connectivity is first restored. Accordingly, we observe in Figure~\ref{subfig:testbed_peak_epsilon2} that \sysname reaches an order of magnitude lower peak $\epsilon$ than PTP. This can be attributed to the reduced time for fast recovery in \sysname from re-establishing connectivity with a suboptimal flooding tree.

We validate our simulator by finding the simulation results match with testbed results, as illustrated in Appendix~\ref{app:testbed-simulator-validation}.

\subsection{Large-scale simulation on individual tasks}
\label{subsec:eval-w-state-of-the-art}

Next, we evaluate \sysname's performance for each use case via simulation in large-scale data center topologies (Table ~\ref{tab:dc-topologies}), including $k=64$ FatTree, Jupiter~\cite{Jupiter}, and Jellyfish~\cite{Jellyfish}. 


\vspace{-0.2cm}
\subsubsection{Clock synchronization}
\vspace{-0.05cm}
\label{subsubsec:eval_on_clock}

\subp{Baseline solutions}. We compare \sysname with the state-of-the-art solutions PTP~\cite{PTP} and Sundial~\cite{Sundial}.

\subp{Experiment setup.} We generate one or two switch failures while limiting the per-link bandwidth of \sysname to be either $100$Mbps or $10$Mbps. 
For PTP, we set the announce interval to be $100 \mu s$. See further details in Appendix~\ref{appendix:exp-setups}.
\begin{figure}
    \vspace{0.05in}
    \raggedright
    \begin{minipage}[t]{0.44\linewidth}{
    \includegraphics[width=\textwidth, ]{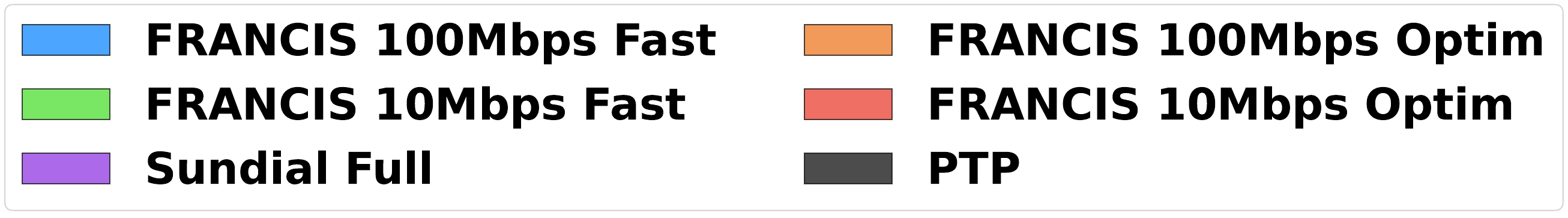}
    \vspace{-0.11in}
    }
    \end{minipage}
    
    \hspace{0.02in}
	\subfigure[Reaction time.]{
		\begin{minipage}[t]{0.42\linewidth}{
		\vspace{-0.15165in}
		\begin{center}
		\includegraphics[width=\textwidth, ]{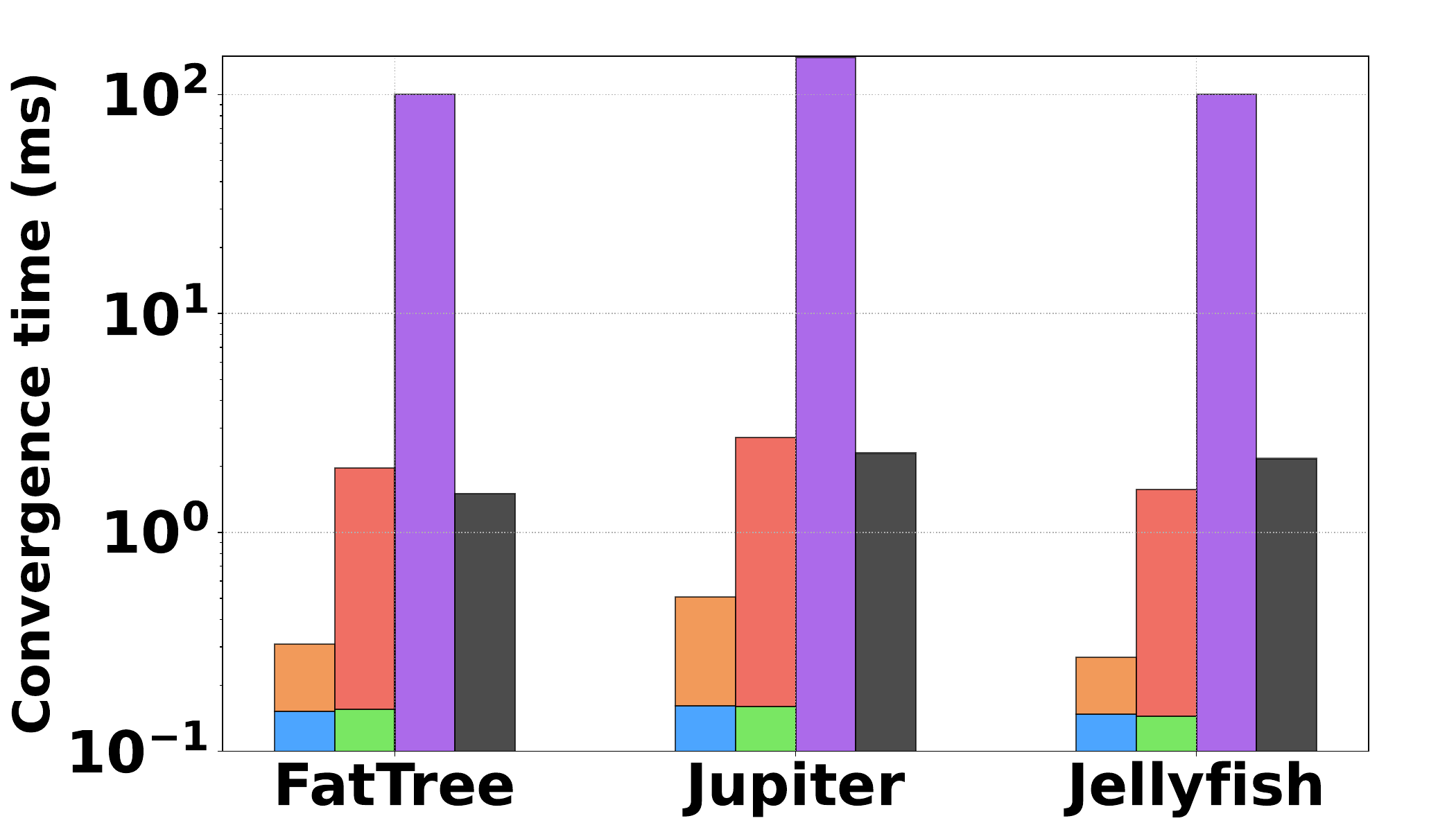}
		\end{center}
        \vspace{-0.10165in}
		}
		\label{subfig:clock_conv_time_2fail}
		\end{minipage}
	}
    \hspace{0.05in}
    \subfigure[Peak $\epsilon$.]{
		\begin{minipage}[t]{0.44\linewidth}{
		\vspace{-0.360in}
		\begin{center}
		\includegraphics[width=\textwidth, ]{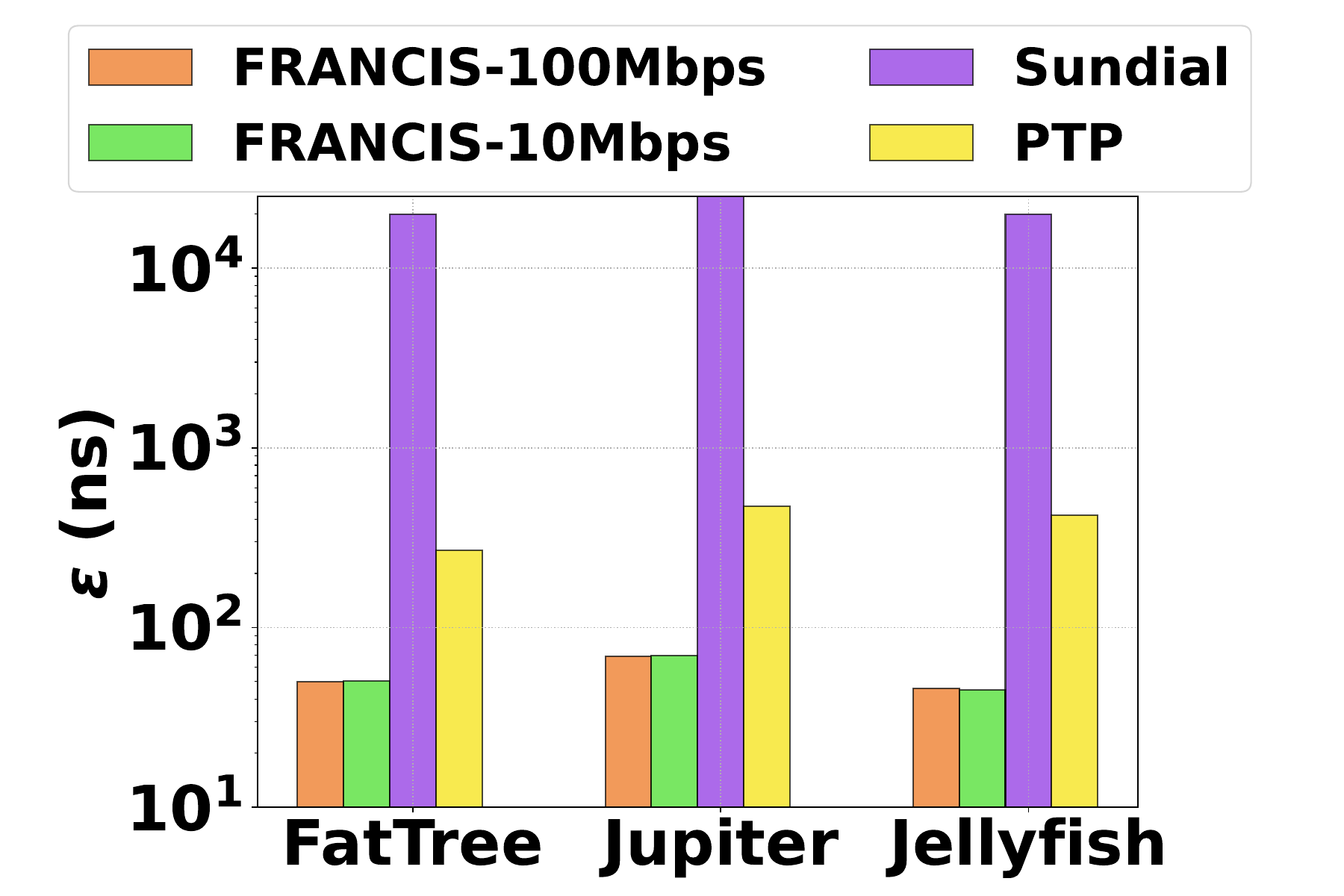}
		\end{center}
            \vspace{-0.10165in}
		}
		\label{subfig:clock_sync_dataset_epsilon_2f}
		\end{minipage}
	}
	\vspace{-0.2in}
	\caption{Performance of clock synchronization compared with Sundial~\cite{Sundial} and PTP~\cite{PTP} with two switch failures.}
    \vspace{-0.2cm}
    \label{fig:clock-sync-performance}
\end{figure}

\begin{figure}
	\centering
	
         \begin{minipage}[t]{0.7\linewidth}{
		\vspace{-0.00in}
		\begin{center}
		\includegraphics[width=\textwidth, ]{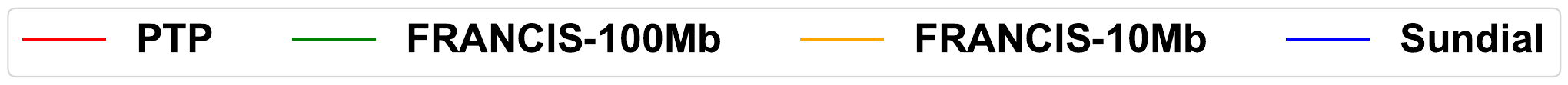}
		\end{center}
		}
		\end{minipage}
	\subfigure[Time uncertainty bound ($\epsilon$).]{
		\begin{minipage}[t]{0.43\linewidth}{
		\vspace{-0.08in}
		\begin{center}
		\includegraphics[width=\textwidth, ]{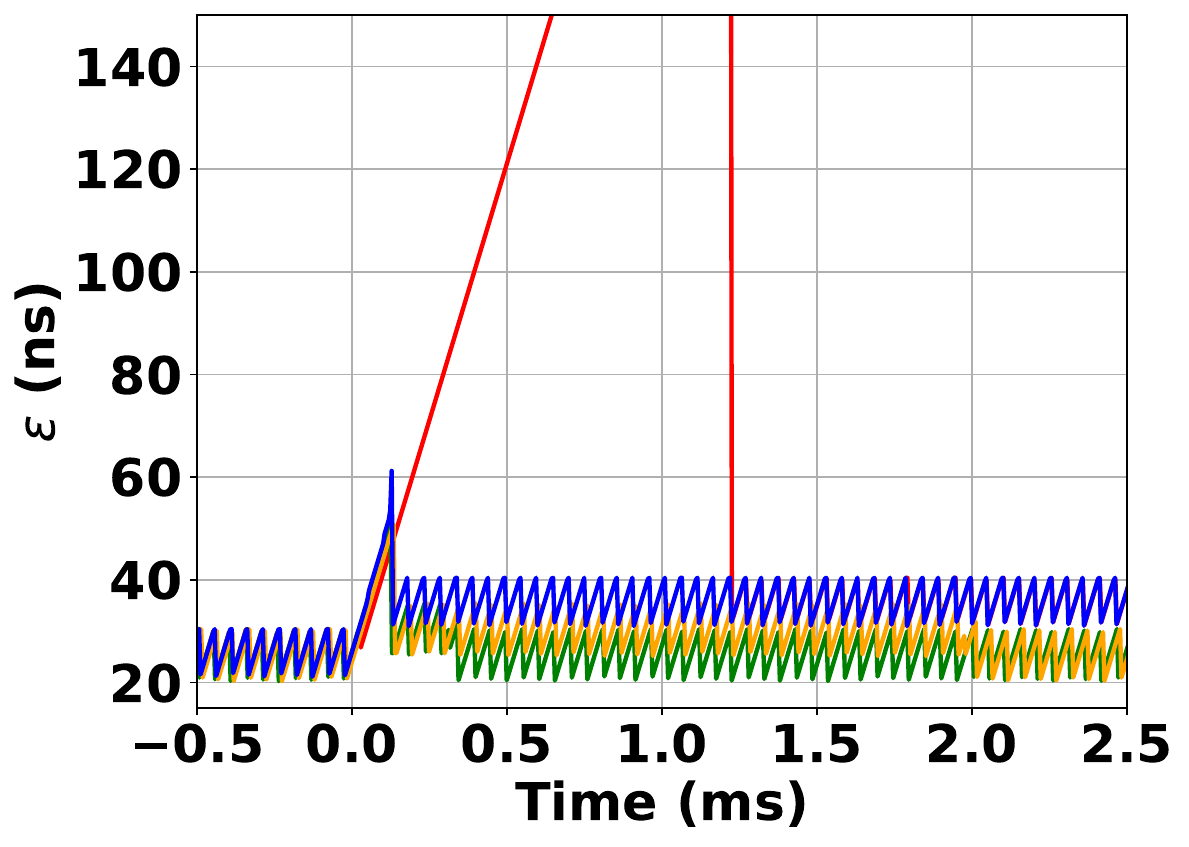}
		\end{center}
            \vspace{-0.10165in}
		}
		\label{subfig:sync_epsilon_series}
		\end{minipage}
	}
	\hspace{-0.1in}
	\subfigure[Message overhead per link.]{
		\begin{minipage}[t]{0.42\linewidth}{
		\vspace{-0.08in}
		\begin{center}
		\includegraphics[width=\textwidth, ]{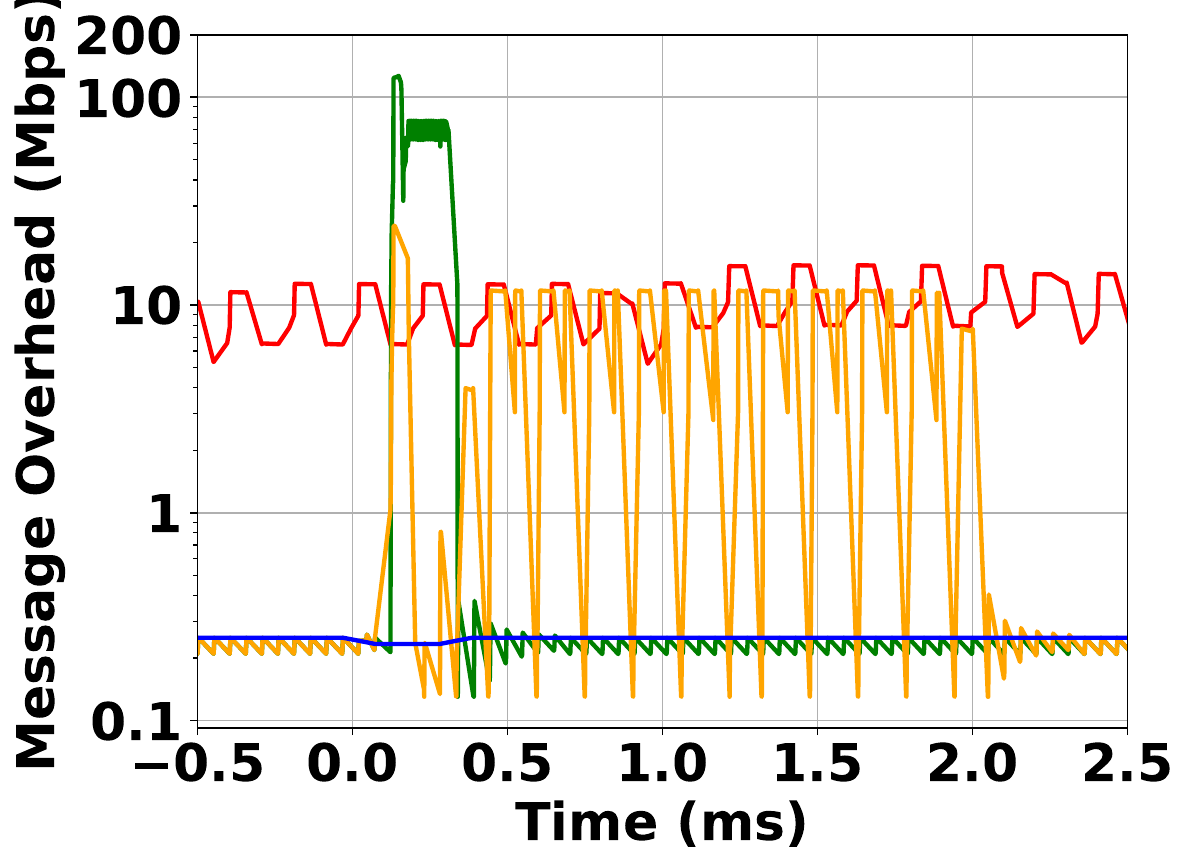}
		\end{center}
            \vspace{-0.10165in}
		}
		\label{subfig:sync-overhead-series}
		\end{minipage}
	}
	\vspace{-0.15in}
	\caption{Time to  $\epsilon$ and per-link message overhead for clock synchronization with one switch failure on the FatTree topology.}
        \vspace{0.2cm}
	\label{fig: bound-overhead-clock-5}
\end{figure}

\subp{Experimental results.} 
We first evaluate how fast \sysname reacts to switch failures, as depicted in Figure~\ref{fig:clock-sync-performance} for two switch failures and Figure~\ref{fig:clock-sync-performance-appendix} in Appendix for one failure. We observe that \sysname restores connectivity in the fast recovery phase within $200 \mu s$ for both $100$Mbps and $10$Mbps bandwidth. For Jupiter, the distributed optimization phase lasts for $506 \mu$s and $2.72$ms with $100$Mbps and $10$Mbps bandwidth respectively. This is orders of magnitude faster than Sundial as its backup plan recovery could not handle multiple failures in different domains, in which case it falls back to the controller.

As shown in Figure~\ref{subfig:clock_sync_dataset_epsilon_2f}, \sysname's has lower peak $\epsilon$ in all topologies when two failures happen. PTP's peak $\epsilon$ is $5.3$-$9.3\times$ higher than \sysname-10Mbps. Sundial incurs a large peak $\epsilon$ ($2 \cdot 10^4$) value in this setting, giving unfavorable performance.

For one switch failure, figure~\ref{fig: bound-overhead-clock-5} depicts how $\epsilon$ and message overhead change over time. We observe that PTP is markedly slower in restoring connectivity than \sysname and has higher $\epsilon$ and message overhead throughout. 
Sundial achieves an optimal $150 \mu$s recovery by switching to the backup plan immediately upon detecting the failure.
However, its $\epsilon$ remains suboptimal until the control plane steps in, as the tree of the backup plan is not optimal.
%

\begin{figure}
    \vspace{0.1in}
    \raggedright
    \hspace{0.1in}
    \begin{minipage}[t]{0.9\linewidth}{
    \begin{center}
    \includegraphics[width=\textwidth, ]{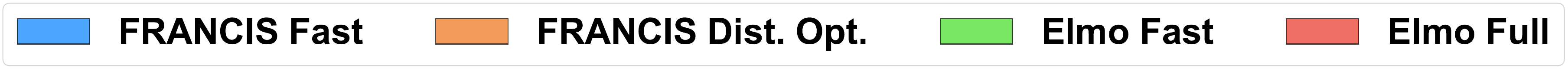}
    \end{center}
    }
    \label{subfig:mcast_conv_time_legend}
    \end{minipage}

    \centering
    \subfigure[1 link failure.]{
        \begin{minipage}[t]{0.4\linewidth}{
        \vspace{-0.22in}
        \begin{center}
        \includegraphics[width=\textwidth, ]{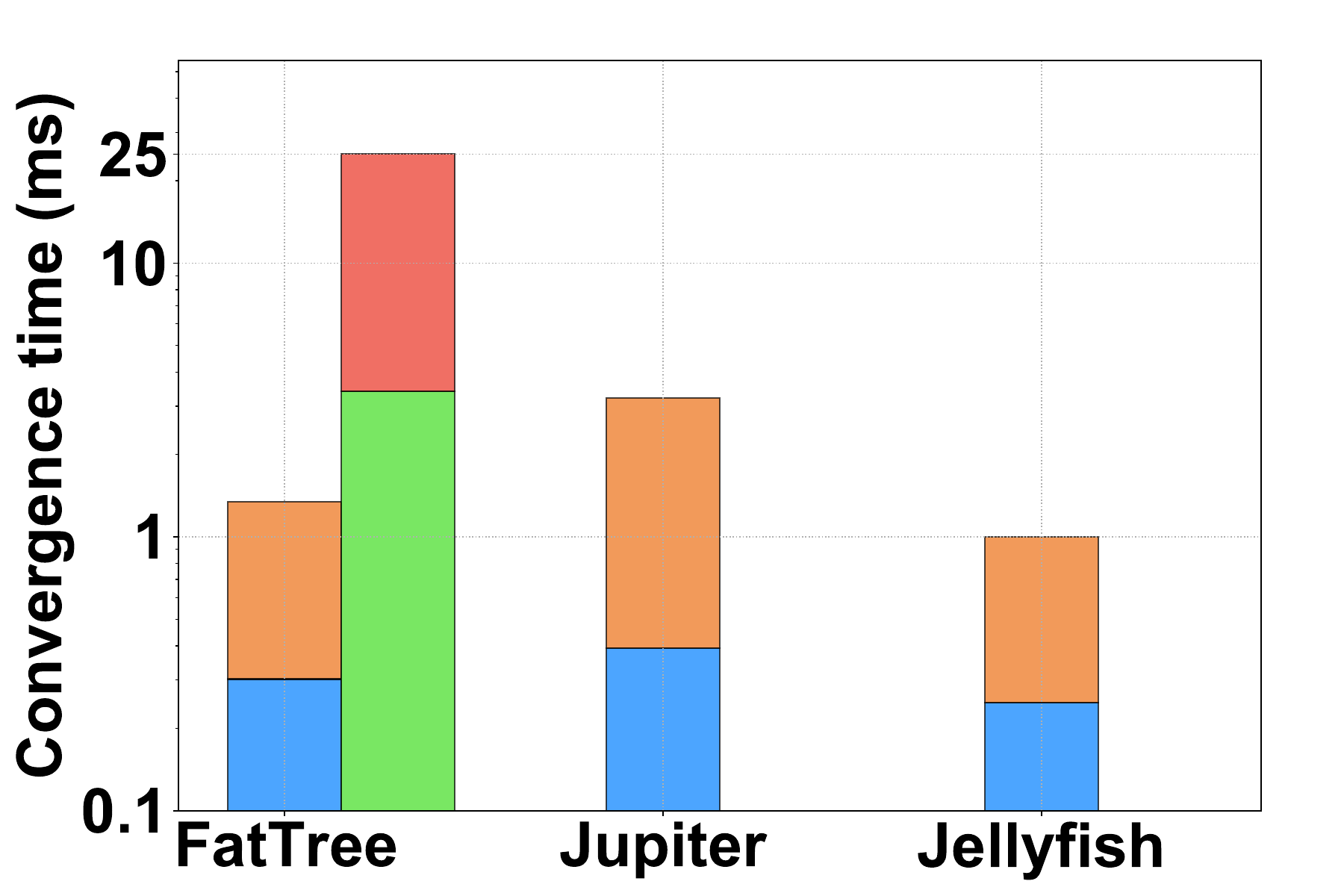}
        \end{center}
        }
        \label{subfig:mcast_conv_time_1fail}
        \vspace{-0.10165in}
        \end{minipage}
    }
    \subfigure[2 link failures.]{
        \begin{minipage}[t]{0.4\linewidth}{
        \vspace{-0.22in}
        \begin{center}
        \includegraphics[width=\textwidth, ]{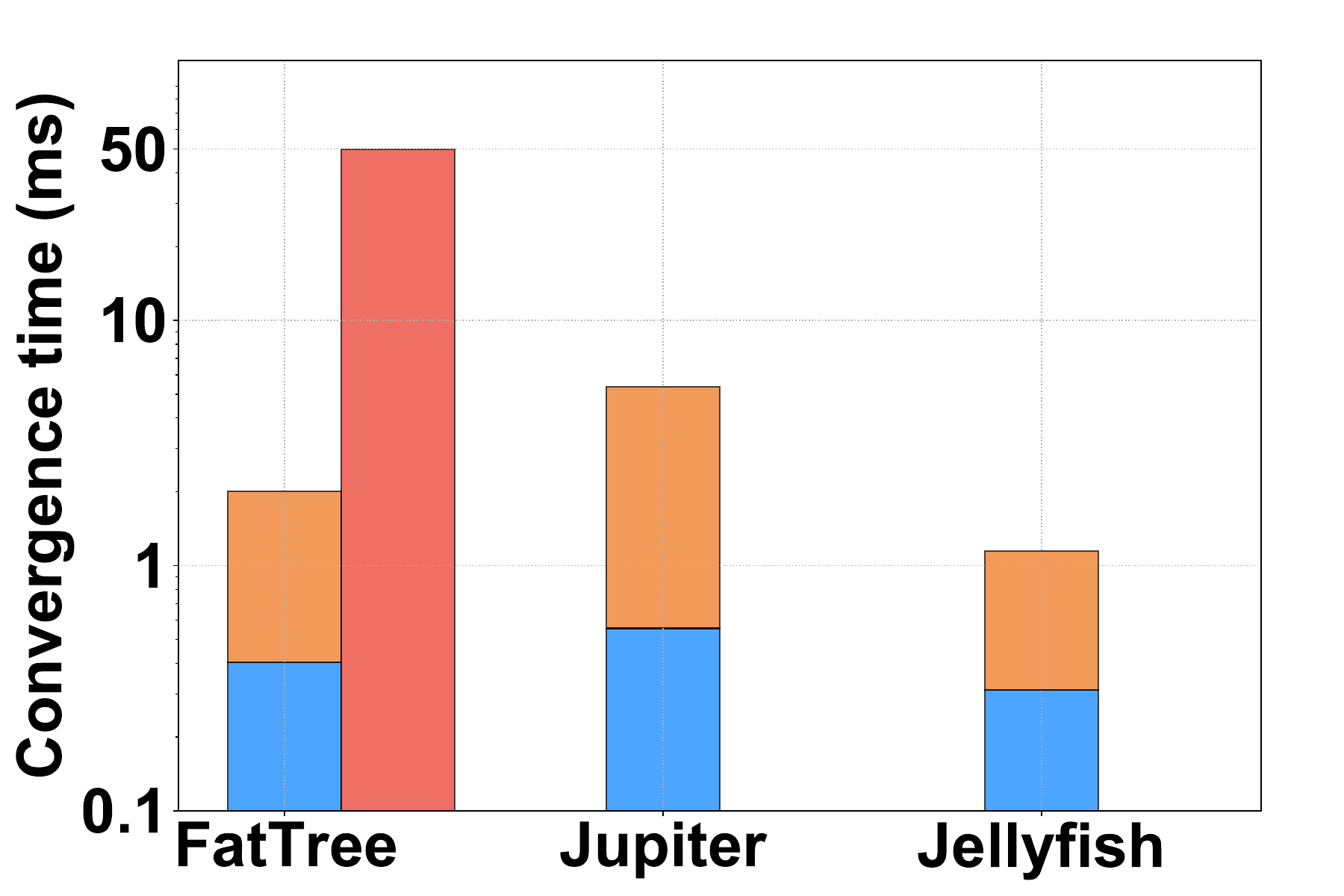}
        \end{center}
        \vspace{-0.10165in}
        }
        \label{subfig:mcast_conv_time_2fail}
        \end{minipage}
    }
    \vspace{-0.18in}
	\caption{Reaction time (log-scale) of \sysname for source-routed multicast compared with the FatTree-specific Elmo~\cite{Elmo}.}
	\label{fig:mcast-conv-time}
\end{figure}

\begin{figure}
    \raggedright
    \hspace{0.8in}
    \begin{minipage}[t]{0.5\linewidth}{
    \vspace{-0.05in}
    \begin{center}
    \includegraphics[width=\textwidth, ]{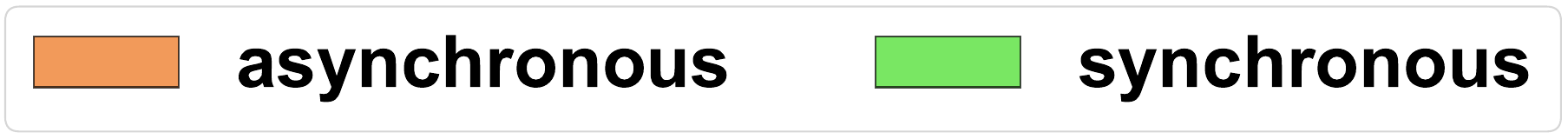}
    \end{center}
    }
    \label{subfig:async_sync_legend}
    \end{minipage}

    \centering
    \subfigure[1 link failure.]{
        \begin{minipage}[t]{0.4\linewidth}{
        \vspace{-0.22in}
        \begin{center}
        \includegraphics[width=\textwidth, ]{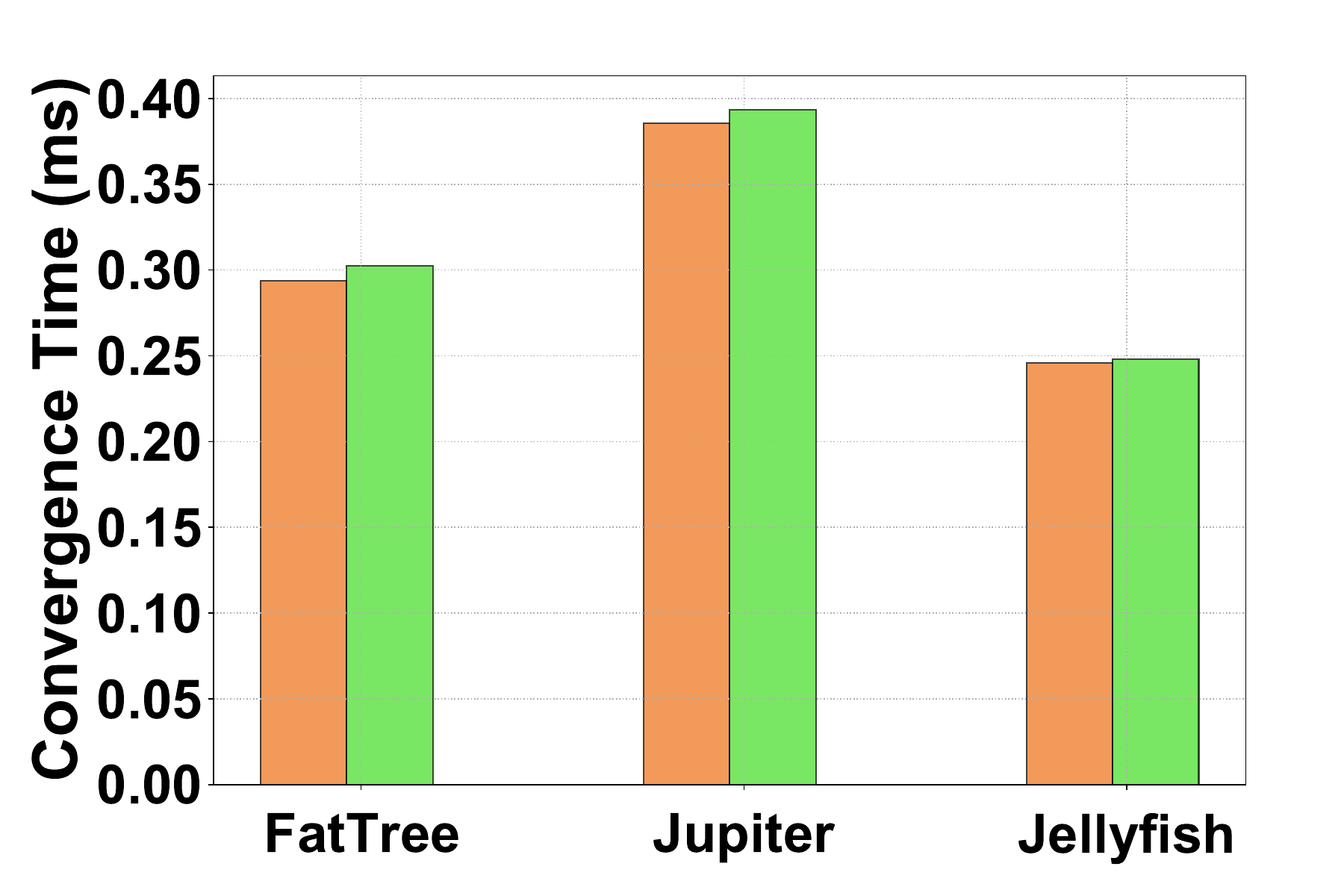}
        \end{center}
        \vspace{-0.10165in}
        }
        \label{subfig:async_vs_sync_1fail}
        \end{minipage}
    }
    \subfigure[2 link failures.]{
        \begin{minipage}[t]{0.4\linewidth}{
        \vspace{-0.22in}
        \begin{center}
        \includegraphics[width=\textwidth, ]{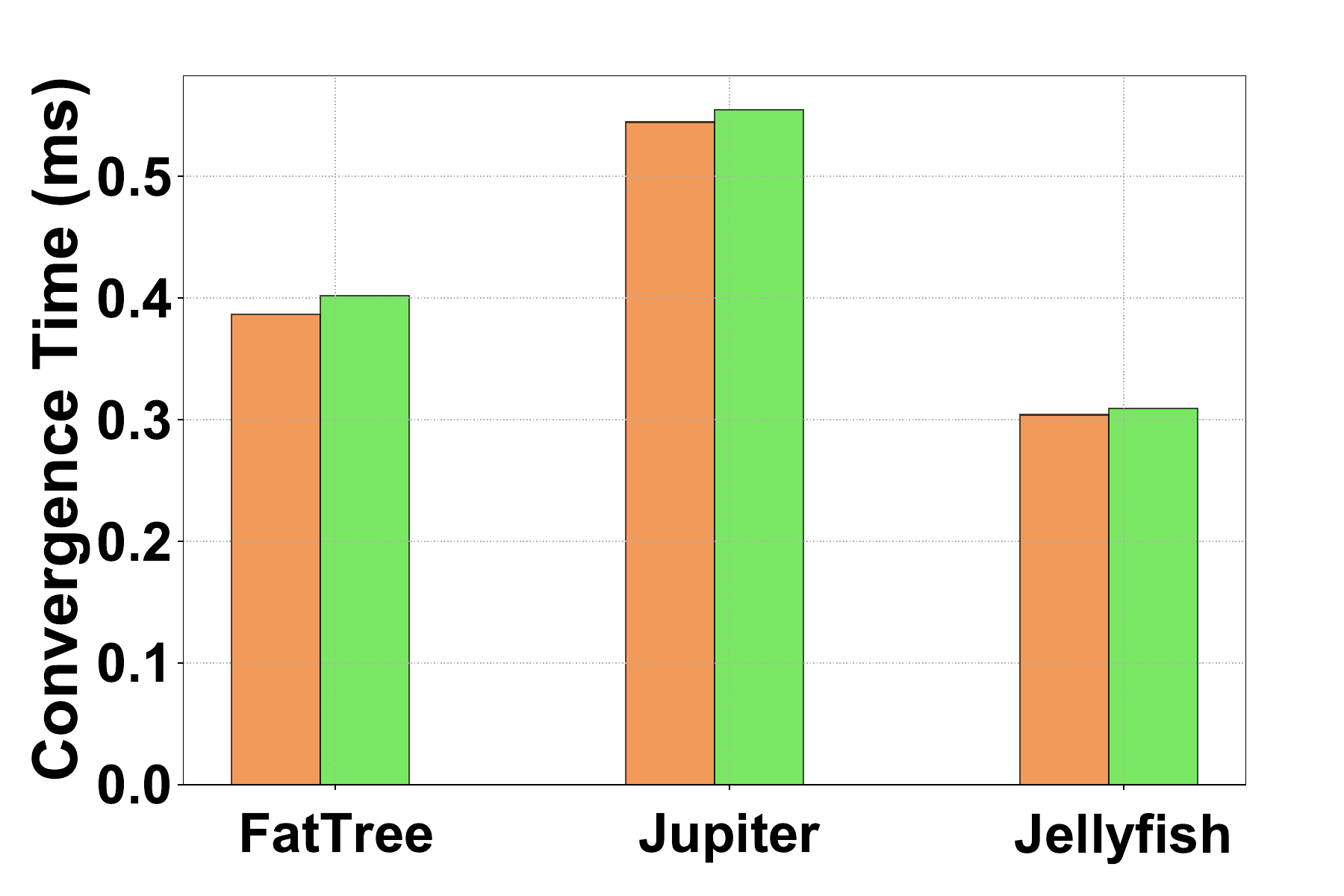}
        \end{center}
        \vspace{-0.10165in}
        }
        \label{subfig:async_vs_sync_2fails}
        \end{minipage}
    }
    \vspace{-0.13in}
	\caption{Reaction time comparison between asynchronous approach and synchronous approach for the fast recovery phase of source-routed multicast.}
	\label{fig:async_vs_sync}
\end{figure}

\vspace{-0.2cm}
\subsubsection{Multicast}\label{subsubsec: eval-multicast}
\vspace{-0.1cm}

\subp{Baseline solutions}. We compare with state-of-the-art source \mbox{routed multicast methods Elmo~\cite{Elmo} and Orca~\cite{orca}.}


\subp{Experiment setup.} We generate 1 or 2 link failures for our experiments and allocate $200$Mbps bandwidth for \sysname. We note that Elmo is designed specifically for FatTree topologies, and we are \textit{unable} to adapt it to other topologies. Other settings are shown in Appendix~\ref{appendix:exp-setups}.

\subp{Experimental results}. 
As depicted in Figure~\ref{fig:mcast-conv-time}, \sysname's fast recovery and distributed optimization are $11\times$ and $18-24\times$ faster than Elmo's fast and (control-plane based) full recovery, respectively.

Next, to evaluate the spanner algorithm for multicast, we compare additional metrics such as the header size and message overheads in~\cref{app:additionalComparisons}. 
As shown, our fast recovery solution has a traffic overhead is $2-3.9\times$ lower than Elmo's unicast-based fast recovery solution, while the distributed optimization solution is on par with both Elmo and Orca~\cite{orca} while reducing the header size by $1.4-2\times$. 

Finally, we also implement the asynchronous version of \texttt{bottom-up tree aggregation} for building fast recovery multicast trees. This is achieved by waiting for all of the children in the flooding tree to complete the aggregation and then executing the bitmap aggregation. Figure~\ref{fig:async_vs_sync} shows that the reaction time difference between the asynchronous version and the synchronous version is negligible ($<5\%$). On the other hand, as mentioned earlier, the benefits for \sysname to focus on synchronous algorithms include easier protocol design and better system-level optimization.

\subsubsection{Contra Routing}
\vspace{-0.1cm}
\label{subsubsec:exp-contra}

\subp{Experiment setup.} We compare Contra's \sysname adaptation against both the original Contra~\cite{hsu2020contra} and ONOS~\cite{berde2014onos}, a control-plane routing system. We also run Contra-\sysname without message packing to study effects of our wait-and-merge and message packing techniques separately. We use the FatTree topology and set the bandwidth limit as $200$Mbps.

\subp{Experimental results.}  Table~\ref{tab:routing-performance} shows that 
Contra-\sysname achieves an order of magnitude faster reaction speed than original Contra which is faster than the control-plane-based ONOS. 
Further, \sysname's wait-and-merge technique reduces the reaction time by $1.8\times$ and the message packing contributes to a further $5.3\times$ reduction. In general, this suggests that deploying existing asynchronous data-plane protocols into \sysname may significantly reduce the recovery time. 

{\small
\begin{table}
\centering
\resizebox{.9\linewidth}{!}{
    \begin{tabular}{|c|c|c|c|} \hline
        Methods & Category & Reaction time & Bandwidth Usage \\ \hline
        \begin{tabular}[c]{@{}l@{}}
        Contra-\sysname \\ Optimized
        \end{tabular} & 
        DP Sync & $1.720$ms & \multirow{5}{*}{$200$Mbps}  \\ \hhline{---~}
        \begin{tabular}[c]{@{}l@{}}
        Contra-\sysname \\ Unoptimized
        \end{tabular} & 
        DP Sync & $9.186$ms &   \\ \hhline{---~}
        \begin{tabular}[c]{@{}l@{}}
        Original Contra
        \end{tabular} & DP Async & $17.15$ms &   \\ \hline
        ONOS~\cite{berde2014onos} & CP & $145.8$ms (expected) & /  \\ \hline
    \end{tabular}
}
    \caption{We compare Contra-\sysname with the original Contra protocol and ONOS. Here optimized and unoptimized refer to if we apply message packing or not. DP and CP refer to data plane  and control plane respectively.}
    \label{tab:routing-performance}
\end{table}
}

\vspace{-0.1cm}
\subsection{Running Concurrent Applications}
\vspace{-0.05cm}

\label{subsec:exp-combined}

\begin{figure}
    \centering
    \vspace{-0.05in}
    \subfigure[Comparing \sysname-individual with \sysname-combined.]{
        \begin{minipage}[t]{0.5\linewidth}{
        \vspace{-0.00in}
        \begin{center}
        \includegraphics[width=\textwidth, ]{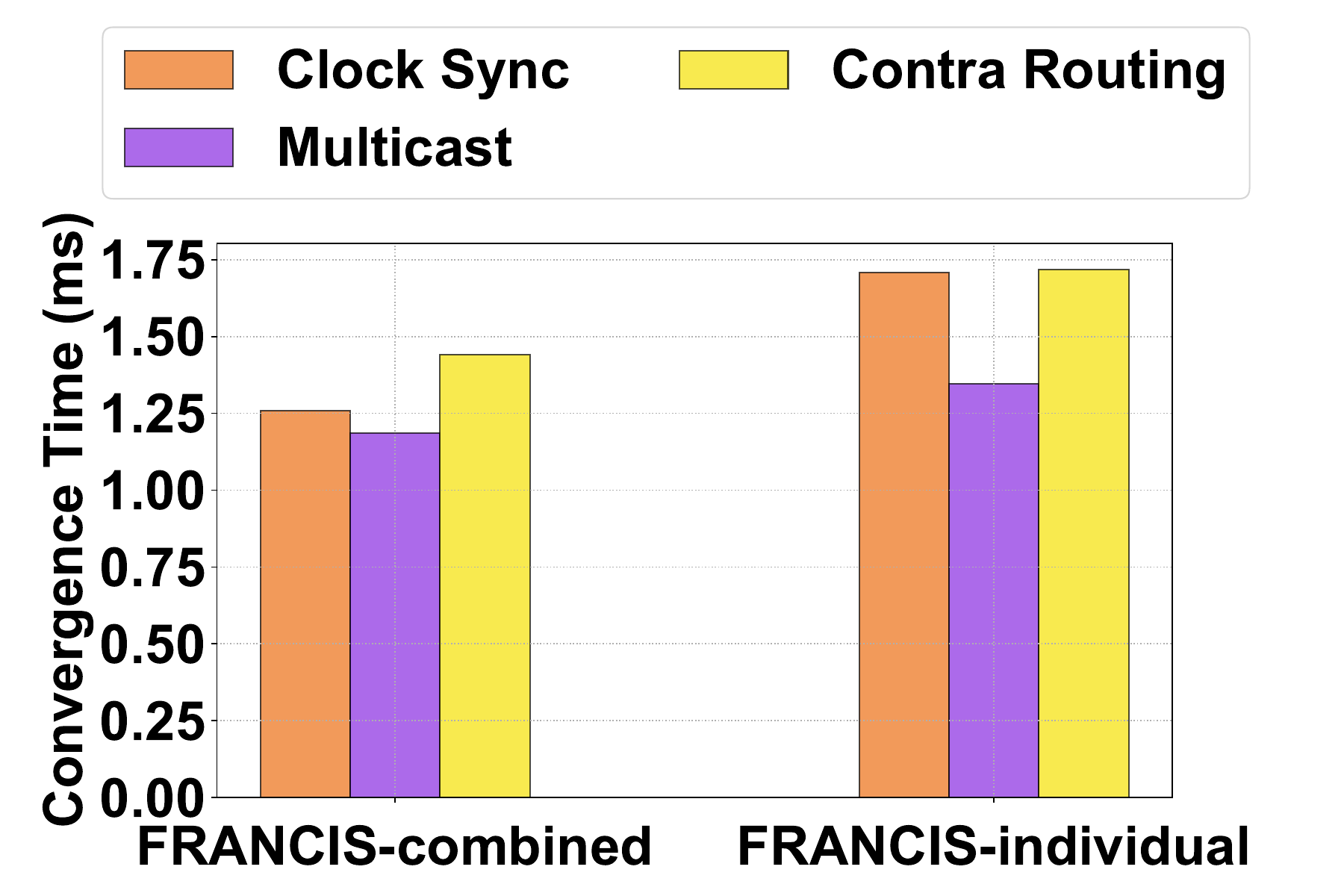}
        \end{center}
        \vspace{-0.02in}
        }
        \label{subfig:overall_w_Contra}
        \end{minipage}
    }
    \hspace{-0.06in}
    \subfigure[Varying total allocated memory.]{
		\begin{minipage}[t]{0.45\linewidth}{
		\vspace{-0.00in}
		\begin{center}
		\includegraphics[width=\textwidth]{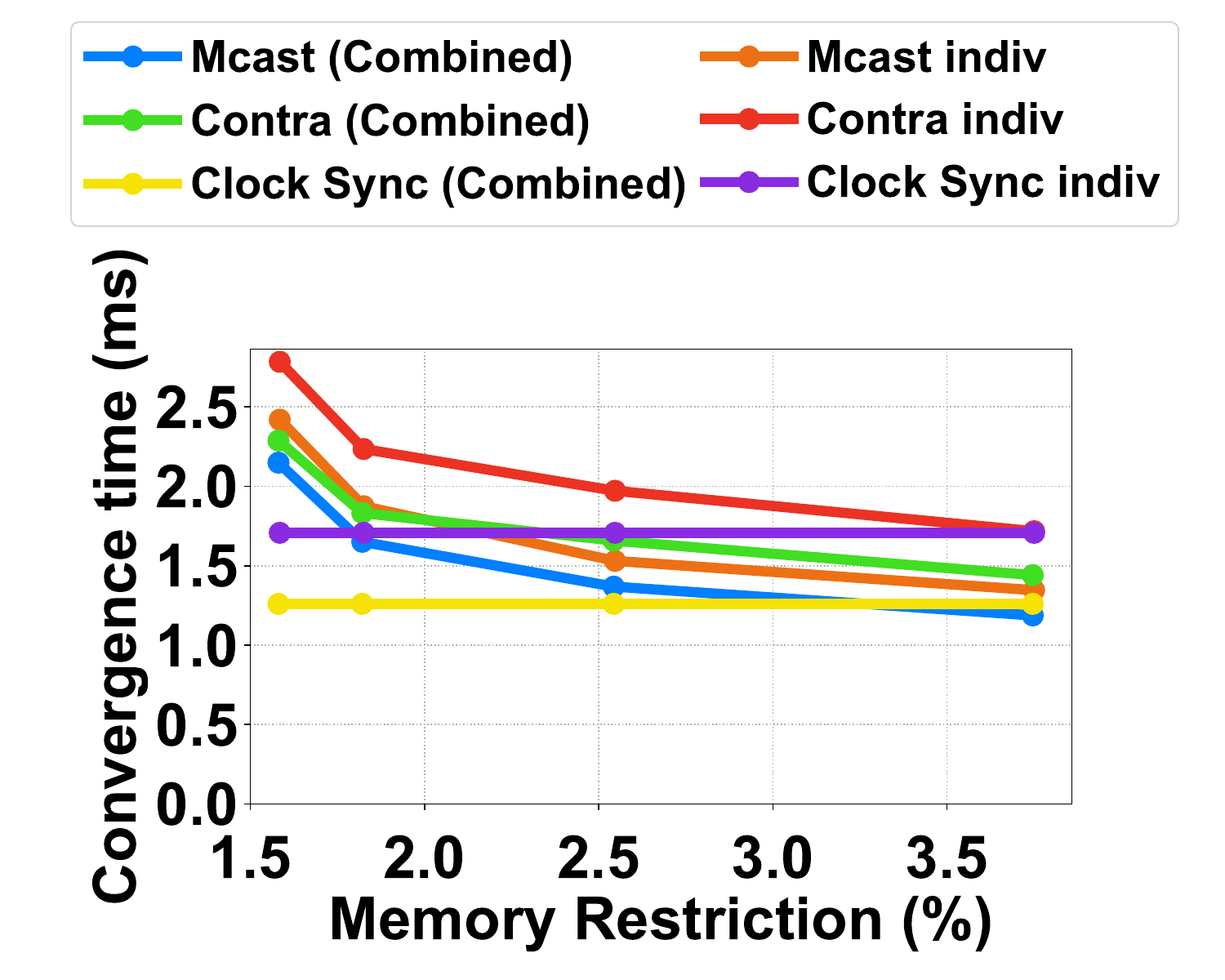}
		\end{center}
            \vspace{-0.1in}
		}
		\label{subfig:overall_varying_memory}
		\end{minipage}
	}
	\vspace{-0.16in}
	\caption{Evaluation on running concurrent applications. Here \sysname denotes \sysname-individual.}
    \vspace{0.2cm}
	\label{fig:clock-sync-topology}
\end{figure}



\subp{Experiment setup}. We next evaluate running multiple concurrent applications.
The baseline solution, ``\sysname-individual'', simply launches multiple \sysname instances, one for each application. Each application is given a fixed amount of bandwidth and memory resources, and different applications do not share algorithm modules. 
Our alternative is ``\sysname-combined'', where one instance runs all applications {while sharing modules and resources across applications. } To implement the combined version, we share the functionally identical instances of \texttt{spanning tree} and \texttt{Shortest path tree} for both clock synchronization and multicast (and remove the redundant instances from the clock synchronization). \mbox{Other details can be found in Appendix~\ref{appendix:exp-setups}}. 

\subp{Experimental results}. 
As Figure~\ref{subfig:overall_w_Contra} shows, the reaction time for all applications in the \sysname-combined is improved compared with \sysname-individual. For clock synchronization, its reaction time is improved by $1.35\times$ as the redundant algorithm components of clock synchronization are removed. For both multicast and Contra routing, \sysname-combined improvements arise from bandwidth sharing that improves the bandwidth utilization efficiency. For instance, if Contra routing is idle on some links, its bandwidth is allocated to \mbox{multicast, speeding up the reaction.}

As Figure~\ref{subfig:overall_varying_memory} shows, because of Tofino's~\cite{Tofino} architecture, \sysname's scheduler cannot adjust the memory usage of the clock synchronization. For other applications, algorithm scheduling (Section~\ref{subsubsec:efficient-communication}) enables some algorithm instances to execute in parallel to fit into different memory constraint settings. Thus, the reaction time of the clock synchronization application remains the same, while the reaction time of multicast and Contra increases as we shrink the allocated switch memory. Finally, we note \mbox{that the gains are consistent for the whole memory range.}

\begin{figure}[tb]
	\centering
        \hspace{0.1in}
		\begin{minipage}[t]{0.6\linewidth}{
		\vspace*{0.0in}
		\begin{center}
		\includegraphics[width=\textwidth, ]{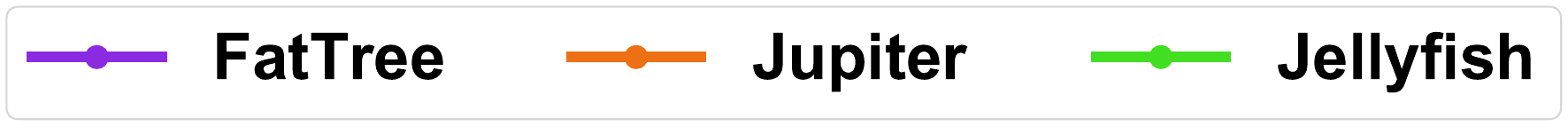}
		\end{center}
		}
		\end{minipage}

    \subfigure[Header size.]{
		\begin{minipage}[t]{0.31\linewidth}{
		\vspace{-0.0510in}
		\begin{center}
		\includegraphics[width=\textwidth, ]{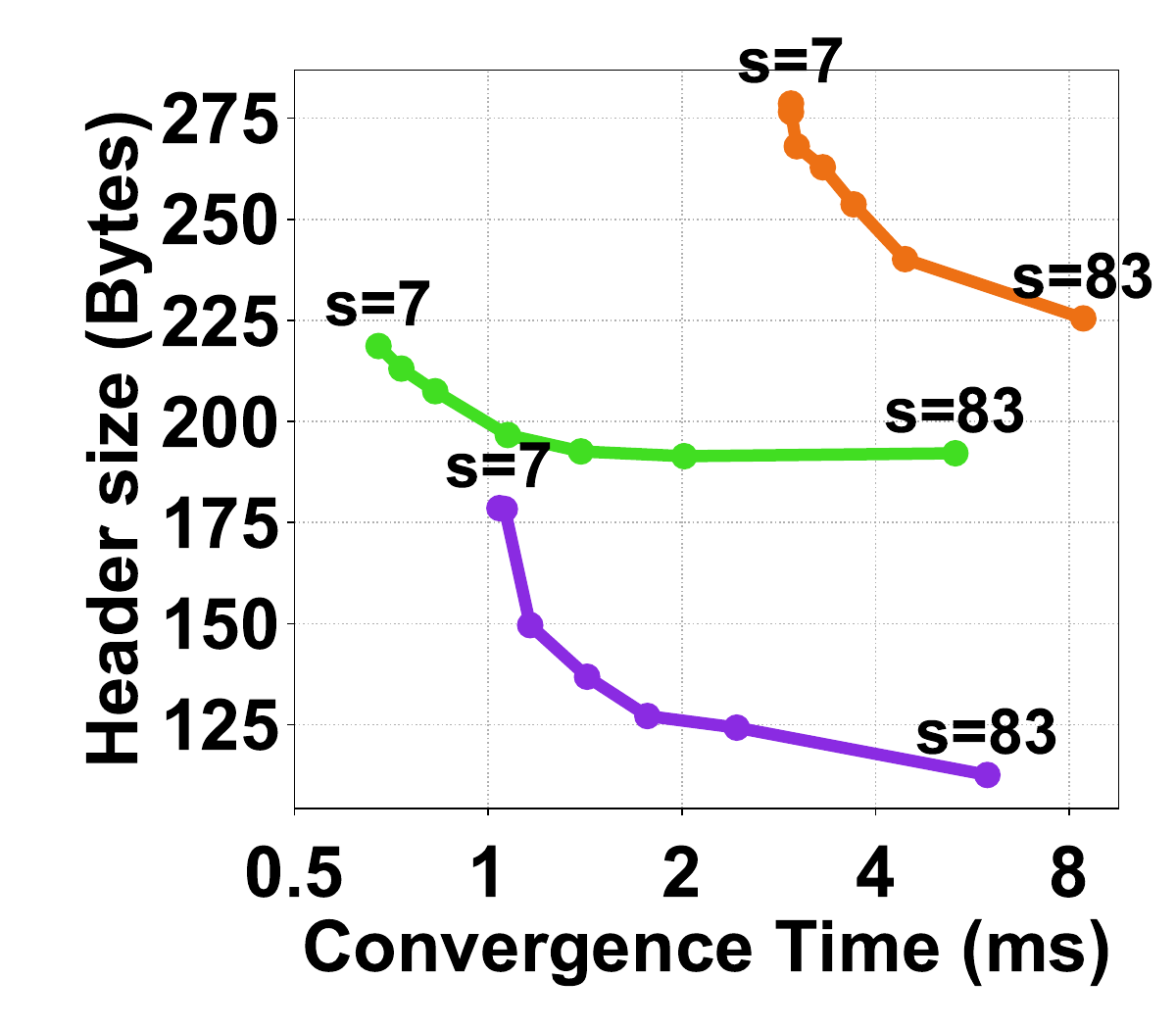}
		\end{center}
		}
		\label{subfig:/mcast_spanner_perform_Header Size (Bytes)}
		\end{minipage}
	}
    \hspace{-0.1in}
    \subfigure[Average latency ($\mu$s).]{
		\begin{minipage}[t]{0.31\linewidth}{
		\vspace{-0.051in}
		\begin{center}
		\includegraphics[width=\textwidth, ]{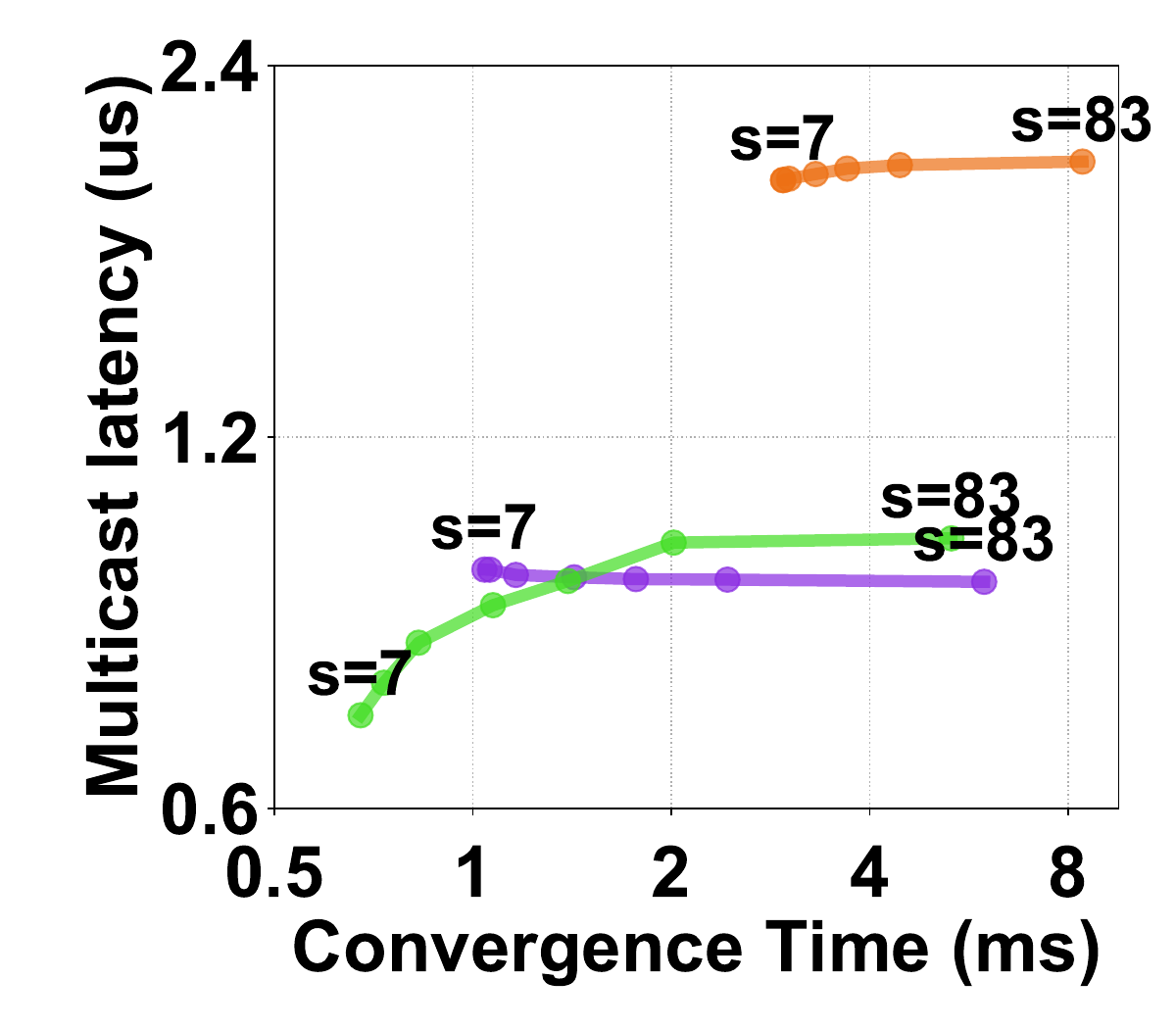}
		\end{center}
		}
		\label{subfig:mcast_spanner_perform_Multicast average latency (us)}
		\end{minipage}
	}
    \hspace{-0.05in}
    \subfigure[Multicast overhead.]{
		\begin{minipage}[t]{0.31\linewidth}{
		\vspace{-0.051in}
		\begin{center}
		\includegraphics[width=\textwidth, ]{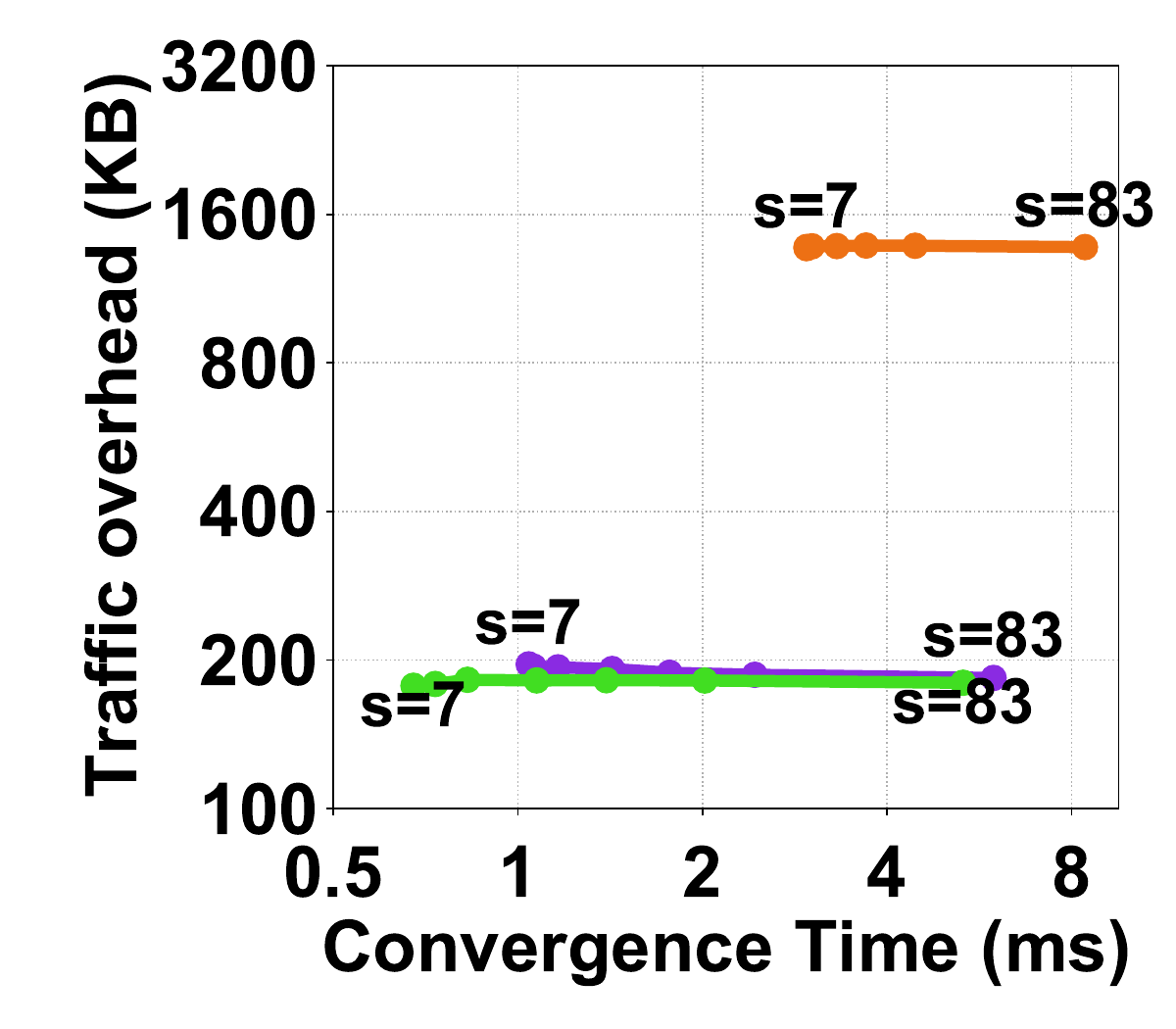}
		\end{center}
		}
		\label{subfig:mcast_spanner_perform_Total incurred bandwidth (KB)}
		\end{minipage}
	}

	\vspace{-0.12in}
	\caption{Evaluation on the spanner algorithm for multicast. Specifically, we vary the stretch $s$ from $7$ to $83$.}
    \vspace{-0.01096205in}
	\label{fig: mcast-varying-k}
\end{figure}

\vspace{-0.1cm}
\subsection{\mbox{Flexibility in achieving various tradeoffs}}

\label{subsec:param-settings}

In this section, we demonstrate how \sysname allows users to flexibly customize their distributed algorithms and \sysname's resource usage, achieving different tradeoffs. We also present our evaluation for bandwidth allocation and reliably handling packet loss in Appendix~\ref{appendix:flex-reliable-exps}.

\smallskip
\subp{Tradeoffs between reaction time and performance.} 

\subp{Experiment setup.}
In our multicast use case, we vary the \texttt{strech} parameter from $7$ (the smallest allowed in~\cite{spanner}'s algorithm) to $83$ and \mbox{follow the previous setup parameters.}

\subp{Experimental results.}
In Figure~\ref{subfig:/mcast_spanner_perform_Header Size (Bytes)}, increasing the stretch value decreases the header size by up to $38\%$. In Figure~\ref{subfig:mcast_spanner_perform_Multicast average latency (us)}, larger stretch adds multicast latency in Jellyfish and Jupiter but slightly reduces the multicast latency in the FatTree. To explain this, the multicast latency is determined by both the packet size and the path length between the root of the multicast tree (source) and the leaves (destinations). Empirically, sparser graphs with higher stretch tend to result in longer path length but lower packet size. Finally, in Figure~\ref{subfig:mcast_spanner_perform_Total incurred bandwidth (KB)}, the stretch has little impact on the total network bandwidth consumption. Accordingly, users can set a suitable \mbox{stretch parameter to obtain the best tradeoffs. }

\begin{figure}[tb]
	\centering
	\vspace{-0.005405in}
    		\begin{minipage}[t]{0.6\linewidth}{
		\vspace{-0.0in}
		\begin{center}
		\includegraphics[width=\textwidth, ]{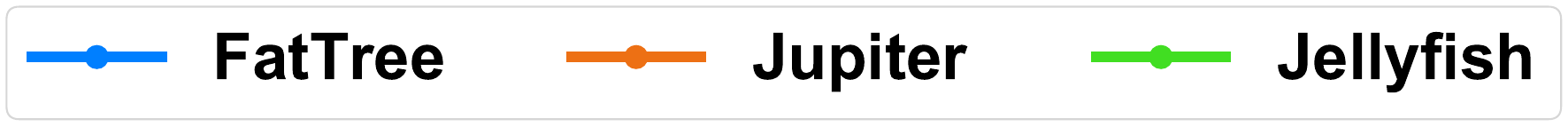}
		\end{center}
		}
		\end{minipage}
 
    \subfigure[Makespan.]{
		\begin{minipage}[t]{0.45\linewidth}{
		\vspace{-0.051in}
		\begin{center}
		\includegraphics[width=\textwidth, ]{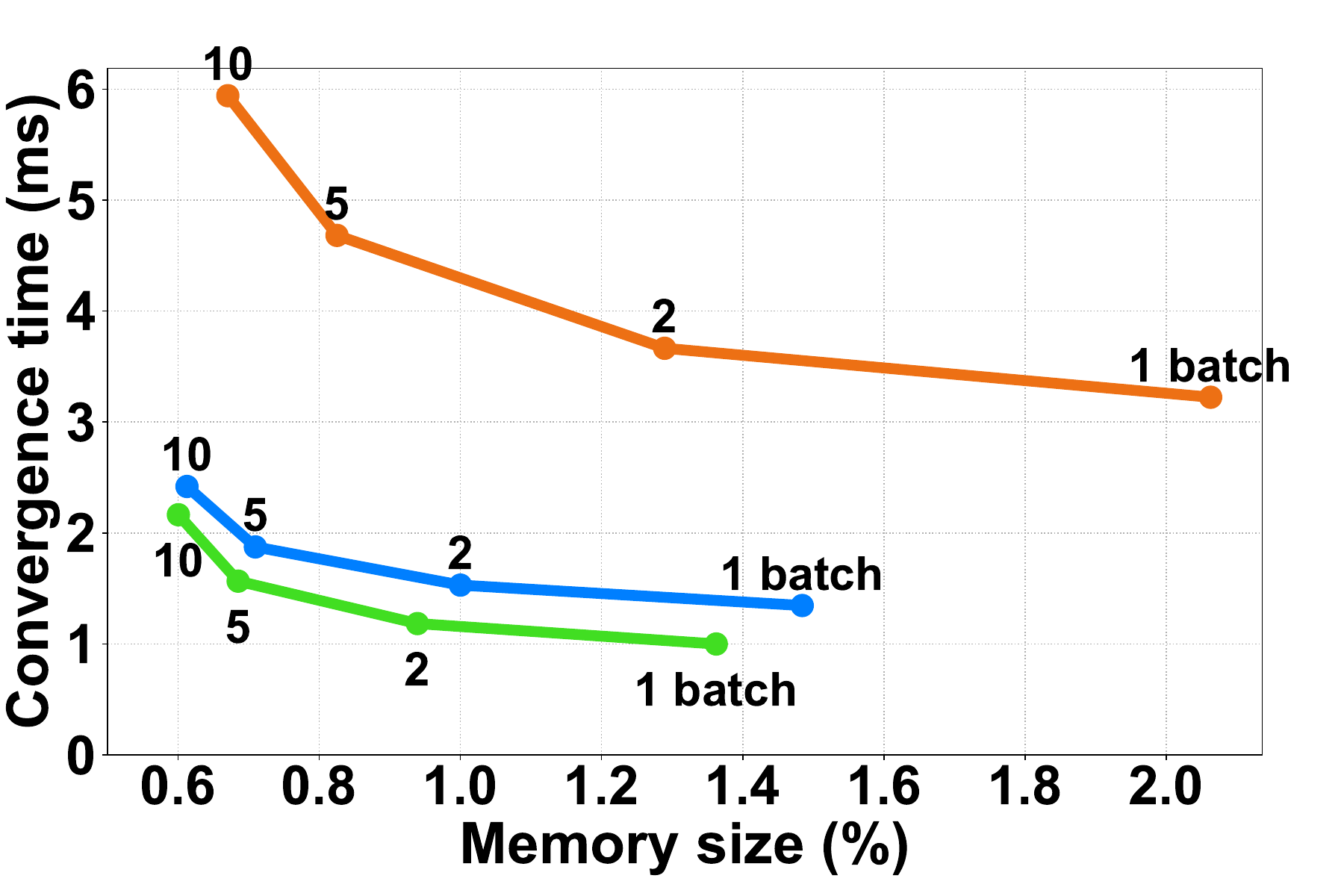}
		\end{center}
		}
            \vspace{-0.3cm}
		\label{subfig:mcast-makespan}
		\end{minipage}
	}
	\subfigure[Median reaction time.]{
		\begin{minipage}[t]{0.45\linewidth}{
		\vspace{-0.051in}
		\begin{center}
		\includegraphics[width=\textwidth, ]{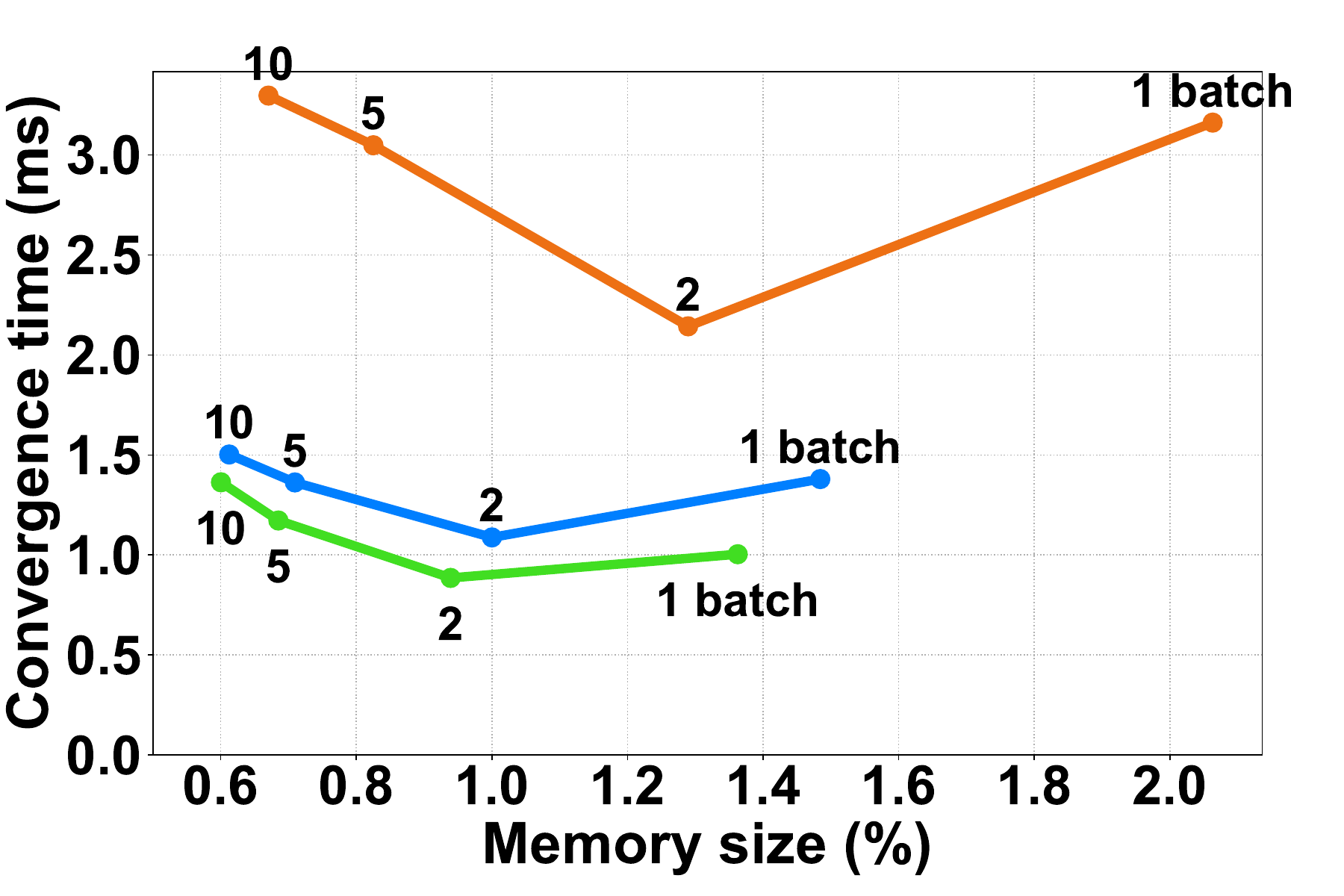}
		\end{center}
		}
            \vspace{-0.3cm}
		\label{subfig:mcast-median-convergence-time}
		\end{minipage}
	}
	\vspace{-0.134in}
	\caption{Evaluation \mbox{for memory limitation in multicast use} case. Here our \mbox{scheduling policy is to optimize Makespan.}}
    \vspace{-0.01in}
	\label{fig: mcast-memory-size}
\end{figure}

\begin{figure}[tb]
	\centering
	\vspace{-0.2cm}
    \subfigure[Time to latency.]{
		\begin{minipage}[t]{0.47\linewidth}{
		\vspace{-0.0in}
		\begin{center}
		\includegraphics[width=\textwidth, ]{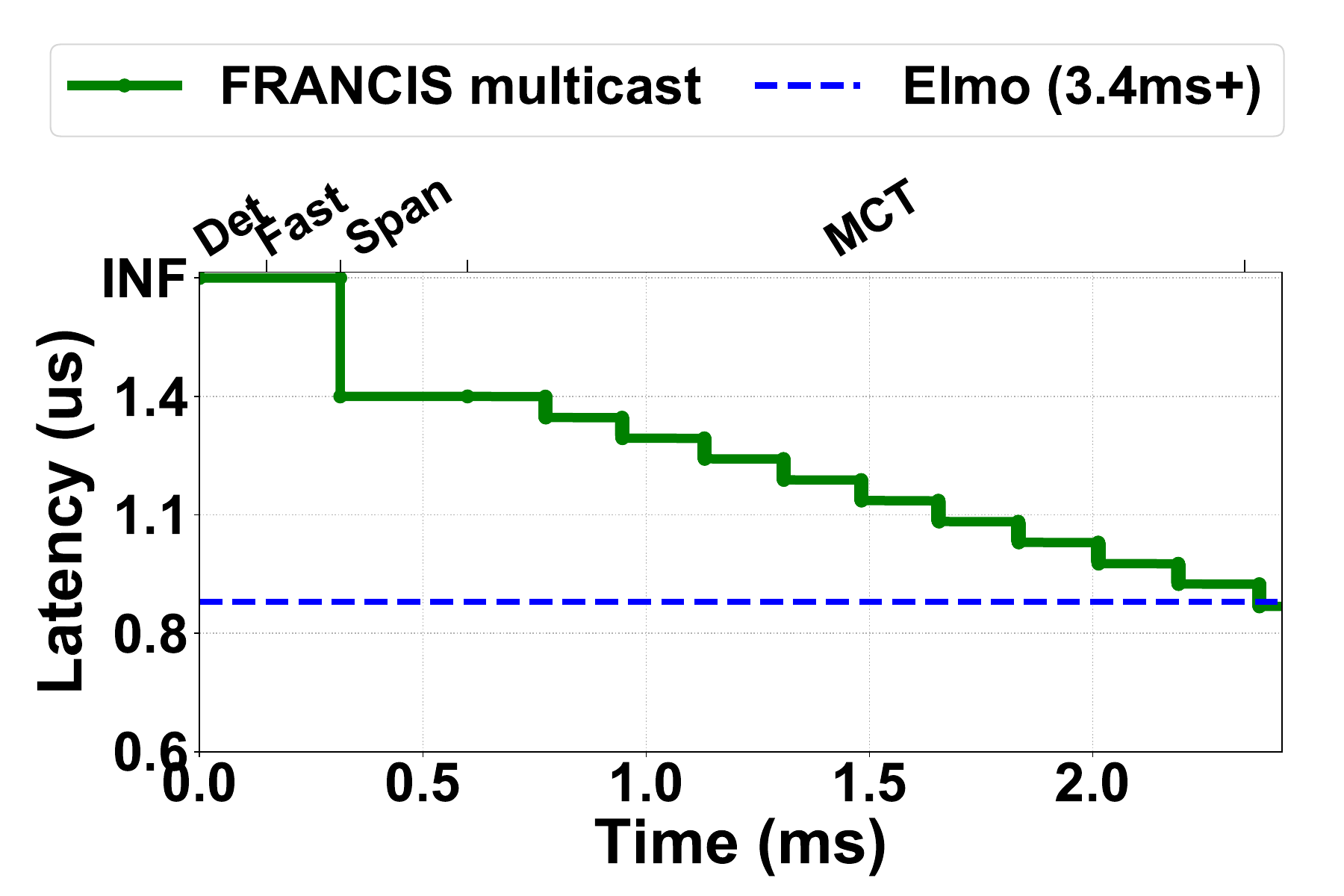}
            \vspace{-0.6cm}
		\end{center}
		}
		\label{subfig:mcast_performance_dynamics_Latency (us)}
		\end{minipage}
	}
    \subfigure[Time to multicast traffic overhead.]{
		\begin{minipage}[t]{0.47\linewidth}{
		\vspace{-0.0in}
		\begin{center}
		\includegraphics[width=\textwidth, ]{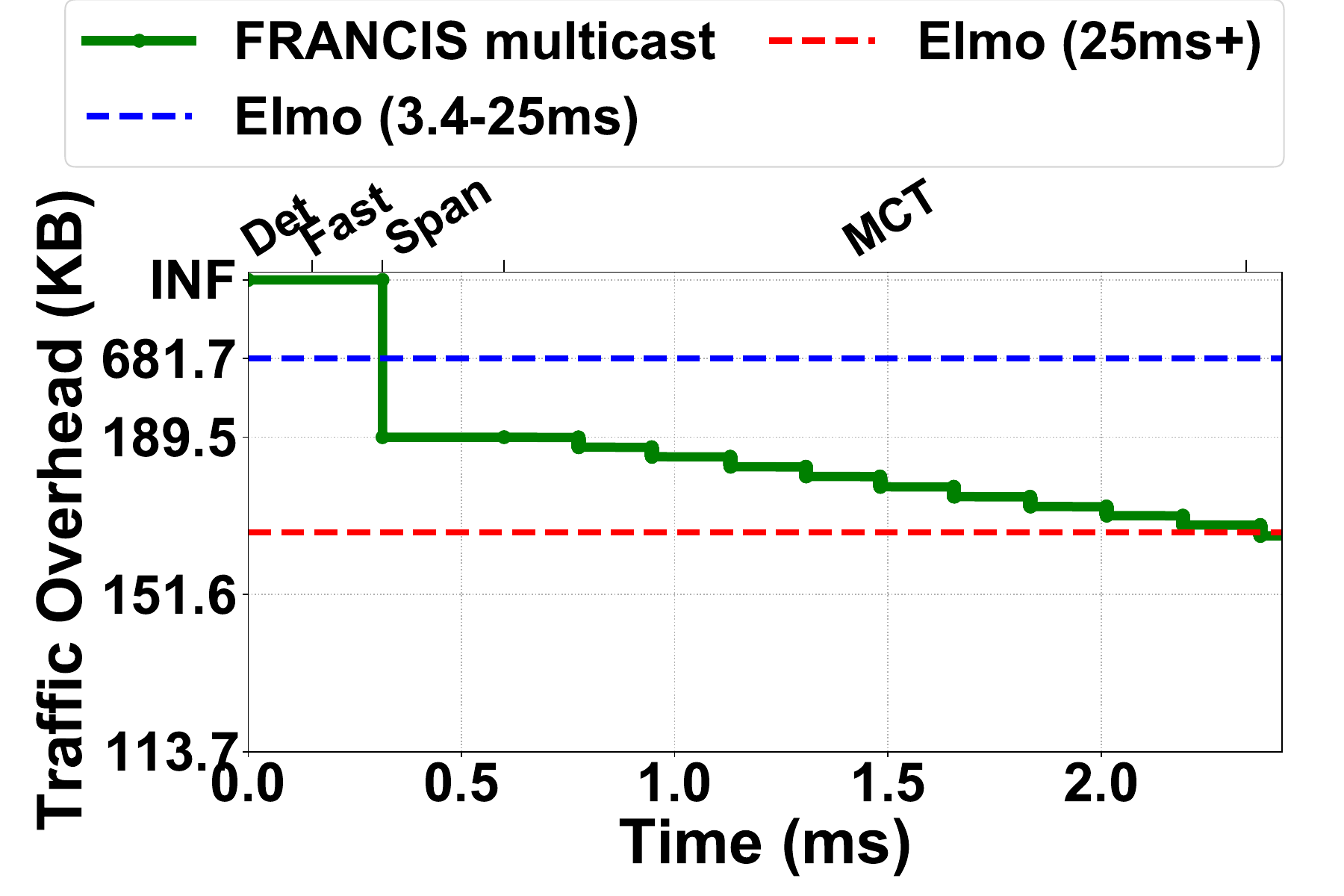}
            \vspace{-0.6cm}
		\end{center}
            
		}
		\label{subfig:mcast_performance_dynamics_Traffic Overhead (KB)}
		\end{minipage}
	}
	\vspace{-0.3cm}
	\caption{Time to performance with limited memory. MCT denotes multicast tree construction.}
	\label{fig: mcast-time-to-performance}
\end{figure}

\smallskip
\subp{Tradeoffs between reaction time and memory.} 

\subp{Experiment setup.} As detailed in Section~\ref{subsubsec:algorithm-scheduling}, \sysname's execution scheduler flexibly adapts to different memory constraints via batch-by-batch execution, which schedules more but smaller batches for lower memory size. In our multicast use case, we adjust the memory size so that the resulting number of batches ranges from $1$ to $10$. 

\subp{Experimental results.} Figure~\ref{fig: mcast-memory-size} shows that \sysname achieves flexible memory adjustment. As expected, the Makespan (the time at which the last algorithm terminates) increases monotonically when shrinking memory sizes. In the Jupiter topology, the reaction time is nearly doubled as we reduce the memory from $2.06\%$ to $0.67\%$.
However, for the median reaction time over failed multicast groups, we observe that the value is minimized with 2 batches. This is because we optimize Makespan, which does not take into account that batches that are scheduled early will be recovered sooner. We leave it as \mbox{future work to optimize for different metrics.}

Indeed, such a batch execution progressive recovery schedule can be observed in other performance metrics, as demonstrated in Figure~\ref{fig: mcast-time-to-performance}. These results show how the fast recovery algorithm restores multicast service with suboptimal performance and that with batch execution, the performance metrics are further optimized progressively.

\section{Related work}\label{sec:related}

\textbf{Clock synchronization.} 
The last decade has witnessed research efforts such as DTP~\cite{lee2016globally}, Huygens~\cite{geng2018exploiting}, and DPTP~\cite{dptp} optimizing against the synchronization precision. These works, however, do not take into account network failures which, if not handled timely, lead to magnitudes higher time uncertainty bound. While Precision Time Protocol (PTP)~\cite{PTP} and Sundial~\cite{Sundial} design heuristic algorithms to handle failures (which \sysname outperforms), \sysname is not tied to any specific algorithm but enables incorporation of more advanced protocols. Recently, Nafaji and Wei proposed an orthogonal technique that minimizes the local drift rate~\cite{Graham}. Indeed, such a method reduces \texttt{max\_drift\_rate} (see Eq.~\ref{eq:epsilon}) and improves the uncertainty bound for \sysname as well.

\noindent\textbf{Source routed multicast.} 
Source-routed multicast achieves good scalability for multicast without requiring storing multicast rules inside switches' SRAM. Elmo~\cite{Elmo} and Orca~\cite{orca} encode multicast trees as a set of multicast rules in each packet at the sender. Both incur considerable traffic overhead and do not support real-time failure recovery.
IP multicast~\cite{ip-multicast-rfc} is often even slower, as its detection is reported to require up to 3 minutes to recover from a failure.

\noindent\textbf{Routing.}  
Liu et al.~\cite{liu2013ensuring} propose to let user packets explore new routes, but this explore-with-detour increases the latency of user traffic. F10~\cite{liu2013f10} designs heuristics to recover from a network event, but their algorithm is specific to the ab-FatTree topology. Contra~\cite{hsu2020contra} proposes an asynchronous protocol to support different routing policies but is less efficient in maintaining low bandwidth consumption than \sysname.

\noindent\textbf{Fault tolerant switch state applications.}
The recently introduced RedPlane~\cite{RedPlane} makes running applications that rely on switch state robust to network failures. Their solution allows replicating switch state to other switches, thereby ensuring functionality in case of a failure.

\section{Discussion}\label{sec:discussion}

%
The message-passing algorithms we employ assume a static network topology, but we utilize them in a \emph{dynamic} network where changes could happen any time. Our solution (Section~\ref{sec:distributed-framework}) is to reinvoke reactions upon every event. 
An alternative is \emph{dynamic} message-passing algorithms~\cite{baswana2012fully,censorhillel_et_al:LIPIcs:2021:13513} that leverage previous solutions to derive a new one on a topology change, which may reduce round complexity. However, these algorithms are often more complex and may require more switch resources, and we found our solution sufficiently fast.

Possible future work includes defining a language of \sysname and implementing a compiler to automatically translate user code into switch implementation. Some components of the language would be an execution graph that specifies the dependency of different algorithms, and an objective function, in terms of convergence time or other performance metrics. The compiler would compute an execution plan to satisfy all constraints and minimize the objective function.

We remark that after completing \sysname's reaction, users may optionally launch the controller to perform some management tasks, such as running a centralized algorithm to obtain a fully optimal solution, or offloading the switches' states back to end hosts to free up switch memory. 


\clearpage

\bibliographystyle{abbrv} 
\begin{small}
\bibliography{nsdi24}
\end{small}

\newpage
\appendix

\section{Theoretical problems and algorithms} 

\textbf{The CONGEST model} is one of the main models studied in the distributed graph algorithms community, and most practical message-passing algorithms follow this model~\cite{Linial92, peleg2000distributed}.
The network is modeled as a graph, $G(V,E)$, where nodes are computational units and edges are communication links. 
We denote the number of vertices by $n=|V|$, the neighbors of $v$ in $G$ by $N(v)$, and the diameter of $G$ by $D$. 
We assume that every node in the network has a unique $O(\log n)$ bit identifier.  
In the CONGEST model, communication takes place in synchronous rounds,
where at the beginning of the round a node receives all messages sent by its neighbors. Next, the node performs some computation and updates its local state. At the end of the round, the node sends (potentially different) messages to all of its neighbors, which are delivered at the beginning of the next round. The size of each message is bounded by $O(\log n)$ bits.
One performance metric used in this model is the \emph{round complexity} of the algorithm, which is the number of synchronous communication rounds required for completing a task. Another metric is the \emph{message complexity} of the algorithm -- the number of total messages sent during the execution.
The focus on communication cost over local computation is meant to model the fact that the main bottleneck in the performance of real-world \mbox{distributed systems is often the communication cost.}

Alternatively, \textbf{asynchronous event-based algorithms} provide another framework for distributed algorithms that has been widely studied. Unlike the CONGEST model, which executes on a per-round basis, the asynchronous event-based algorithms proceed on a per-packet basis: for each incoming packet, the switch may decide to update its stateful memory and multicast messages to some neighbors. As discussed in Section~\ref{subsec:why-synchronous-algorithms}, since asynchronous algorithms can typically be converted to synchronous CONGEST algorithms and they are typically more bandwidth-consuming than CONGEST algorithms, CONGEST algorithms are a better design choice for the distributed optimization phase. However, for the fast recovery phase, asynchronous algorithms can be a better choice because they enable immediate reactions to incoming packets without waiting for all packets in a round, which can reduce the convergence time.

\textbf{Graph problems.}
While we have presented three use cases to show the effectiveness of our framework in improving the performance of networking applications, we also list several other theoretical problems and algorithms (e.g., maximal independent set, minimum vertex cover, etc.) that are potentially useful for networking. The relevant theory problems are given in Table~\ref{tbl:algs}.

\begin{table*}[t]
\centering
\resizebox{.95\linewidth}{!}{
\begin{tabular}{|l|l|c|c|l|l|}
\hline
\textbf{Problem name} & \textbf{Description} & \textbf{Complexity} & \textbf{Deterministic}& \textbf{Example application} & \textbf{Reference} \\ \hline\hline
$(2k-1)$-Spanner & 
\begin{tabular}[c]{@{}l@{}}An edge set $S\subseteq E$ of size \\$|S|=O(kn^{1+1/k})$ such that for all $u,v$:\\ $d_{(V,S)}(u,v)\le (2k-1) d_{(V,E)}(u,v)$.\end{tabular}& $O(k^2)$ & N & \begin{tabular}[c]{@{}l@{}}Source-routing\\multicast\end{tabular} & Baswana and Sen~\cite{spanner}\\ \hline
\begin{tabular}[c]{@{}l@{}}Shallowest spanning tree\\ 2-approx (shortest path tree).\end{tabular} &
\begin{tabular}[c]{@{}l@{}}A rooted spanning tree\\ of minimum depth.\end{tabular}& $O(D)$ & Y & \begin{tabular}[c]{@{}l@{}}Clock synchronizaiton\end{tabular} & Section~\ref{subsec:clock-sync}\\ \hline
Leader election & \begin{tabular}[c]{@{}l@{}}Elect one vertex as "leader".\end{tabular} & {\normalfont $O(D)$} & Y & Clock synchronization & Folklore \\ \hline\hline
\begin{tabular}[c]{@{}l@{}}Min spanning tree\end{tabular} & A tree with minimum weight. & {\normalfont $O(n \log n)$ } & Y & Loop prevention & Boruvka~\cite{boruuvka1926jistem} \\ \hline
\begin{tabular}[c]{@{}l@{}}Maximal \\ Independent Set\end{tabular} & \begin{tabular}[c]{@{}l@{}}A set of vertices without edges between \\ them such that any other \\ vertex has neighbor in the set.\end{tabular} & { $O(\log n)$} & N & \begin{tabular}[c]{@{}l@{}}Concurrent wireless\\ transmission\end{tabular} & Luby~\cite{Luby86}, Alon et. al~\cite{AlonBI86} \\ \hline
\begin{tabular}[c]{@{}l@{}}Min vertex cover\\$(2+\epsilon)$-approx.\end{tabular} & \begin{tabular}[c]{@{}l@{}}A minimal weight vertex set such that\\  each edge has an endpoint in the set.\end{tabular} & $O\left(\frac{\log \Delta}{\log \log \Delta}\right)$ & Y & \begin{tabular}[c]{@{}l@{}}Assign edge-monitoring\\responsibilities \end{tabular} &  Ben-Basat et al.~\cite{ben2021optimal} \\ \hline
\end{tabular}
}
\caption{{A summary of the theory problems either applied to our use cases or potentially being useful for other networking problems.}
}
\label{tbl:algs}
\vspace{-0.15 in}
\end{table*}

\section{Pseudo code for message-passing algorithm specification}\label{sec:pseudocodes}

\begin{algorithm}
\caption{Workflow of a message-passing algorithm module. Users only need to implement the highlighted parts, and the blue-color code is function calls of our algorithmic primitives.}\label{alg:dis-congest-alg}
\begin{algorithmic}[1]
\Procedure{Synchronous CONGEST algorithms}{}
\EndProcedure
\State {\textbf{Configs}}:
\Indent
    \State {\textcolor{red}{Synchronizer $\in \{\alpha, \beta, \cdots\}$}}
    \State {\textcolor{red}{Priority: Integer}}
    \State {\textcolor{red}{Round\_number ($rcnt$): Integer expression}}
    \State {\textcolor{red}{Bandwidth and memory for this module.}}
    \State {\textcolor{red}{Parallel instance IDs $[l, r)$: $l, r \in$ Integer expression}}
    \State {\textcolor{red}{Algorithm\_parameters: Numbers and functions}}
\EndIndent
\For{\textbf{parallel} instance\_id $\in [l, r)$}
    \State{\textbf{Option 1:}} \Comment{algorithmic primitives provided by \sysname}
    \Indent
    \State{\textcolor{blue}{Primitive\_calls(params)}}
    \EndIndent
    \State{\textbf{Option 2:}} \Comment{Streaming Computation}
    \Indent
    \For{round\_id $\in$ \{$0, \cdots$, $rcnt - 1$\}}
    \State{Synchronization()}
    \State{\textbf{Upon} receiving a packet $p$:}
    \Indent
    \State{\textcolor{red}{Stream-processing($p$)}}
    \EndIndent
    \State{\textbf{Until} \textcolor{red}{Completion()} of the current round}
    \Indent
    \If {\textcolor{red}{Termination()}}
        \State{Term\_notify\_and\_exit}
    \Else {\quad \textcolor{red}{Send\_to\_neighbors()}}
    \EndIf
    \EndIndent
    \EndFor
    \EndIndent
    \State{\textbf{Option 3:}} \Comment{Computation at the end of the round}
    \Indent
    \For{round\_id $\in$ \{$0\cdots$rcnt-1\}}
        \State{Synchronization()}
        \State{Receive and stash packets from neighbors}
        \State{\textbf{Until} \textcolor{red}{Completion()} of the round:}
        \Indent
            \State{\textcolor{red}{Computation($pkts$)}}
        \EndIndent
        \If {\textcolor{red}{Termination()}}
            \State{Term\_notify\_and\_exit}
        \Else {\quad \textcolor{red}{Send\_to\_neighbors()}}
        \EndIf
    \EndFor
    \EndIndent
\EndFor
\end{algorithmic}
\end{algorithm}

\textbf{The distributed algorithm modules}. Implementation of a distributed algorithm module begins with a wide range of user configurations. As Algorithm~\ref{alg:dis-congest-alg} show, the configurations consist of 6 components. \texttt{Synchronizer} specifies which synchronizer to choose for the CONGEST algorithm. Different synchronizers achieve different performance tradeoffs and we elaborate them in Section~\ref{subsec:backend-synchronization}. \texttt{Priority} is an integer to determine the priority of allocating resources to algorithm instances in this module and scheduling their execution. \texttt{Round\_number} refers to the number of rounds in this CONGEST algorithm module. \texttt{Bandwidth and memory for this module} impose further resource limitations to this particular module. A single algorithm module may correspond to multiple algorithm instances, and users may manually decide to spread out the execution of these instances by running only a batch of instances at a time and another batch later on. Users thus may optionally specify \texttt{Parallel instance ID}, which is the algorithm instance IDs to run in parallel in this module. 
Finally, users can also add other \texttt{Algorithm\_parameters} flexibly to further configure their algorithms.

Then comes the body of distributed algorithms, which consists of instances of algorithms to be run in parallel, following the user implementation. To implement, one option is to directly call the primitives in \sysname's library. Table~\ref{tbl:primitive-calls} lists some of our algorithmic primitives as well as their potential applications in this paper. We note that users can supply their own parameters and even their own functions to be executed in addition to what the primitive executes. In Section~\ref{sec:use_cases}, the use cases provide more contexts and explanations for these algorithmic primitives. Another option is to implement custom user algorithms. For CONGEST algorithm modules, users can choose the streaming computation scheme, corresponding to option 2. In this case, \sysname aggregates incoming packets until \texttt{Completion()} of the round is met, when the module may terminate or continue to \texttt{send\_to\_neighbors} next round, as implemented by users. Alternatively, users may opt for doing computation at the end of the round, where users implement their own computation functions. 

\section{Pseudocode of some algorithmic primitives and use cases} \label{app: pseudocode}

\begin{algorithm}[H]
\caption{Set cover algorithm on a shortest path tree.}\label{alg:set-cover}
\begin{algorithmic}[1]
\Procedure{Set cover}{}
\BState {Round $3k$ for layer $D - k$}: \Comment{D: depth of the SPT}
    \State{Send msg() to all the neighbors on layer $l$}
\BState {Round $3k + 1$ for layer $D - k - 1$}:
    \State{self.count = \# messages received}
    \State{\textit{Multicast} msg(.count = self.count) back to neighbors that sent a packet at round $3k$}
\BState {Round $3k + 2$ for layer $D - k$}:
    \State{self.parent = $\arg \max_{\text{nb}}$ msg[nb].count}
    \State{Send msg(.father = "yes") to self.parent}
\EndProcedure

\end{algorithmic}
\end{algorithm}


\begin{algorithm}[H]
\caption{Conditional broadcast and aggregation, and its application to leader election.}\label{alg:cond-broadcast-agg}
\begin{algorithmic}[1]
\Procedure{Conditional broadcast and aggregation}{}
\EndProcedure
\BState{}\Comment{Each switch initially owns a value self.value}
\State{}\Comment{The goal is to aggregate the value and}
\State{}\Comment{broadcast them across the network.}
\State{Stream-processing(p):}\Comment{Line 16 in Algorithm~\ref{alg:dis-congest-alg}}
\Indent
    \State{self.value = arithmetic(self.value, p.value)}
\EndIndent
\State{Completion:}\Comment{Line 17 in Algorithm~\ref{alg:dis-congest-alg}}
\Indent
    \State{The current switch has received one message per each neighbor}
\EndIndent
\State{Termination:}\Comment{Line 18 in Algorithm~\ref{alg:dis-congest-alg}}
\Indent
    \State{A total of $D$ rounds have passed, where $D$ is the estimated diameter of the graph.}
\EndIndent
\State{Send\_to\_neighbors:}\Comment{Line 20 in Algorithm~\ref{alg:dis-congest-alg}}
\Indent
    \If{A condition() is met.}
    \State{Broadcast self.value to all neighbors.}
    \EndIf
\EndIndent

\Procedure{Leader election}{}
\EndProcedure
\State{arithmetic(): $\text{min(self.value, p.value)}$}
\State{condition(): always true.}

\end{algorithmic}
\end{algorithm}

\begin{algorithm}\caption{Reaction algorithms for clock synchronization.}\label{alg:pseudocode-clock-sync}
{\small
\begin{algorithmic}[1]
\Procedure{Fast Recovery}{}
\EndProcedure
\State{Use the Asynchronous global flooding primitive}
\smallskip
\Procedure{Distributed Optimization}{}
\EndProcedure
\State{First round:}
\Indent
\State{Randomly sample $r$ switches as candidate roots.}
\EndIndent
\State{Flooding(record\_candidate\_roots)}
\For{each candidate root}
    \State{Use the Shortest path tree primitive}
    \State{Use the Bottom up tree aggregation primitive}
    \State{Move the root to the center of the tree.}
\EndFor
\State{Pick the tree with the minimal depth among these.}
\end{algorithmic}
}
\end{algorithm}

\begin{algorithm}[H]
\caption{The modified Contra algorithm}\label{alg:contra}
\begin{algorithmic}[1]
\Procedure{Synchronous shortest path (Contra)}{}
\EndProcedure
\State{At each round:}
\Indent
\State{Receive and stash packets from all neighbors in PG.}
\State{Until all the updates have been received:}
\Indent
\State{Merge packets corresponding to the same PG}
\State{Update the routing table.}
\EndIndent
\State{Send:}
\For{Merged packet that contributes to an update}
\State{Find the downstream states and multicast.}
\EndFor
\EndIndent
\end{algorithmic}
\end{algorithm}

\begin{table*}[!htbp]
\centering
\resizebox{.95\linewidth}{!}{
\begin{tabular}{|l|l|l|l|}
\hline

\textbf{Algorithmic primitives} &\textbf{\begin{tabular}[c]{@{}l@{}}Category\end{tabular}} & \textbf{Description} & \textbf{Applications} \\ \hline

\textbf{\begin{tabular}[]{@{}l@{}} \texttt{Asynchronous global flooding}\end{tabular}} & Asynchronous & \begin{tabular}[c]{@{}l@{}} Use flooding to 1) broadcast a short message to all switchs \\ 2) construct a flooding spanning tree, \\ 3) initializing the synchronizer throughout the network.
\end{tabular} & \begin{tabular}[c]{@{}l@{}}Fast recovery for clock sync and multicast; \\ bootstrapping for Contra routing. \end{tabular} \\ \hline

\textbf{\begin{tabular}[]{@{}l@{}} \texttt{Bottom-up tree aggregation} \\ \end{tabular}} & Asynchronous & \begin{tabular}[c]{@{}l@{}} Aggregate values from the leaves of a spanning tree to the  root. \end{tabular} & \begin{tabular}[c]{@{}l@{}}  Distributed optimization for clock sync, \\ fast recovery for multicast. \end{tabular} \\ \hline

\textbf{\begin{tabular}[]{@{}l@{}} \texttt{Conditional broadcast} \\ \texttt{and aggregation} \\ (as detailed in Algorithm~\ref{alg:cond-broadcast-agg}) \end{tabular}} & CONGEST & \begin{tabular}[c]{@{}l@{}} Maintain a single value $v$ for each switch. Aggregate $v$ \\ with those received from neighbors for each round. Then \\ broadcast the updated value next round provided that \\ the condition function is met. \end{tabular} & \begin{tabular}[c]{@{}l@{}}Leader election for clock sync, \\ Spanner for multicast. \end{tabular} \\ \hline

\textbf{\begin{tabular}[]{@{}l@{}} \texttt{Shortest path tree}\end{tabular}} & CONGEST & \begin{tabular}[c]{@{}l@{}} Construct a spanning tree layer-by-layer in a BFS fashion. \\ Each round the current leaf switches look for unvisited \\ switches to be added as a new layer of leaf switches. \end{tabular} & \begin{tabular}[c]{@{}l@{}} Distributed optimization for clock sync, \\ multicast tree for multicast. \end{tabular} \\ \hline

\textbf{\begin{tabular}[]{@{}l@{}} \texttt{Set cover on a} \\ \texttt{shortest path tree} \\ (as detailed in Appendix~\ref{app:set_cover})\end{tabular}} & CONGEST & \begin{tabular}[c]{@{}l@{}} In the shortest path tree, greedily select a minimal number 
\\ of switches in layer $i$ to cover layer $i + 1$, \ie, every switch \\ in layer $i + 1$ has a selected neighboring switch in layer $i$. \end{tabular} & \begin{tabular}[c]{@{}l@{}} Multicast tree for multicast. \end{tabular} \\ \hline

\textbf{\begin{tabular}[]{@{}l@{}} \texttt{Synchronous shortest path} \\ \end{tabular}} & CONGEST & \begin{tabular}[]{@{}l@{}} Synchronously compute shortest path routes w.r.t. a routing policy. \end{tabular} & \begin{tabular}[c]{@{}l@{}} Contra routing. \end{tabular} \\ \hline


\end{tabular}
}
\caption{Descriptions of some \sysname's algorithmic primitives.}\label{tbl:primitive-calls-full}
\vspace{-6mm}
\end{table*}

 \section{The Choice of Synchronizers} \label{app:sync}
 The CONGEST model assumes a synchronous network, but real-world networks are usually asynchronous. Various theoretical synchronization algorithms have been proposed, such as the $\alpha$-synchronizer and $\beta$-synchronizer~\cite{Awerbuch85}. The $\alpha$-synchronizer is essentially a local synchronizer that only ensures that a node does not begin round $t + 1$ until its neighbors have completed round $t$, while $\beta$-synchronizer is a global synchronizer that guarantees full synchronization, so round $t + 1$ does not begin until all nodes have completed round $t$. As the CONGEST model operates locally, a local synchronizer suffices. However, global synchrony may be needed for scenarios such as launching a CONGEST algorithm after an asynchronous algorithm. We have implemented both $\alpha$ and $\beta$ synchronizers and leave the implementations of other synchronizes (such as $\gamma$ synchronizer~\cite{Awerbuch85}) as future works.

We now briefly present the message-passing process of the $\alpha$ and the $\beta$ synchronizer.
In the $\alpha$-synchronizer, every message is tagged with the round number. For each round, every node sends the message to each neighbor (we send a "no-msg" if there is no message-passing to a neighbor). When a node receives messages from all its neighbors, it performs local computation and proceeds to the next round. 
In the $\beta$-synchronizer, we construct a spanning tree before running the CONGEST algorithm. Before the execution of round $t$, the root of the spanning tree must confirm that all nodes complete their execution of round $t - 1$. This is achieved by sending notification messages from every node to the root in a bottom-up manner. Similarly, to launch round $t$, the root replies to every node through the tree that round $t$ has begun in a top-down manner. 
One can think of the $\alpha$- and $\beta$-synchronizers as two extremes, where the first sends more messages ($O(n^2)$ on each link) but the messages only travel for one hop, while the latter is more message-efficient ($O(n)$ messages through the tree) but the latency is potentially longer because messages travel for multiple hops.

\section{Hardness of min-size shortest path tree}\label{app:SPT_hardness}
In this appendix, we show that finding the minimal shortest path tree is at least as hard as set cover, thus implying its NP-hardness.

Formally, the set cover problem is defined as follows:
Let $\mathcal S$ be a set of subsets of the universe $U=\{1,\ldots, n\}$ (i.e., each $S\in\mathcal S$ satisfies $S\subseteq U$).
The set cover problem asks what is the smallest set $\mathcal C\subseteq\mathcal S$ such that $\bigcup_{S\in\mathcal C} S = U$.

Recall that our min-cast shortest path tree problem is defined as: Given a graph $G=(V,E)$, a sender $s\in V$ and receivers $D\subseteq V$, find a shortest path tree $T$ (i.e., $\mathit{dist}_G(s,d)=\mathit{dist}_T(s,d)$) of minimal size.
Here, we show that an algorithm to the min-cast shortest path tree problem implies an algorithm for set-cover, and therefore the above hardness result holds for our problem as well.

\newcommand{\set}[1]{\left\{#1\right\}}

Given a set cover instance $(U,\mathcal S)$, construct the following graph:
$$
G = \left(\{s\} \cup \mathcal S \cup U, \{\{s,S\} \mid S\in\mathcal S\}\cup \set{\set{S, u} \mid u\in S, S\in\mathcal S}\right).
$$
Further, let $s$ be the source and $D=U$ be the destinations.
That is, in the graph, the source $s$ is connected to all sets $S\in\mathcal S$, and each element $u\in U$ is connected to the sets that it is a member of.

We have that the optimal solution must include $s$ and all elements $u\in U$, in addition to some cover $\mathcal C\subseteq \mathcal S$. Thus, the solution's size is $n+1+|\mathcal C|$, and finding the optimal shortest path tree reveals the optimal cover to the underlying problem.

\section{Multicast Algorithms}

\subsection{Constructing Fast Recovery Multicast Trees on top of a Spanning Tree}
\label{appendix:construct-fast-recovery-multicast-tree}

We mentioned in Section~\ref{subsec:multicast} that we need to construct fast recovery multicast trees that are subtrees of $T_0$, the spanning tree we constructed earlier. The fast recovery multicast tree should connect together the source and the destinations. Instead of directly constructing the trees independently, we propose to build the trees altogether in a bottom up manner, thanks to our message packing technique.
We define the bitmap is\_son[child][j] as whether the multicast tree of group $j$ contains an edge linking itself to the child. We also define a bitmap in\_tree, the $j$'th entry of which denotes whether the multicast tree of group $j$ contains the current switch. Apparently, in\_tree[j] indicates if a switch is in a multicast tree of group $j$ or not, and is\_son[$\cdot$][j] indicates the ports corresponding to the set of children in the tree. Both are boolean arrays that can be compressed into a single bitmap. We have the following equation:
$$\text{in\_tree(v)} = \lor_{\text{child}} \text{is\_son(v)[child]}$$
$$\text{is\_son(parent(v))}[v] = \text{in\_tree(v)}$$
, where $v$ is the current switch.
Using this equation, we carry the in\_tree bitmap (with $< 2000$-bit) into a single $< 2000$-bit packet, compute is\_son and in\_tree bitmaps with bitwise operations and send the bitmaps in a bottom-up manner from the leaf switches to the root switch. Apparently, the message packing reduces the bandwidth consumption and thus the convergence time, compared with the approach where we dedicate one upstream packet per multicast group, resulting in thousands of $64$-byte Ethernet packets to be sent per hop.

\subsection{Efficient Set Cover for Multicast Tree Construction}\label{app:set_cover}
Once we have executed the shortest path tree algorithm for finding shortest path length between the sender (source) switch and all the switches in the network, we start to build the multicast tree from leaf switches to the source switch from the leaf layer to the source progressively. We would like to minimize the number of nodes in the multicast tree so as to minimize the multicast traffic. We have abstracted this algorithm as one of our algorithmic primitives, and we present its pseudocode in Algorithm~\ref{alg:set-cover}. 

Now we elaborate on the details of the algorithm. Initially, all the receivers are marked as active switches. For all the active switches in layer $l$ we run the distributed set cover to find as few parent switches as possible to cover them.  These parent switches are marked as active switches in the next layer, and we repeat the process. We propose a greedy approach for set cover within three rounds:
\begin{enumerate}[align=left, leftmargin=0mm, labelindent=0\parindent, listparindent=0\parindent, labelwidth=0mm, itemindent=!,itemsep=3pt,parsep=0pt,topsep=0pt]
    \item Nodes on layer $l$ send one message to each neighbor on layer $l - 1$, and each node $v$ on layer $l{-}1$ counts the total number of received messages as $count[v]$.
    \item Each node $v$ on layer $l{-}1$ \textit{multicasts} $count[v]$ back to nodes that sent messages in the previous round in layer $l$, and each node in layer $l$ picks up the maximum value of $count[v]$ and identifies the corresponding node on layer $l{-}1$ as its parent.
    \item Each Node $v$ in layer $l$ sends a message to its parent to confirm, and its parent add $v$ as one of its children.
\end{enumerate}
Specifically for our multicast application, once the trees has been constructed, our last step is to transmit the encoding rules (the bitmaps) into the source tor switch and store the rules on it. We continue to use the encoding information from the tor switch for multicast of the affected groups until the controller later fully installs the multicast rules on end hosts so that we switch back to the source-routed multicast again.

\subsection{Installing Multicast Rules into Tor Switches}
During the period between the completion of the distributed optimization phase (when we construct multicast trees) and the time when the controller computes and installs multicast rules on end hosts, we propose to temporarily cache \textit{all} the source-routed multicast rules of the affected groups in the tor switches. When a multicast packet from an affected group reaches the tor switch, it appends the multicast rules stored on the switch. For later hops, switches look up the multicast rules on the header. We note that such a source-routed multicast solution guarantees a bounded memory overhead per switch, compared with the solution that we store the multicast rules separately on all the switches. This is because the number of multicast groups such that their sources are within a certain tor $t$ is generally a bounded quantity.

\section{Implementation details}\label{appendix:impl-details}
\noindent\textbf{Our detection phase for clock synchronization}. As a starting point, Sundial~\cite{Sundial} detects failures by passively listening to the synchronization messages,
so that if several consecutive losses of synchronization packets occur, the switch concludes that some upstream links or switches have failed. However, such an approach cannot detect exactly which devices have failed. Therefore, in \sysname's implementation, 
we add a technique called ping-on-timeout. In this approach, if a switch does not hear the synchronization messages for two timeouts, it actively pings its parent, and if the parent does not reply with an ACK message then the switch can conclude that the failure stems directly from its parent rather than from its parent's ancestor.

\section{Supplemental experimental settings and results}
\label{appendix:exp}

\subsection{Evaluation setups}\label{appendix:exp-setups}
\subp{Testbed experiment setups in Section~\ref{subsec:testbed}.} To configure \sysname's reaction algorithm for clock synchronization, we reconstruct 4 candidate trees and select the best one in terms of the tree depth in the distributed optimization phase. The synchronization interval is set as $50\mu$s for \sysname prototype and $61\mu$s for PTP, the lowest possible interval we could achieve for PTP. To measure the time uncertainty bound, we follow Sundial~\cite{Sundial}'s setup and configure $\epsilon_0=5$ns and $\texttt{max\_drift\_rate}=0.0002$ or equivalently $200$ppm.

\begin{figure}
    \centering
    \includegraphics[width=0.7\linewidth]{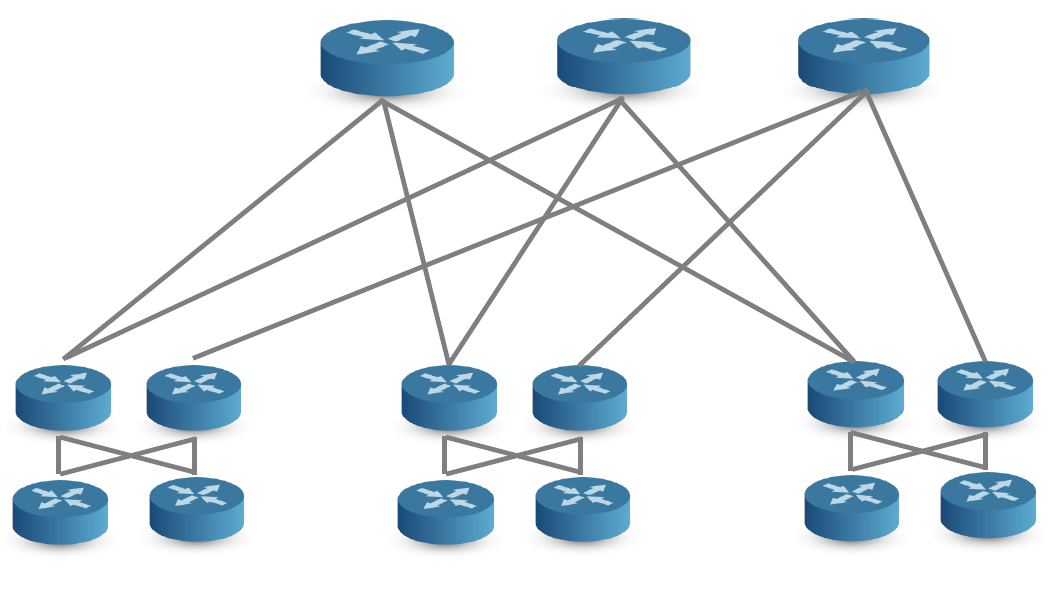}
    \caption{The 3-4-ary FatTree.}
    \label{fig:3-4-ary-fattree}
\end{figure}

\subp{The 3-4-ary FatTree.} We here define the 3-4-ary FatTree topology used in our testbed experiments. As depicted in Figure~\ref{fig:3-4-ary-fattree}, the 3-4-ary FatTree is created from a standard $4$-ary FatTree but with the last core switch and the last pod (containing 2 aggregation switches and 2 ToR switches) removed. The number of switches remaining is therefore $n=15$. Links connected to the removed switches are also deleted accordingly.

\noindent\textbf{Simulation setups.}  We configure Ethernet links to have a propagation delay randomly within $100 \pm 10$ns. 
The bandwidth of links is set to be $100$Gbps; regardless of the bandwidth capacity of links, we always allocate a $<1$Gbps bandwidth for \sysname (as detailed in each application) identically for all links, which is far below the bandwidth capacity of modern Ethernet links. 
The packet loss rate for each link is set to be $0.1\%$ independently at random for each packet, except in Section~\ref{appendix:flex-reliable-exps} where we study how the packet loss rate affects the convergence time.

\subp{Experimental setups in Section~\ref{subsubsec:eval_on_clock}.} We generate one or two switch failures for our experiments. With multiple failures, we deliberately set the failed switches to be in different domains so that Sundial switches back to the control-plane solution. We set the interval between two consecutive synchronization messages to be $50\mu s$ for both our solution and Sundial, and detect a timeout on 3 loss of synchronization messages. This corresponds to $<1$Mbps per-link bandwidth consumption (Figure~\ref{subfig:sync-overhead-series}). To configure \sysname, we set the number of sampled candidate roots in the distributed optimization phase to be $4$. We add the synchronous messaging proposed by Sundial~\cite{Sundial} technique to PTP~\cite{PTP} to further reduce its $\epsilon$ value.

\subp{Experimental setups in Section~\ref{subsubsec: eval-multicast}}. Following the setup from Elmo~\cite{Elmo} and Orca~\cite{orca}, we generate $10^6$ multicast groups; for each multicast group, we generate several VMs and randomly place them on end hosts. One of the VMs serves as the sender and the rest of them serve as the receivers. The number of VMs per group (\ie, group size) follows the IBM WebSphere Virtual Enterprise (WVE)~\cite{scaling-ip-multicast} distribution, and the average group size is set to be $0.05n$, where $n$ is the number of switches in the network. To configure \sysname, we allocate $200$Mbps bandwidth (Figure ~\ref{subfig: mcast-message-overhead}) and $1.5\%$ memory of the Tofino's total switch memory, which is enough to enable failure recovery algorithms for all the failed groups to run in parallel. We also set $k=13$ for the spanner algorithm~\cite{spanner}. 
For Elmo, we choose the redundancy value $R=3$. Since Orca does not adapt to Jupiter and Jellyfish topologies and does not report its failure recovery time, we only evaluate its packet header size and total multicast traffic on FatTree.

\subp{Experiment setups of running concurrent applications in Section~\ref{subsec:exp-combined}}. We choose the FatTree as our topology, and inject one link failure for our experiments. The bandwidth for \sysname-individual is allocated to be $200$Mbps, $200$Mbps, $10$Mbps for multicast, Contra, and clock synchronization respectively. Correspondingly we allocate the sum of \sysname-individual's bandwidth 
to \sysname-combined. For Figure~\ref{subfig:overall_w_Contra}, we do not limit the memory usage, and the \sysname-combined with Contra consumes $3.75\%$ memory. For Figure~\ref{subfig:overall_varying_memory} we gradually shrink the memory size from $3.75\%$ to $1.58\%$ for \sysname-combined, and proportionally adjust the memory for each application in \sysname-individual.

\subsection{Validating the simulator}\label{app:testbed-simulator-validation}

\begin{figure}
	\centering
         \begin{minipage}[t]{0.75\linewidth}{
		\vspace{-0.00in}
		\begin{center}
		\includegraphics[width=\textwidth, ]{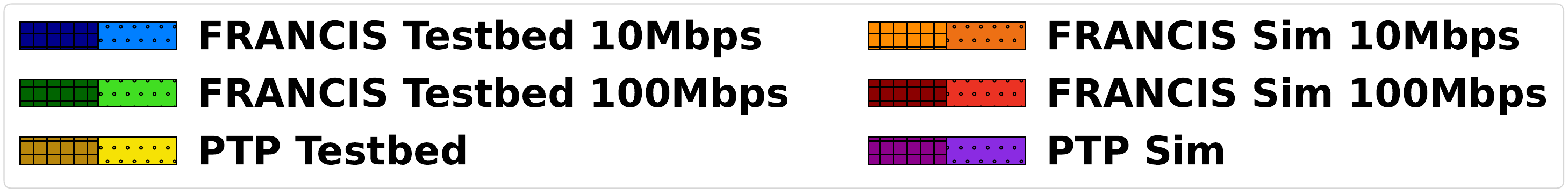}
		\end{center}
		}
		\end{minipage}

        \hspace{-0.2cm}
	\subfigure[Convergence time]{
		\begin{minipage}[t]{0.39\linewidth}{
		\vspace{-0.08in}
		\begin{center}
		\includegraphics[width=\textwidth, ]{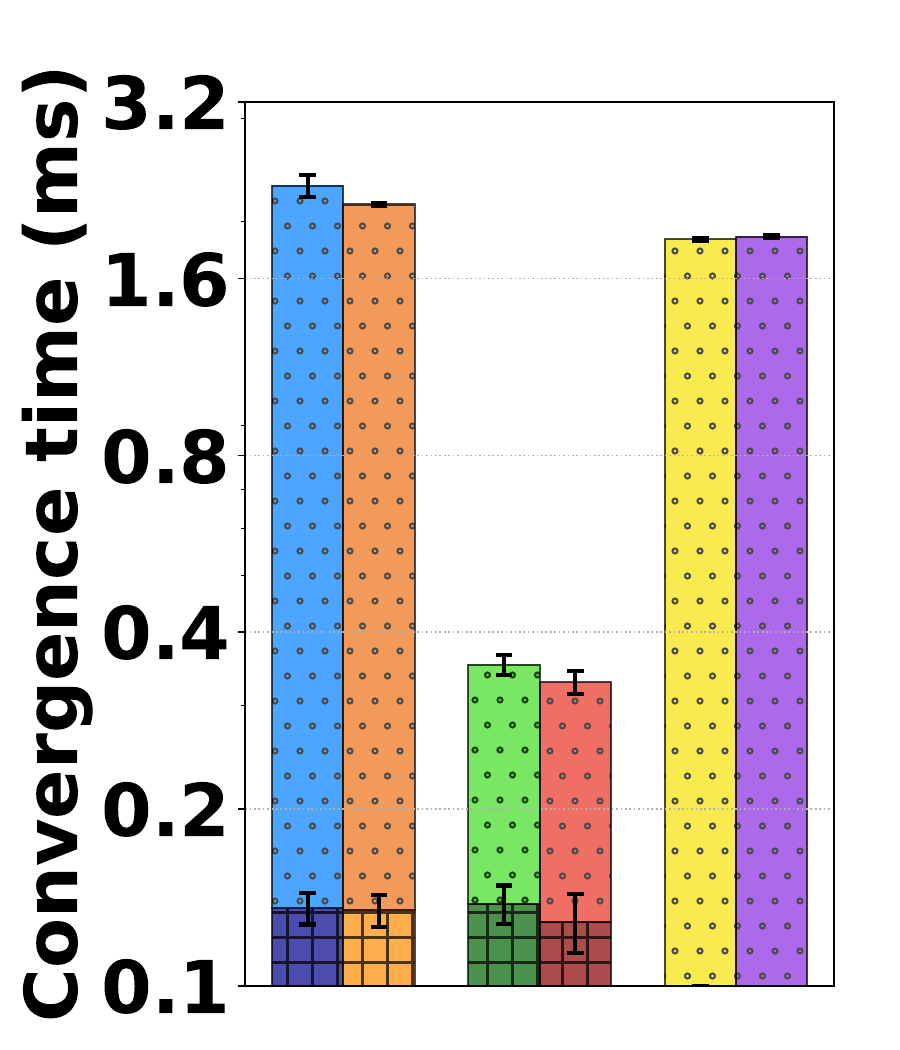}
		\end{center}
            \vspace{-0.10165in}
		}
		\label{subfig:testbed_conv_time}
		\end{minipage}
	}
	\hspace{-0.5cm}
	\subfigure[Peak $\epsilon$]{
		\begin{minipage}[t]{0.39\linewidth}{
		\vspace{-0.08in}
		\begin{center}
		\includegraphics[width=\textwidth, ]{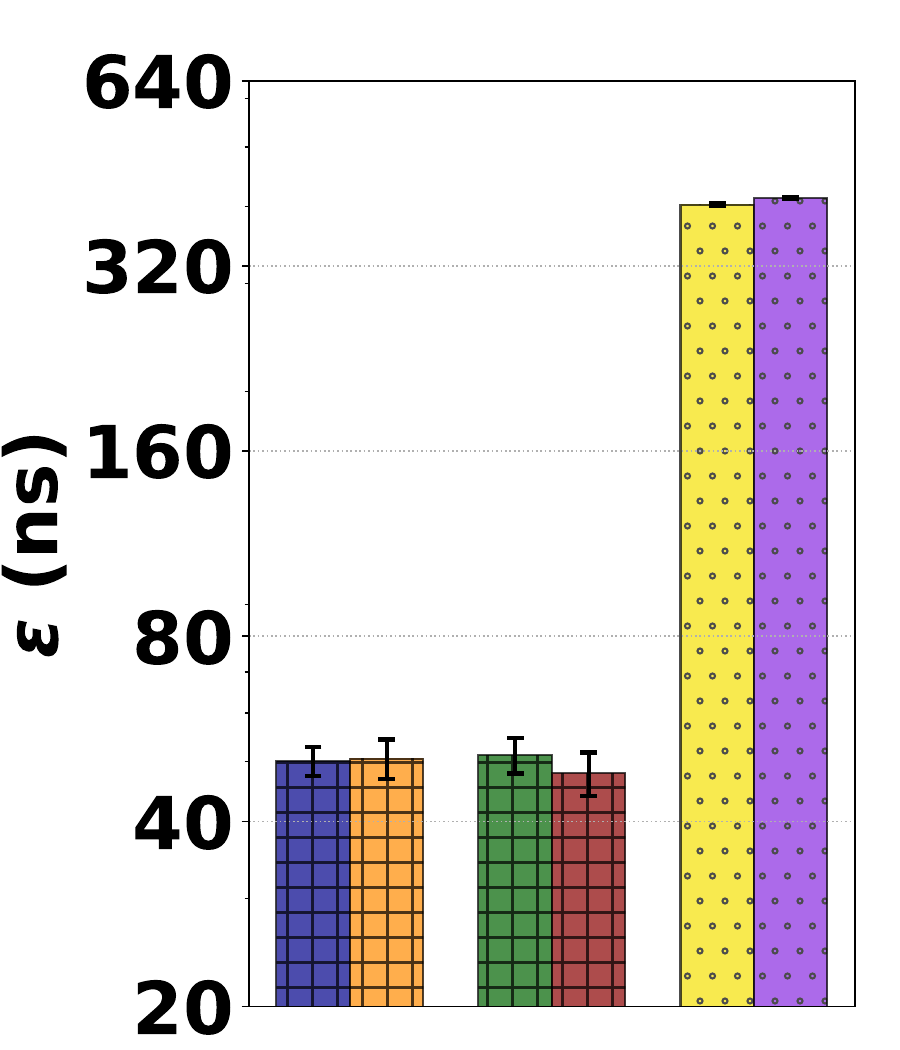}
		\end{center}
            \vspace{-0.10165in}
		}
		\label{subfig:testbed_peak_epsilon}
		\end{minipage}
	}
        %
	\vspace{-0.195in}
	\caption{Performance of \sysname's Tofino prototype for clock synchronization comparing against simulation results. The dark, cross-patched bars represent the fast recovery phase while the light, dotted bars correspond to the distributed optimization phase.}
	\label{fig: testbed}
\end{figure}

In order to evaluate how accurately our simulator could model behaviors of a real-world distributed programmable-switch networks, we compare results conducted on the simulator against testbed results against testbed results. As shown in Figure~\ref{fig: testbed}, the testbed results regarding the convergence time and time uncertainty bound overall match with simulation results, which validates the utility of our simulation model. \sysname Tofino's convergence time is less than $10\%$ slower than that derived from simulation, primarily due to us conservatively implementing the bandwidth limitation component on Tofino to meet with the hardware restrictions (Section~\ref{subsubsec:algorithm-scheduling}). That is, the allocated bandwidth may not be fully utilized, and accordingly extra delays are incurred so that the convergence time becomes longer. 

\subsection{Additional experiments for comparisons with state-of-the-arts}\label{app:additionalComparisons}

\begin{figure}[!htbp]
    \raggedright
    \begin{minipage}[t]{0.4\linewidth}{
    \vspace{-0.00in}
    \includegraphics[width=\textwidth, ]{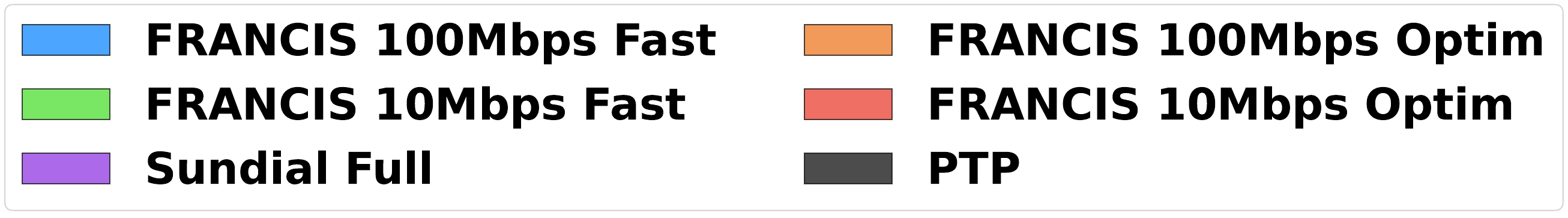}
    \vspace{-0.11in}
    }
    \end{minipage}
    
    \vspace{-0.1in}
	\subfigure[Convergence time.]{
		\begin{minipage}[t]{0.4\linewidth}{
		\vspace{-0.04in}
		\begin{center}
		\includegraphics[width=\textwidth, ]{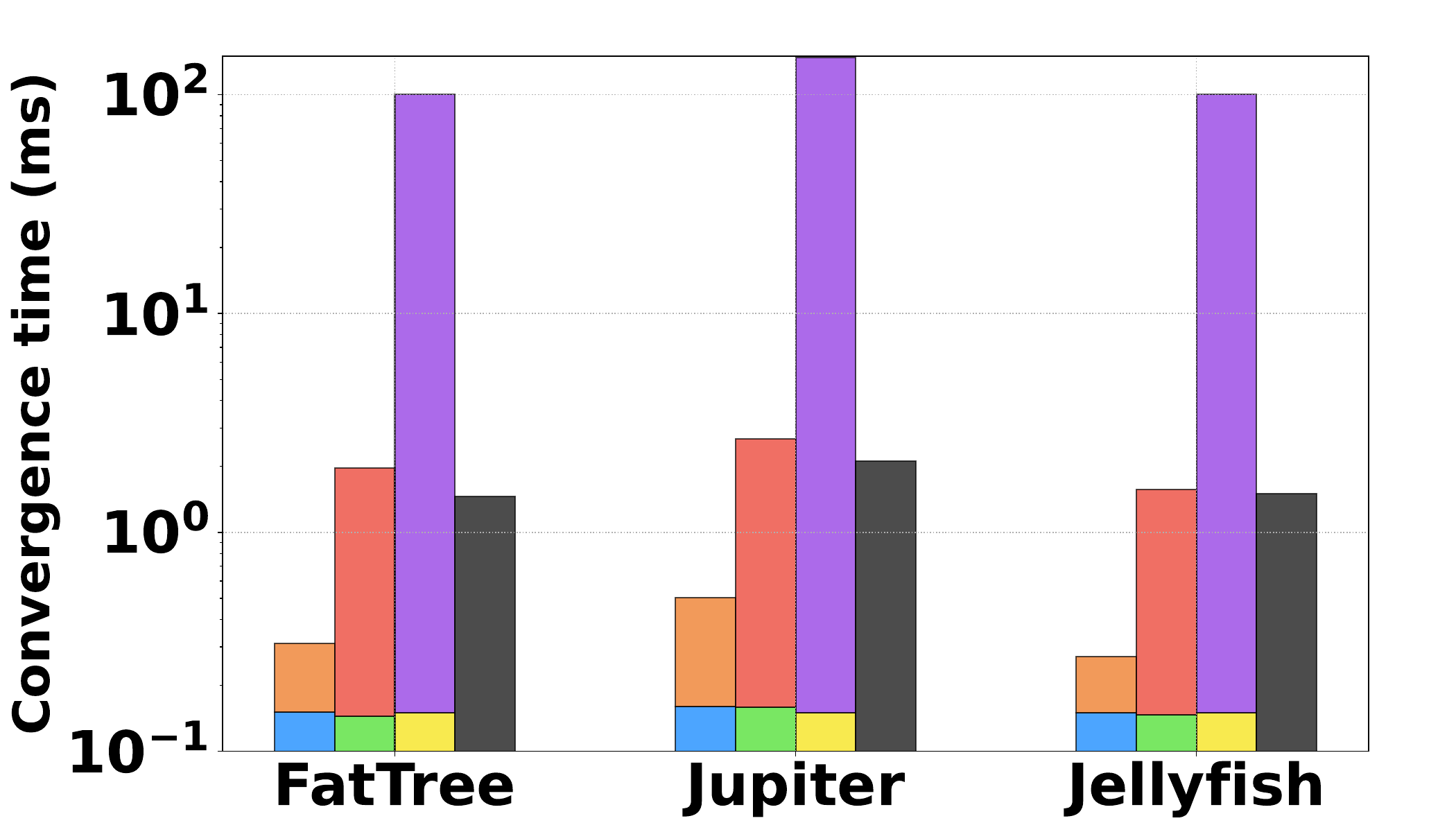}
		\end{center}
		}
		\label{subfig:clock_conv_time_1fail}
		\end{minipage}
	}
    \hspace{0.05in}
    \subfigure[Peak $\epsilon$.]{
		\begin{minipage}[t]{0.42\linewidth}{
		\vspace{-0.2in}
		\begin{center}
		\includegraphics[width=\textwidth, ]{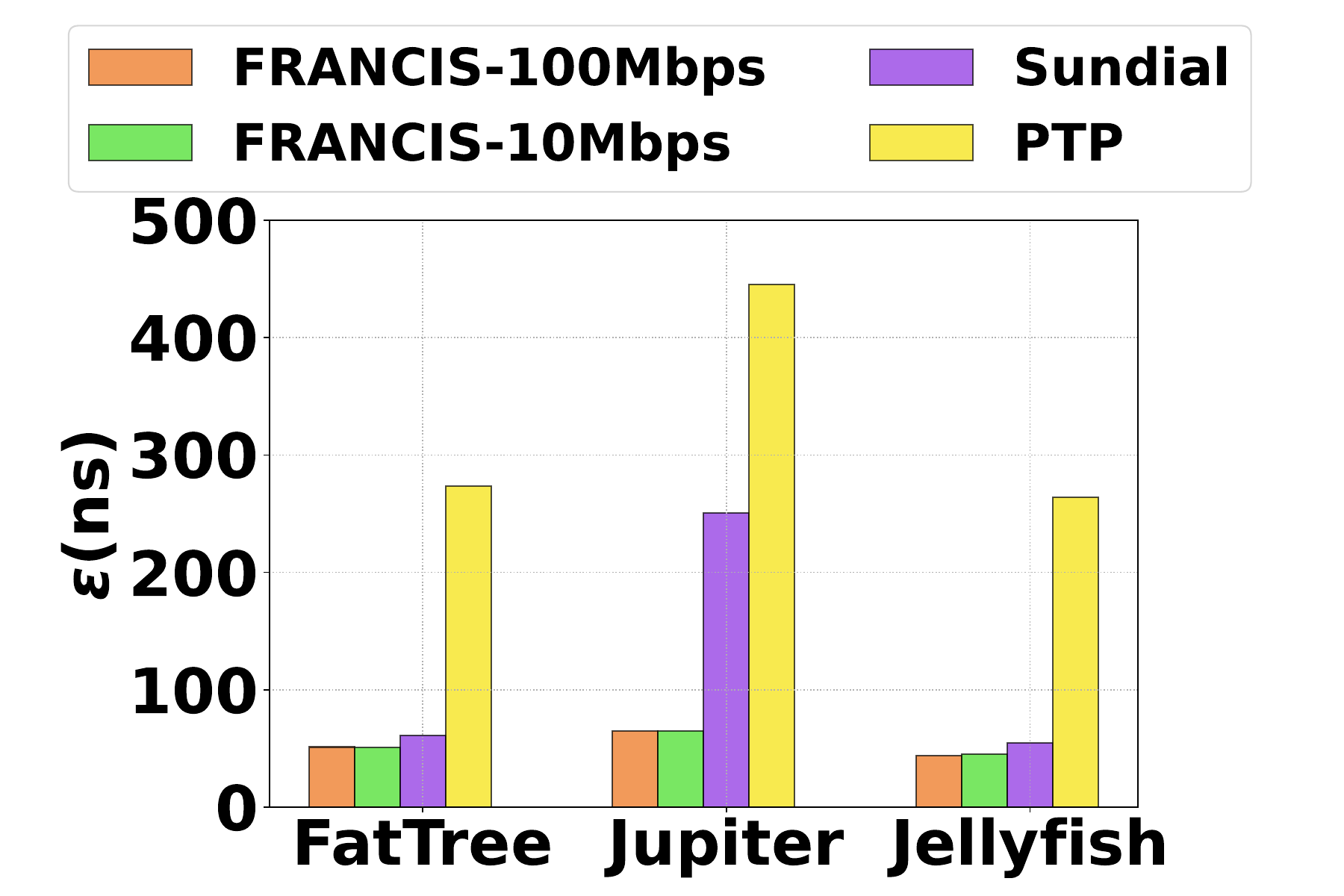}
		\end{center}
		}
		\label{subfig:clock_sync_dataset_epsilon_1f}
		\end{minipage}
	}
	\vspace{-0.2in}
	\caption{Performance for clock synchronization compared with Sundial~\cite{Sundial} and PTP~\cite{PTP} with one switch failure.}
    \vspace{-0.1in}
    \label{fig:clock-sync-performance-appendix}
\end{figure}

\noindent\textbf{Clock synchronization}. In this section, we provide more experimental results for clock synchronization. Figure~\ref{fig:clock-sync-performance-appendix} shows the convergence time and peak $\epsilon$ results with one switch failure. Recall that Sundial~\cite{Sundial} handles one failure by directly switching to the pre-computed backup plan within each switch. Therefore, Sundial's fast recovery plan achieves the best convergence time $150\mu$s, outperforms \sysname-100Mb's 186.7$\mu$s fast recovery time and $359.5\mu$s distributed optimization time for the FatTree topology. However, as shown in Figure~\ref{subfig:clock_sync_dataset_epsilon_1f} and Figure~\ref{subfig:sync_epsilon_series}, both the peak $\epsilon$ and the back-to-normal $\epsilon$ value ($\epsilon$ after the fast recovery of Sundial or the distributed optimization phase of \sysname) of Sundial are worse than \sysname-100Mb. In fact, the suboptimality of Sundial's backup tree is to be blamed for. For the peak $\epsilon$, although the reaction time of \sysname is slightly faster than \sysname-100Mb, its resulting backup tree is 3 hops deeper than \sysname's, and, according to Equation~\ref{eq:epsilon}, the peak $\epsilon$ becomes larger. The same explanation also applies to the back-to-normal $\epsilon$ value. As previously argued, due to Sundial's slow control-plane based full recovery, the suboptimal $\epsilon$ continues to have effect on the clock synchronization application for the duration of $\approx 100$ms full recovery time. Finally, to compare with PTP, although it achieves a similar level of convergence time than \sysname-10Mb (which is several times longer than \sysname-100Mb), both the peak $\epsilon$ and the back-to-normal $\epsilon$ are worse than our solution. 

\begin{figure}[!htbp]
    \subfigure[Packet header size.]{
        \begin{minipage}[t]{0.44\linewidth}{
        \vspace{-0.10in}
        \begin{center}
        \includegraphics[width=\textwidth, ]{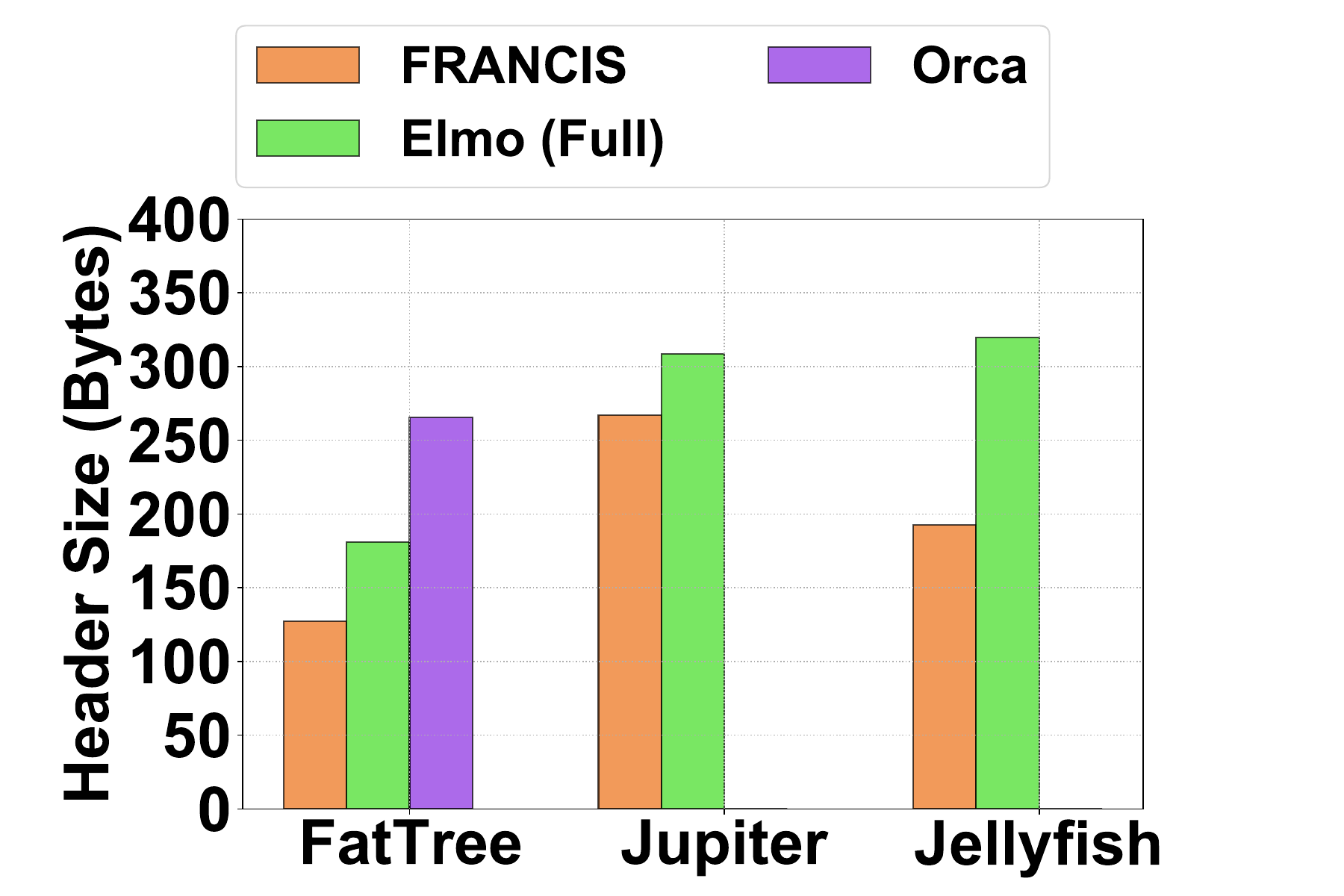}
        \end{center}
        }
        \label{subfig:mcast_header_size}
        \end{minipage}
    }
    \subfigure[Total volume of multicast traffic per $1500$B packet.]{
        \begin{minipage}[t]{0.44\linewidth}{
        \vspace{-0.10in}
        \begin{center}
        \includegraphics[width=\textwidth, ]{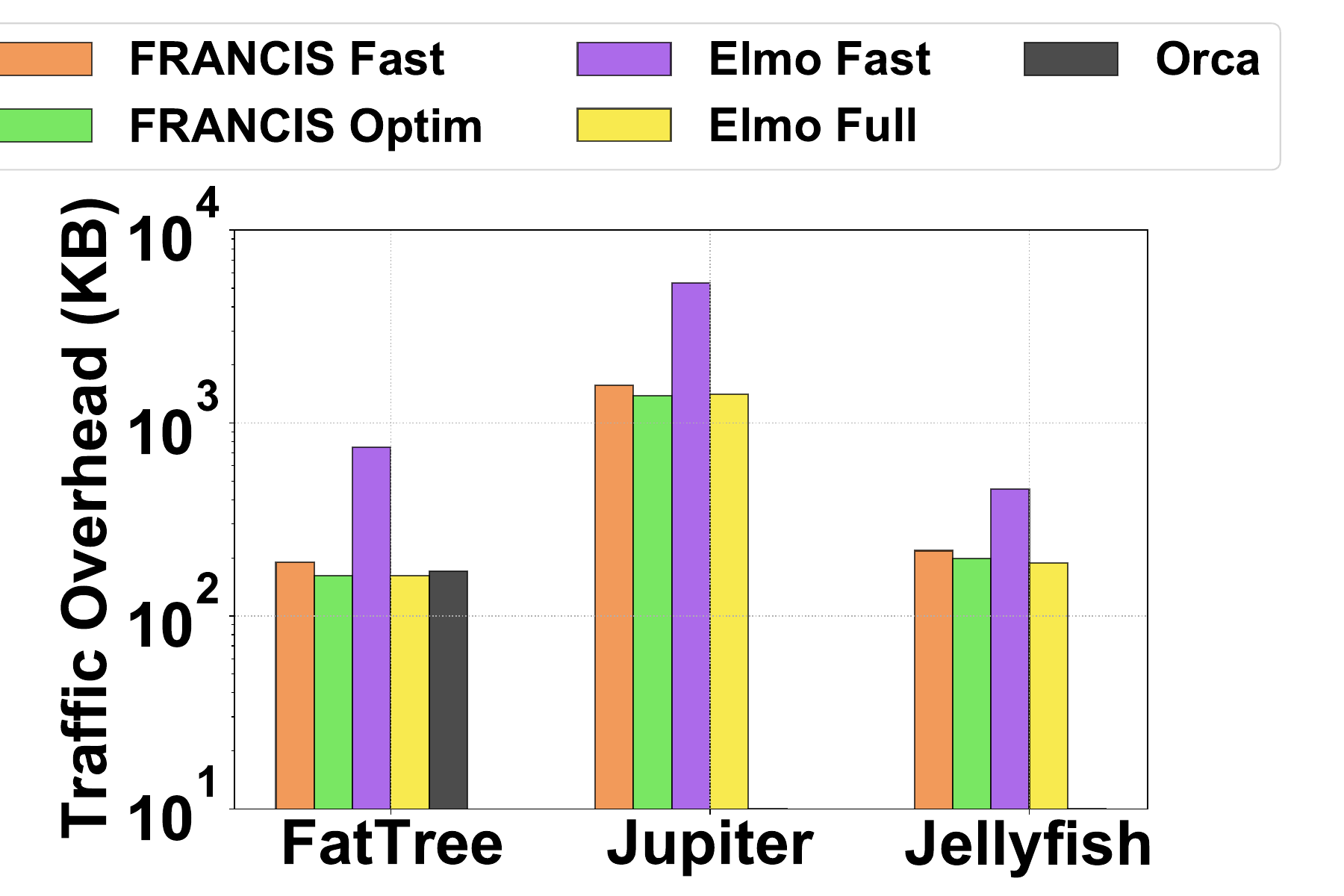}
        \end{center}
        }
        \label{subfig:mcast_traffic_overhead}
        \end{minipage}
    }
    \vspace{-0.1in}
	\caption{Evaluation on the Spanner algorithm for \sysname's multicast. Here "Optim" refers to results after the distributed optimization phase, and "Full" refers to full recovery for Elmo.}
    \label{fig:mcast-performance-appendix}
\end{figure}

\begin{figure}[!htbp]
	\centering
	\vspace{-0.135105in}
	\subfigure[Anatomy of failure recovery time]{
		\begin{minipage}[t]{0.47\linewidth}{
		\vspace{-0.00in}
		\begin{center}
		\includegraphics[width=\textwidth, ]{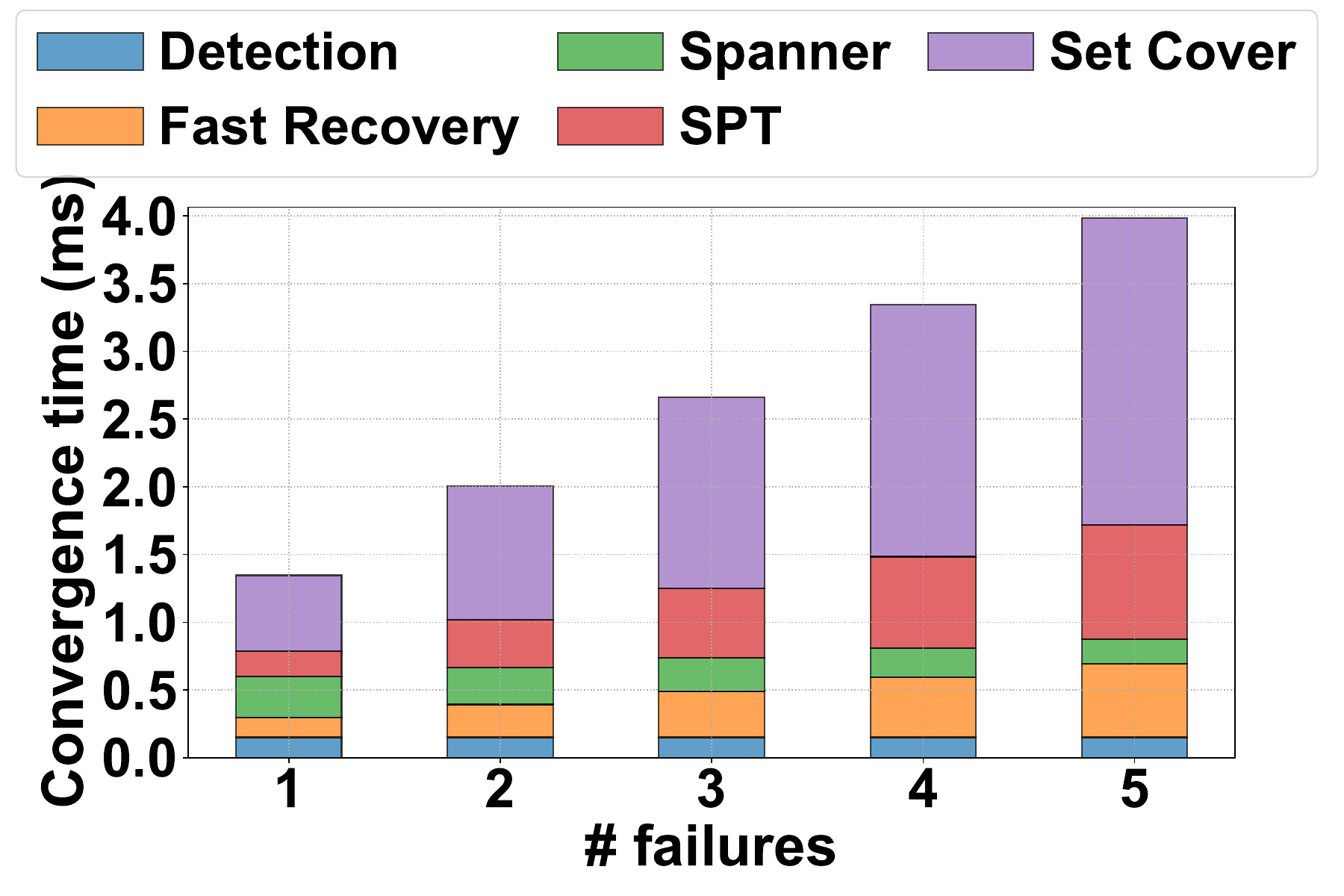}
		\end{center}
		}
		\label{subfig:mcast-time-decompose}
		\end{minipage}
	}
	\subfigure[Time to message overhead]{
		\begin{minipage}[t]{0.47\linewidth}{
		\vspace{-0.00in}
		\begin{center}
		\includegraphics[width=\textwidth, ]{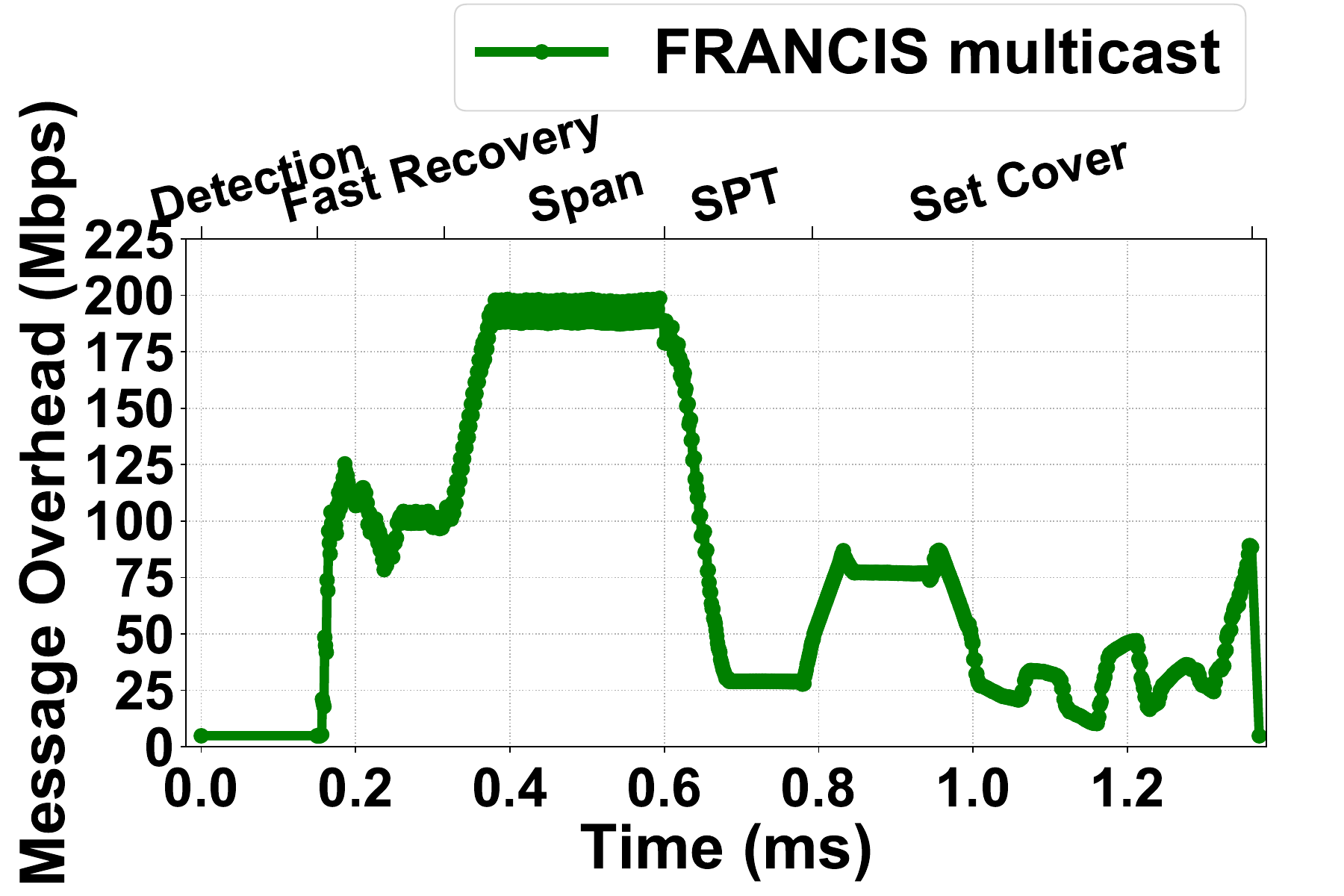}
		\end{center}
		}
		\label{subfig: mcast-message-overhead}
		\end{minipage}
	}
	\vspace{-0.1578975in}
	\caption{The recovery time and message overhead for multicast. SPT refers to the shortest path tree algorithm. Note that we build multicast trees by first running the shortest path tree and then the set cover algorithm.}
	\label{fig: recovery-time-6}
	
\end{figure}

\noindent\textbf{Multicast}. In Section~\ref{subsubsec: eval-multicast}, we take the FatTree as an example to show the results of packet header size and the total volume of multicast traffic. Figure~\ref{fig:mcast-performance-appendix} illustrates all of our results in the three topologies. For packet header size, we can derive a straightforward conclusion that \sysname's Spanner algorithm reduces the source-routed packet header size by a significant amount. For evaluating the multicast traffic overhead, we measure how much traffic it generates in the whole topology to forward a $1500$B multicast packet. As described in Section~\ref{subsubsec: eval-multicast}, the results clearly show how \sysname's full recovery achieves comparable traffic overhead than Elmo's final solution and Orca, and Elmo's fast recovery attempt results in an overwhelmingly high traffic overhead. After all, Elmo's proposes to switch to the highly inefficient unicast as its fast recovery approach. Finally, we provide a decomposition of the convergence time of each \sysname's recovery algorithm in Figure~\ref{subfig:mcast-time-decompose}. We also illustrate \sysname's time-to-message-overhead results in Figure~\ref{subfig: mcast-message-overhead} to provide more intuitions of \sysname's multicast reaction algorithms.

\subsection{Deploying F10 into \sysname}
\label{appendix:f10}

To demonstrate more use cases, we have also implemented F10~\cite{liu2013f10}'s re-routing protocol into \sysname. F10 proposes a local-rerouting protocol as a fast recovery protocol to re-establish an ad-hoc route in place. When the local rerouting protocol fails to find a new route, it falls back to the pushback flow redirection protocol, which is essentially an asynchronous algorithm to redirect the affected upstream flows to new routes. Both protocols are designed specifically for their AB FatTree topology. We deploy their pushback flow redirection protocol to \sysname, and study on their local-rerouting protocol is out-of-scope of our paper.

Following the experimental setups elaborated in Section~\ref{subsec:exp-combined}, we evaluate the combined version of \sysname with our clock synchronization and multicast use cases. Again we compare the combined version against the solution that runs each application individually. The bandwidth for \sysname-individual is $200$Mbps, $10$Mbps, $10$Mbps for clock synchronization, multicast, and F10 respectively.

\begin{figure}[!htbp]
	\centering
	\vspace{-0mm}
	\includegraphics[width=0.28\textwidth]{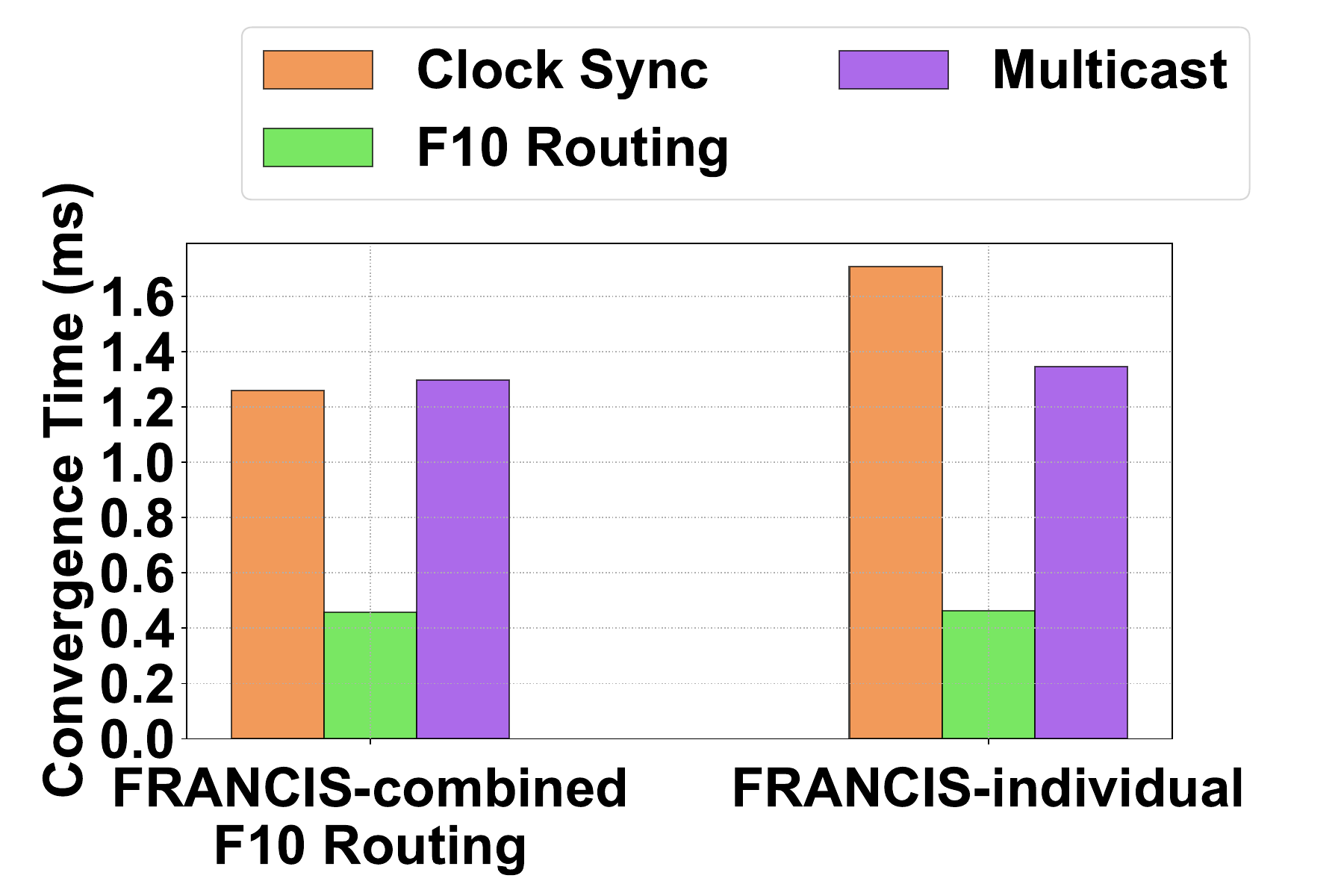}
	\caption{We integrate F10~\cite{liu2013f10} into \sysname as the routing protocol and combine it with reaction algorithms of both clock synchronization and multicast. The figure compares the combined \sysname with \sysname-individual to run each application individually.}
	\label{fig:combined-f10}
\end{figure}

Figure~\ref{fig:combined-f10} shows the results comparing both versions. One direct observation is that \sysname-combined achieves faster convergence for all applications than \sysname-individual. This is because our bandwidth sharing technique utilizes bandwidth resources of idle applications to the full, and we merge the same algorithm components from different applications together to avoid duplicated calculations. Another finding is that, regarding Figure~\ref{subfig:overall_w_Contra}, \sysname-combined with Contra achieves higher improvement for multicast application in convergence time than \sysname-combined with F10. Specifically, \sysname-combined with Contra reduces the convergence time of multicast to $1.187$ms, while \sysname-combined with F10 only reduces it to $1.298$ms. We note that Contra is higher in bandwidth complexity than F10. Therefore, the results suggest that combining multiple bandwidth-intensive protocols together achieves a better bandwidth utilization rate than combining one bandwidth-intensive protocol with non-intensive protocols and thus exhibits a faster convergence time.

\subsection{Evaluations on \sysname's flexibility and reliability}
\label{appendix:flex-reliable-exps}

\begin{table}[!htbp]
\centering
\small
\resizebox{\linewidth}{!}{
\begin{tabular}{|c|c|c|c|c|c|}
\hline\hline
\textbf{Number of trees} & 1 & 2 & 3 & 4 & 5\\ \hline
Convergence time (ms) & 1.687 & 1.743 & 1.805 & 1.859 & 1.973\\ \hline
Expected $\epsilon$ (ns) & 26.054 & 25.131 & 25.018 & 25.003 & $25.0005$ \\ \hline
\end{tabular}
}
\caption{Convergence time and $\epsilon$ varying the number of clock synchronization trees to be constructed in the FatTree topology. The bandwidth is limited as $10$Mbps.}\label{tab:clock-sync-param-settings}
\vspace{-0.2 in}
\end{table}

For the clock synchronization application, one critical parameter is the number of shortest path trees (SPTs) to be constructed. Constructing more SPTs takes longer convergence time but allows the leader election algorithm to yield better clock synchronization tree and thus lower time uncertainty bound ($\epsilon$). In Table~\ref{tab:clock-sync-param-settings}, we vary the number of SPTs from 1 to 5 evaluate both the convergence time and the converged $\epsilon$ value on the FatTree topology. We observe that with more trees constructed, the converged $\epsilon$ value reduces from expected $26.054$ns to an optimal $25$ns, at the cost of a slight increase of convergence time from $1.687$ms to $1.973$ms.

\begin{figure}[tb]
	\centering
 \vspace{0.2cm}
    \subfigure[Clock Synchronization]{
		\begin{minipage}[t]{0.45\linewidth}{
		\vspace{-0.00in}
		\begin{center}
		\includegraphics[width=\textwidth, ]{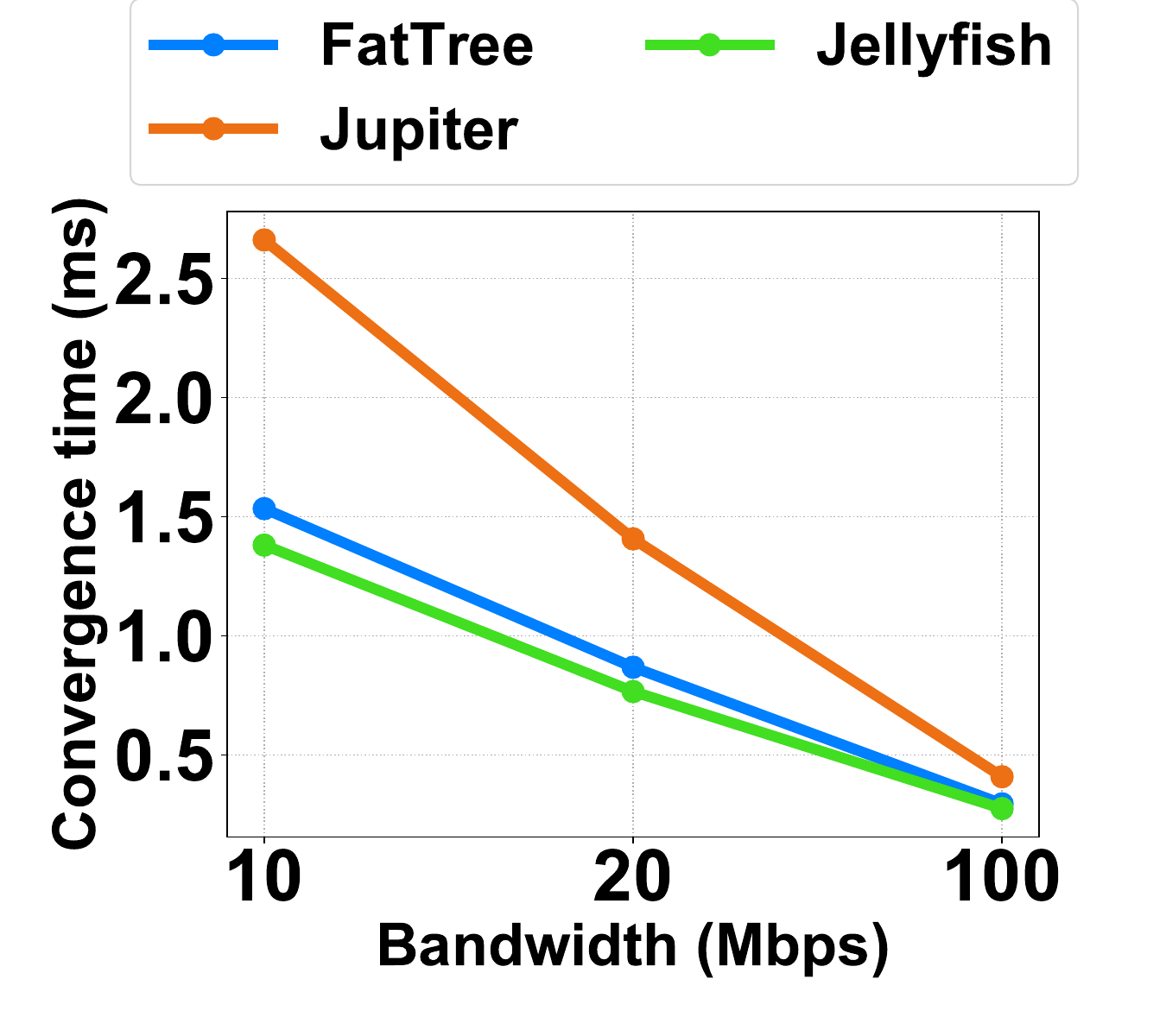}
		\end{center}
		}
		\label{subfig:clock_bandwidth_time}
		\end{minipage}
	}
    \subfigure[Multicast]{
		\begin{minipage}[t]{0.45\linewidth}{
		\vspace{-0.00in}
		\begin{center}
		\includegraphics[width=\textwidth, ]{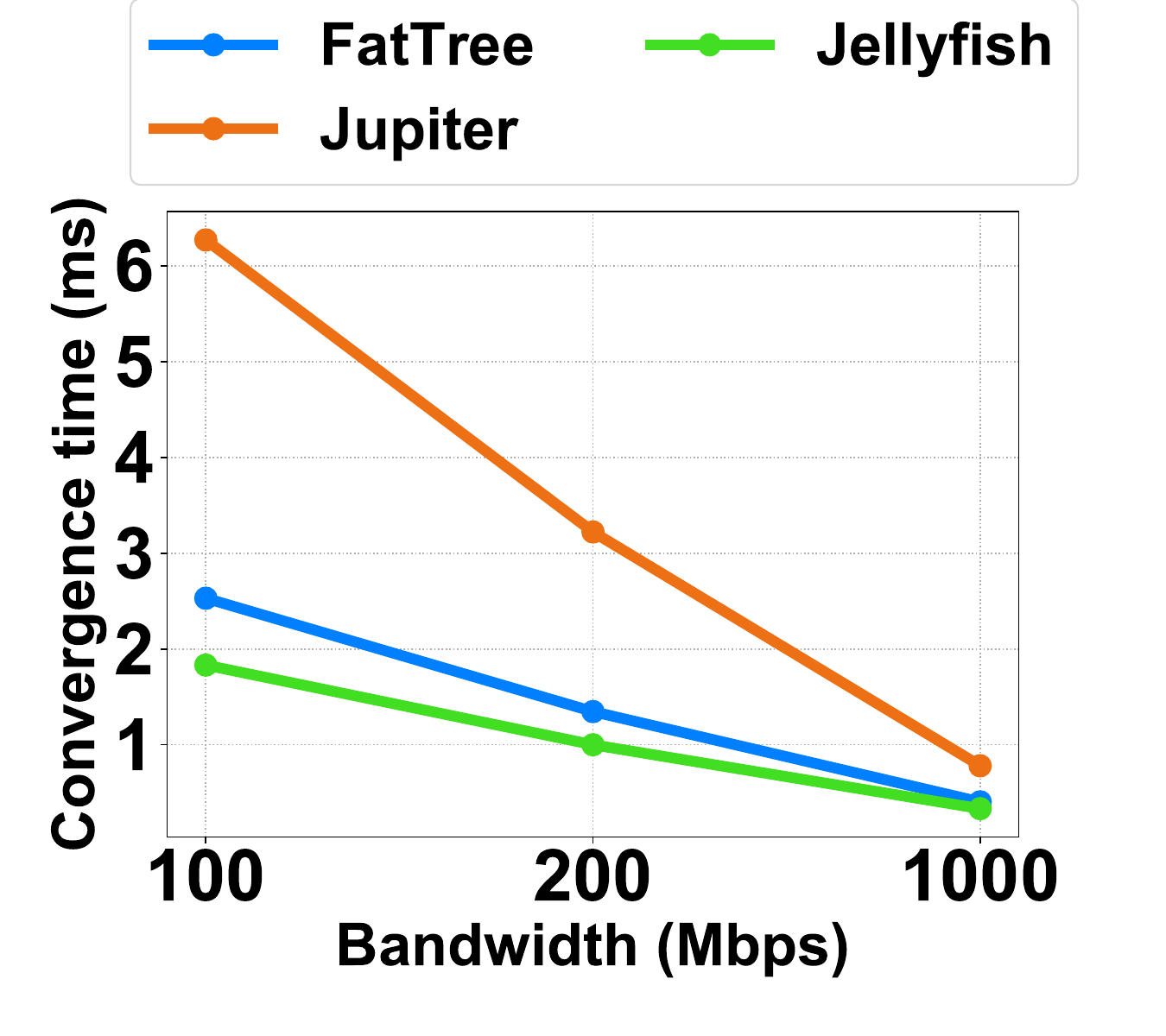}
		\end{center}
		}
		\label{subfig:mcast_bandwidth_time}
		\end{minipage}
	}
	\vspace{-0.1578975in}
	\caption{The convergence time with respect to different bandwidths for both clock synchronization and multicast.}
	\label{fig: bandwidth-to-convergence-time}
\end{figure}

\subp{Convergence time with respect to different bandwidth settings.} Next, we vary the bandwidth configuration to evaluate its impact on the convergence time. We range the bandwidth from $10$Mbps to $100$Mbps for the clock synchronzation application and $100$Mbps-$1000$Mbps for multicast. The results in Figure~\ref{fig: bandwidth-to-convergence-time} verify that the convergence speed (1$/$Convergence-time) grows proportional with respect to the bandwidth allocated to \sysname. 

\begin{figure}[!htbp]
	\centering
	\subfigure[Clock Synchronization]{
		\begin{minipage}[t]{0.48\linewidth}{
		\vspace{-0.00in}
		\begin{center}
		\includegraphics[width=\textwidth, ]{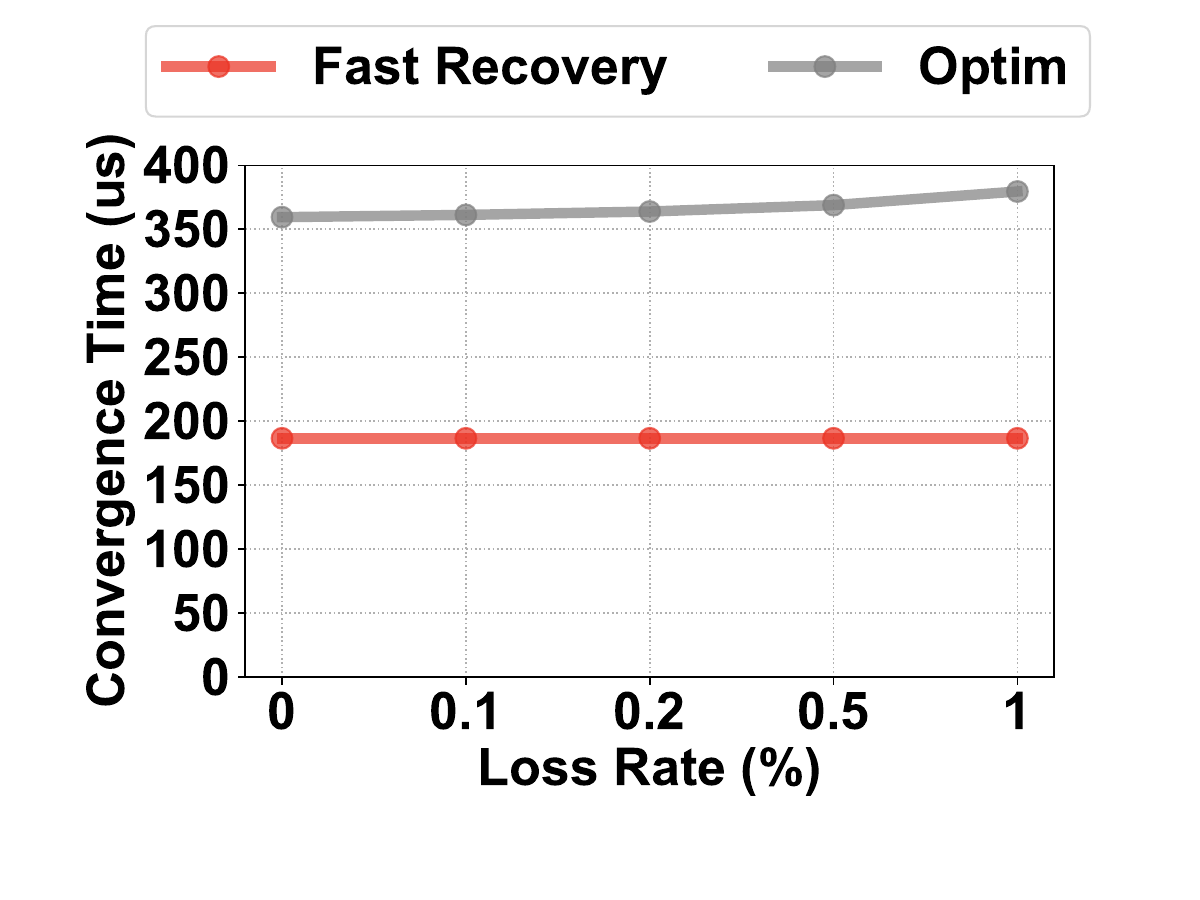}
		\end{center}
		\vspace{-0.06in}
		}
		\label{subfig:sync_epsilon_series_loss_rate}
		\end{minipage}
	}
	\hspace{-0.1in}
	\subfigure[Multicast]{
		\begin{minipage}[t]{0.48\linewidth}{
		\vspace{-0.00in}
		\begin{center}
		\includegraphics[width=\textwidth, ]{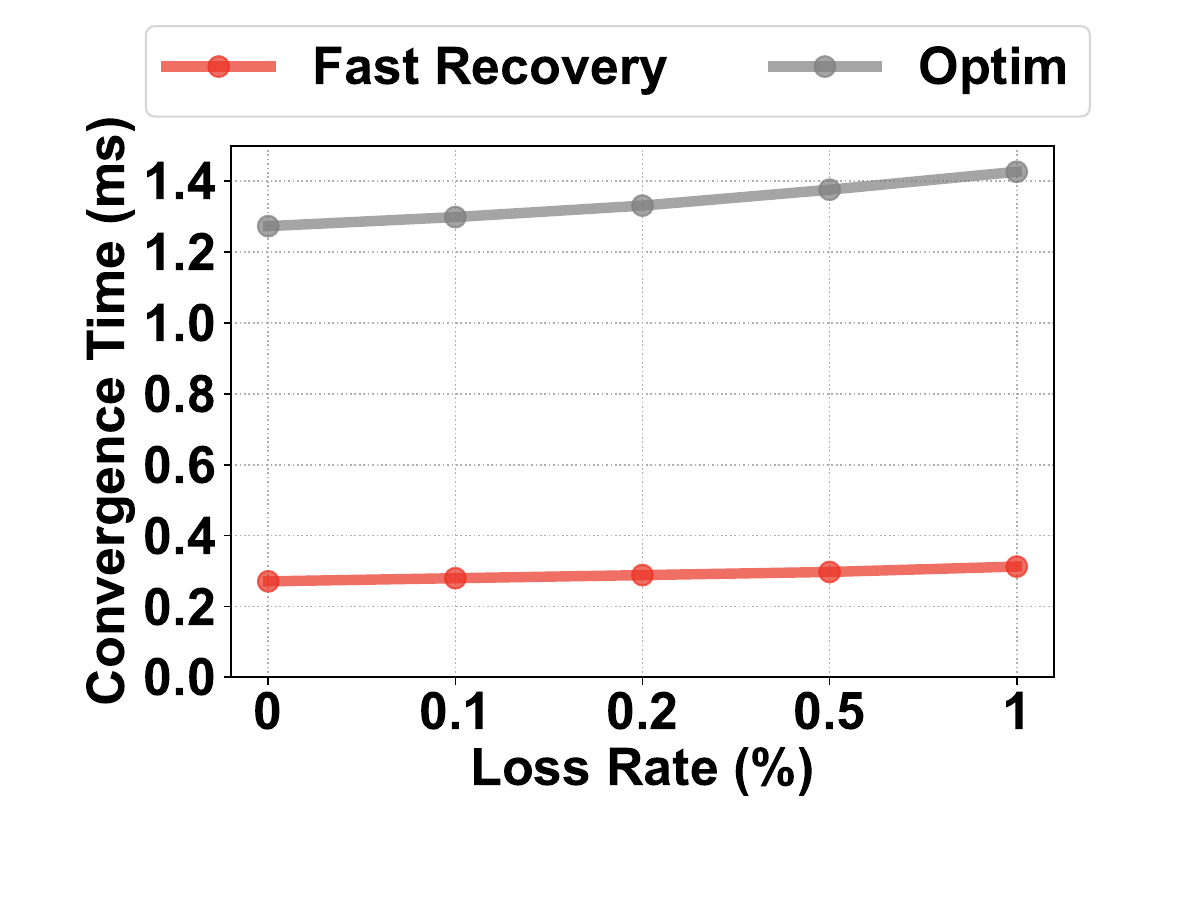}
		\end{center}
		\vspace{-0.06in}
		}
		\label{subfig:mcast_epsilon_series_loss_rate}
		\end{minipage}
	}
	\vspace{-0.172015in}
	\caption{The recovery time with respect to different loss rates. Optim denotes the convergence time after the completion of distributed optimization.}
	\label{fig:loss}
	
\end{figure}

\subp{\sysname ensures that the convergence time is hardly affected by packet losses.} To demonstrate \sysname's reliability with respect to packet losses, we evaluate if a large loss rate may lead to much slower convergence. The loss rate in data center networks is typically less than $0.1\%$ \Wenchen{Find out the relevant works regarding packet loss rate in the data center.}, but in our simulations, we further raise the loss rate to be $1\%$. As shown in Figure~\ref{fig:loss}, compared with an ideal lossless network, the convergence time with $1\%$ loss rate for both tasks inflates within $15\%$ for fast recovery and within $12\%$ for full recovery. We argue that a $1\%$ loss rate is rare in data centers, and such inflation is even more negligible for $<0.1\%$ loss rates. 

\section{Discussion: Controller Participation}
After the completion of the distributed optimization phase, we have obtained either an optimal or a near-optimal solution to recover user applications. The remaining tasks left are to clean up the memory and launch a centralized algorithm to obtain a better solution. Clean-ups of some temporary stateful objects can be done entirely in the data plane. Some other data structures in the data plane, however, should be kept in the switches’ memory to keep functioning after the distributed optimization until they are offloaded into the end hosts. Therefore, we propose to periodically launch the controller to do the management tasks such as offloading switches' stateful objects to hosts to prevent accumulation of memory consumption and executing a centralized algorithm to further optimize the system performance. For example, the controller may offload the multicast rules stored in tor switches back to end hosts; for clock synchronization, with a negligible $0.0025\%$ (Table~\ref{tab:clock-sync-param-settings}) probability that \sysname fails to find an optimal clock synchronization tree, the controller is responsible to compute an optimal one. For both tasks, we argue that the need of launching the controller happens much more infrequently than occurrences of \mbox{network events, adding minimal overhead}.

\end{document}